\documentclass[sigchi]{acmart}
\renewcommand\footnotetextcopyrightpermission[1]{} 

\acmSubmissionID{462}

\usepackage{booktabs} 
\long\def\ignorethis#1{}

\usepackage{amsmath}
\usepackage{mathptmx}
\usepackage{url}
\usepackage{xfrac}

\usepackage{calrsfs}
\DeclareMathAlphabet{\pazocal}{OMS}{zplm}{m}{n}

\usepackage{booktabs}

\usepackage{soul}
\soulregister\ref{7}
\soulregister\cite{7}
\soulregister\refFig{7}

\usepackage{color}
\definecolor{gray}{rgb}{0.35,0.35,0.35}
\definecolor{blue}{rgb}{0,0,1}
\definecolor{white}{rgb}{1,1,1}

\usepackage{booktabs}

\usepackage{epsfig}
\usepackage{epstopdf}
\usepackage{subfig}
\usepackage{multirow}
\usepackage{array}

\usepackage{xcolor}

\newbox\jsavebox
\newcommand{\jsubfig}[2]{%
	\sbox\jsavebox{#1}%
	\parbox[t]{\wd\jsavebox}{\centering\usebox\jsavebox\\#2}%
	}

\newcommand{\Norm       } [1] {{\left\| #1 \right\|}}
\newcommand{\Pnorm      } [2] {\Norm{#1}_{#2}}

\author{Matan Shoef}
\affiliation{%
  \institution{Tel-Aviv University}
}
\author{Sharon Fogel}
\affiliation{%
  \institution{Tel-Aviv University}
}
\author{Daniel Cohen-Or}
\affiliation{%
  \institution{Tel-Aviv University}
}
\renewcommand\shortauthors{Shoef et al.}
\usepackage[ruled]{algorithm2e} 

\SetAlFnt{\small}
\SetAlCapFnt{\small}
\SetAlCapNameFnt{\small}
\SetAlCapHSkip{0pt}
\acmJournal{TOG}

\setcopyright{none}

\RequirePackage{fancyhdr}
\begin{document}
\title{PointWise: An Unsupervised Point-wise Feature Learning Network}



%
%


\begin{abstract}
We present a novel approach to learning a point-wise, meaningful embedding for point-clouds in an unsupervised manner, through the use of neural-networks.

The domain of point-cloud processing via neural-networks is rapidly evolving, with novel architectures and applications frequently emerging. Within this field of research, the availability and plethora of unlabeled point-clouds as well as their possible applications make finding ways of characterizing this type of data appealing. Though significant advancement was achieved in the realm of unsupervised learning, its adaptation to the point-cloud representation is not trivial.

Previous research focuses on the embedding of entire point-clouds representing an object in a meaningful manner. We present a deep learning framework to learn point-wise description from a set of shapes without supervision.

Our approach leverages self-supervision to define a relevant loss function to learn rich per-point features. We train a neural-network with objectives based on context derived directly from the raw data, with no added annotation. We use local structures of point-clouds to incorporate geometric information into each point's latent representation. In addition to using local geometric information, we encourage adjacent points to have similar representations and vice-versa, creating a smoother, more descriptive representation.

We demonstrate the ability of our method to capture meaningful point-wise features through three applications. By clustering the learned embedding space, we perform unsupervised part-segmentation on point clouds. By calculating euclidean distance in the latent space we derive semantic point-analogies. Finally, by retrieving nearest-neighbors in our learned latent space we present meaningful point-correspondence within and among point-clouds.   
\end{abstract}

\maketitle

\section{Introduction}
\begin{figure}
\includegraphics[width=\columnwidth]{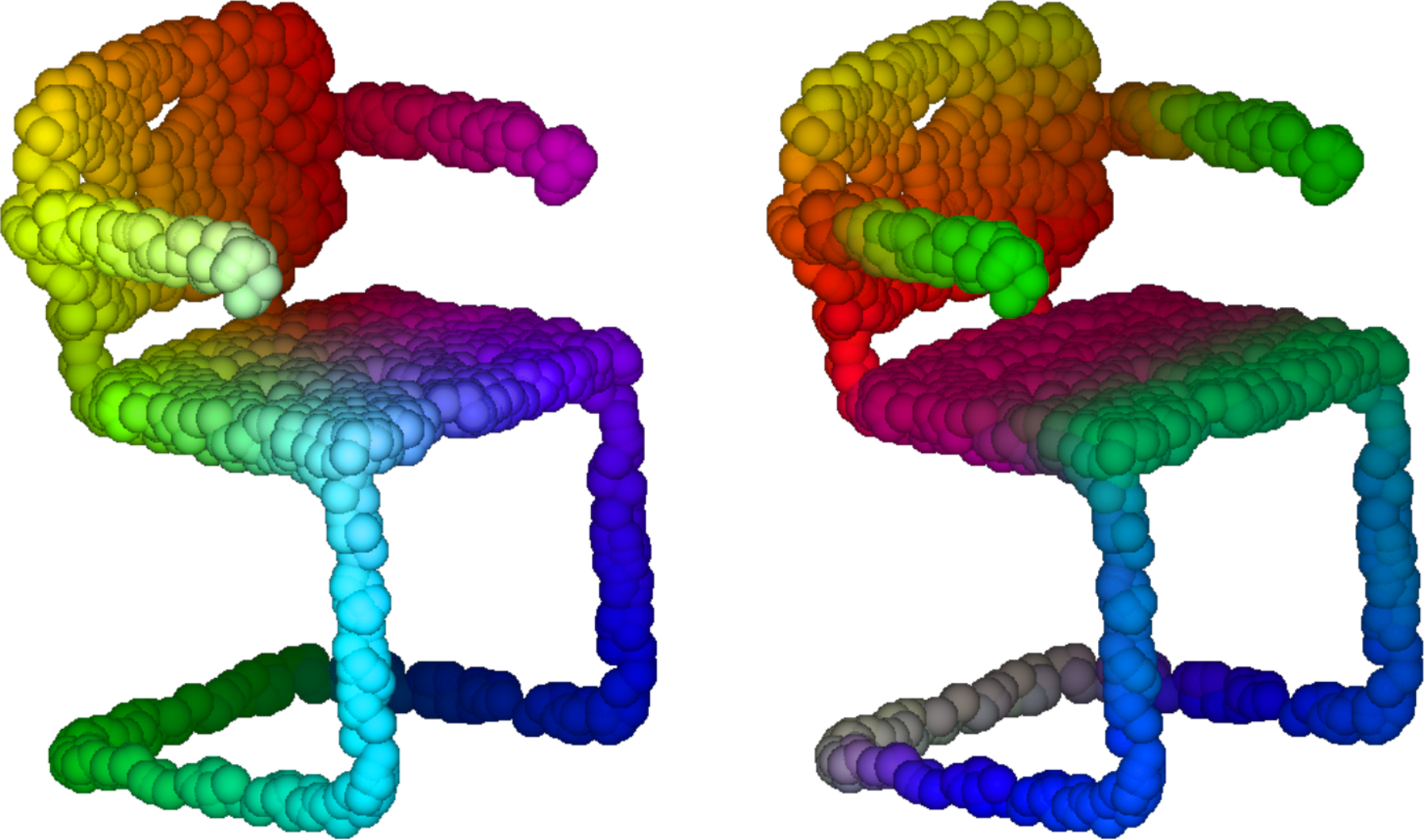}
\caption{Point-cloud colored according to point-coordinates (left) and 3D PCA values of the point-wise features (right). Each axis is normalized separately across the entire model and composes one RGB component.}
\label{fig:PCA_compare}
\end{figure}

The use of three-dimensional (3D) representations is ever-growing  in applications such as computer-vision, augmented-reality, autonomous-driving and many more. In this realm, point-clouds are an efficient representation from a learning perspective. 
They are compact, easy to manipulate and scale, and compose the output of many modern 3D scanning devices.  
The use of point-clouds, however, is technically challenging for several reasons. Points-clouds are set on an irregular grid, preventing the performance on regular convolution. They are also unordered, that is, a single given point-cloud consisting of \(N\) points can be represented by \(N!\) different permutations - all equally valid. Significant progress has been made in the field of direct-processing of point-clouds using neural networks, with architectures such as PointNet \cite{qi2017pointnet} and PointNet++\cite{qi2017pointnet++} that use global pooling operations that bypass the lack of order, with an architecture that transforms the points into a canonical regular configuration that allows performing convolutions \cite{li2018pointcnn}, or with architectures such as \cite{wang2018dynamic} that offer a generalization of the regular-grid convolution to the case of continuous grids. These networks are trained with supervision, i.e they are supplied with pre-annotated data consisting of man-made labels, from which a loss function is derived, yielding task-specific features.

As large as labeled datasets may be, the vast majority of available data remains unlabeled and so, cannot be utilized by the networks described above. In order to overcome this obstacle one may utilize unsupervised learning. Moreover, learning highly-descriptive features without supervision leads to more generic features that can be used in various applications, rather than being task-specific.

In this paper, we present a neural-network that learns point-wise, meaningful features with no supervision. We show that our learned point-wise features constitute semantic meaning and can therefore be used to achieve meaningful segmentation of the dataset.

Our approach leverages architectures which compute internal per-point features. 
The training is self-supervised, where the losses are derived from the geometry of the training set itself. 

To learn a meaningful latent space we base our losses on two main assumptions: 
The first is that to achieve a semantically meaningful embedding, the local geometry around each point should be taken into account. Therefore we incorporate reconstruction of local patches as part of the loss term.
The second assumption is that the latent space learned should be smooth, i.e., that generally nearby points in the original 3D space should also have a similar embedding. Distant points should have a different embedding in the latent space, unless they have similar local geometry.

We show that the point-wise features learned by our network are highly-descriptive, and carry more meaningful information than otherwise the bare coordinates themselves.
Figure \ref{fig:PCA_compare} shows an example of a chair colored according to the Cartesian coordinates and according to the 3D PCA of our learned representation. The learned representation is more descriptive and learns meaningful attributes of the chair such as embedding the two armrests in a similar manner, while still keeping a smooth transition between representation of close points.  

We demonstrate the competence of the point-wise features through several applications: 
We perform unsupervised part segmentation by applying clustering to our learned latent space, we show point-analogies inside a given model buy performing arithmetic operation on the latent space and we present semantic correspondence between points within a given model and among different models.

\section{Related Work}
\subsection{Neural Point-cloud processing}
In recent years, neural networks for point-cloud processing are rapidly emerging. PointNet \cite{qi2017pointnet} is a pioneering work, which presents an architecture that elegantly tackles the unordered nature of point-clouds. PointNet calculates  point-wise descriptors using a shared multi-layer perceptron (MLP), followed by a symmetric, channel-wise pooling function (specifically - max pooling) on all these descriptors simultaneously. In practice, the network first computes a feature vector for every point in a model based only on that point's coordinates and then uses a single, simultaneous pooling operation on all points' feature-vectors to produce one global feature-vector describing the entire cloud. For the purpose of segmentation the network later embeds the entire model's feature vector in that of each point. PointNet is robust to the irregularity of the data by default, as it does not directly take into account any relations between points. This architecture's main shortcoming lies in the fact that it does not apply any local filtering in the form of convolution or any other. This means that, when producing a single point's feature vector, PointNet takes a rather dichotomous approach of considering only that specific point's coordinates, and the structure of the entire cloud to which the point belongs.

More advanced solutions were later suggested.    
In PointNet++ \cite{qi2017pointnet++}, the authors apply PointNet recursively on neighboring sub-sets at various scales in order to produce valuable local information for varying model densities. In \cite{wang2018dynamic}, the autors introduce a new layer named EdgeConv to generalize the regular-grid convolution operator and capture local geometric features of point-clouds. The authors of \cite{li2018pointcnn} propose PointCNN, an architecture that aspires to bring neighboring point subsets to a canonical form before applying a standard weight-based convolution. \cite{atzmon2018point} finds a way to apply convolutional neural nets on point-clouds as well by  mapping point cloud functions to volumetric functions and vice-versa.

All the above suggest different methods of aggregating local-neighborhood based features into some reference point. These architectures are invariant to point-order and robust to data irregularity. 
They are all trained to perform classification and segmentation in a supervised manner, using a loss based on comparison between the estimated scores  and a given ground-truth. Consequently, they don't take advantage of most of the available data, which is unlabeled.

\subsection{Unsupervised feature learning}
Unsupervised feature learning is an attractive approach since it leverages the abundance of available, unlabeled data. To produce a semantically meaningful latent-space without relying on  annotations, networks are trained to perform tasks based on some context derived from the data itself. The assumption is that such tasks encourage the learning of rich, descriptive features. A notable use of such approach was presented by the authors of \cite{doersch2015unsupervised}. They train a  network to predict the relative positions of image patches in order to encourage the learning of semantically meaningful features. They show that with these features it is possible to cluster large, unlabeled datasets into classes.

In \cite{danon2018unsupervised}, the authors use a triplet margin loss on image patches to establish a latent space in which semantically similar patches are adjacent, and semantically different patches are not.

In the realm of point-clouds, \cite{achlioptas2018learning}, \cite{deng2018ppf}, \cite{yang2018foldingnet} and \cite{li2018so} train various architectures to learn an embedding for whole sets of points. That is, the point-cloud representation of an entire shape is embedded to a single vector in latent space. In particular, \cite{achlioptas2018learning} train a deep auto-encoder for whole point-clouds and utilize the learned latent-space for training several generative models composed of generative adversarial networks (GANs) and Gaussian mixture models (GMMs).
Similarly, \cite{sun2018pointgrow} train a network to generate new point clouds by training on existing point-clouds without annotation. In contrast to previous methods, we learn a point-wise representation. 

\cite{guerrero2018pcpnet} train a network to estimate local 3D shape geometric properties such as point-normals. In comparison we use local shape properties to achieve meaningful point-wise features. 
\section{Unsupervised Point-Wise Feature Learning}

\begin{figure}[t]
\includegraphics[width=\columnwidth]{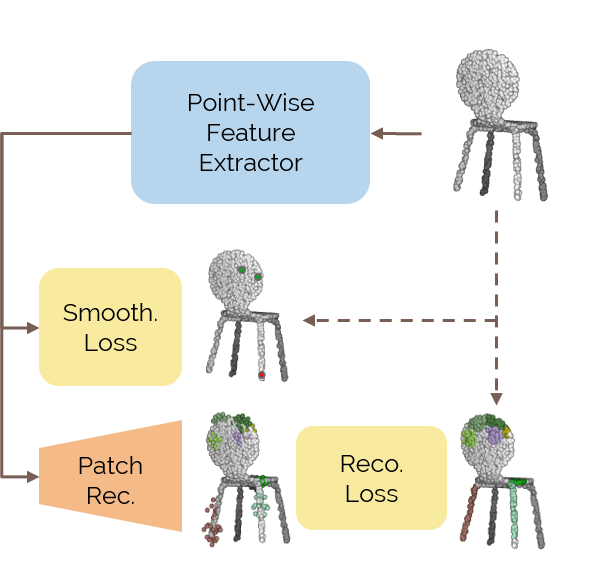}
\centering
\caption{An overview of our method. For each point in a set, point-wise features are extracted using a deep neural network. The features extracted are optimized by two loss terms: a reconstruction and a smoothness loss. We use an additional network in order to reconstruct a surrounding spatial patch out of each point's feature-vector. Smoothness is achieved through a triplet margin loss over the latent space  according to context derived from spatial coordinates.}
\label{fig:Architecture}
\end{figure}

The goal of this work is to learn a meaningful, unsupervised representation of each of the points that compose a given point-cloud.\\
In order to produce a descriptive latent-space in an unsupervised manner, our architecture includes a sense of "self-supervision" as a means to define a relevant loss over our extracted point-wise features. 
The chosen loss terms define a task that encourages the network to learn a semantically meaningful point-wise description. We formulate a loss which yields a latent space that is geometrically-descriptive on one hand, and context-aware on the other. 
We base the feature extraction module of our architecture on an existing network used for point cloud processing.
However, contrary to previous methods which use labeled data for training, we use unlabeled data.
To that end, we add layers to the network and change the training loss function. A schematic illustration of our architecture can be seen in Figure \ref{fig:Architecture}.
 
In the following section we will describe the network architecture and the different loss terms used.

Let us define some key notations. Given a point-cloud \({C_i}\) comprised of points\\ \(\left\{p_{i,j} : j=1,...,N\right\}\), we denote the latent-space descriptor of the point \(p_{i,j}\) by  \(f_{i,j}\). The set of all latent-space descriptors for the points in \(C_i\) is denoted as \(F_i\).  
Let $h$ denote the function that maps a single point $p_{i,j}$ to its associated descriptor: \(f_{i,j}=h\left(p_{i,j},C_i,\theta\right)\) where $\theta$ denotes the set of parameters characterizing the network. 
Let $H$ denote the mapping of an entire point set to its set of point-wise descriptors: \(F_{i}=H\left(C_{i},\theta\right)\).
\begin{figure}[t]
\includegraphics[width=\columnwidth]{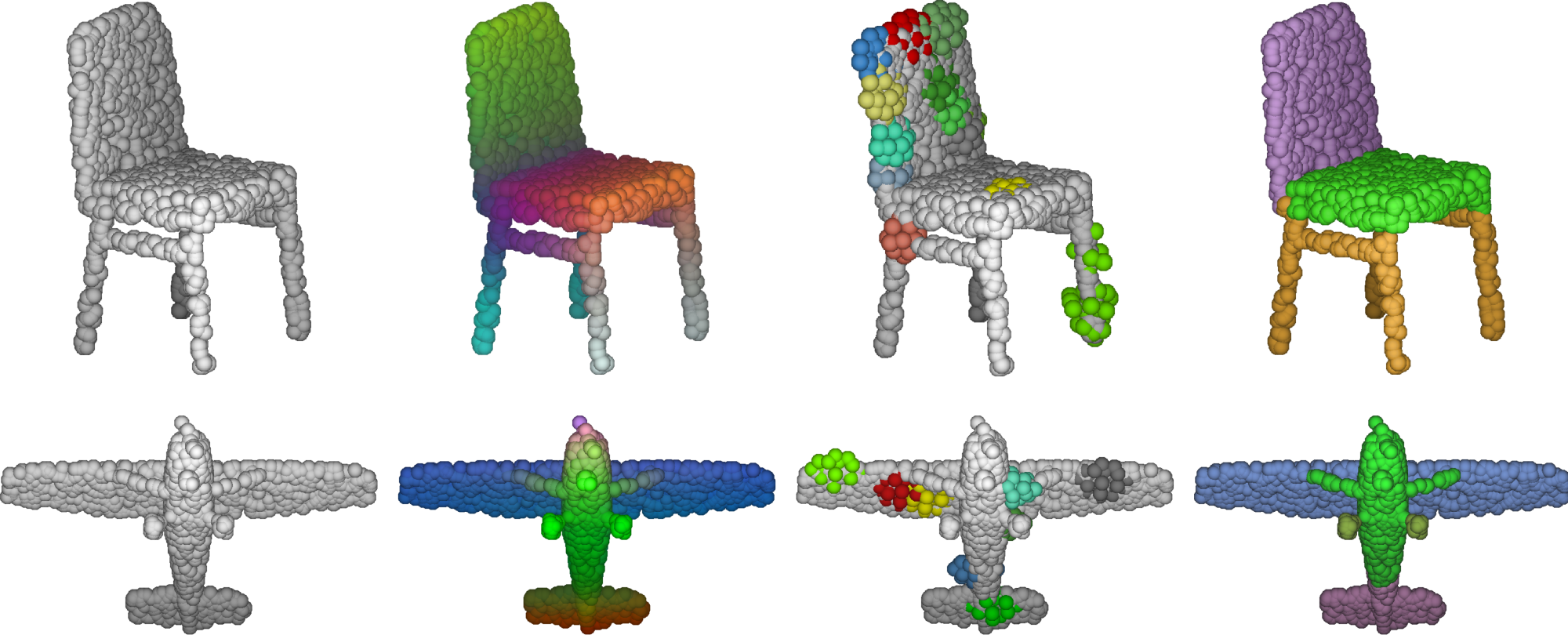}
\caption{An illustration of our results. From left to right: the input shape, the point-cloud colored according to our learned features, patch reconstruction and part-segmentation results.}
\label{fig:life_point_cloud}
\end{figure}
\subsection{Feature Extractor Network}

To achieve a feature representation which has a contextual meaning, the learned embedding of each point should not depend solely on that point's coordinates, but on its context in a broader sense. Therefore, the feature extractor network receives an entire point cloud (i.e. \({C_i}\)) as input and outputs a set of point-wise feature vectors (i.e. \(F_{i}=H\left(C_{i},\theta\right)\)) each of which is a function of both the corresponding point's coordinates and the coordinates of the entire set. 

In this paper, we use a modified, untrained version of the PointNet segmentation network \cite{qi2017pointnet} as a feature extractor network. However, other more complex networks such as PointNet++ \cite{qi2017pointnet++} or PointCNN \cite{li2018pointcnn} may be used as well.

The network is comprised of a point-wise feature extractor which is then used to create a single, global point-cloud feature. 
For a given input point-cloud, every point passes through an MLP that lifts its dimension to extract local features. next, a global, channel-wise pooling layer is applied on all point-wise feature vectors. The pooling layer's output passes two fully connected layers to produce a single global feature of the entire point-cloud. The global feature is then concatenated to all the point-wise features to produce new feature vectors which contain information about the global structure. The subsequent feature vectors pass through a second MLP to create the extracted point-wise features. Both the coordinates set and point features set undergo learned  transformations in the form of matrix multiplication aimed at creating a canonization of the respective space. 

\subsection{Loss Terms}

Our loss function consists of two components:
\begin{equation} 
\label{eq:loss_function}
L = \alpha L_{reconst} + \beta L_{smooth}  \\ 
\end{equation}
The first loss term encodes geometrically meaningful information into the latent space. It encourages the incorporation of information regarding the local environment of each point into the point's representation. The second term utilizes a triplet margin loss to create a similar embedding for close-by points in the original set and a different representation for distant points. Thus creating a smooth latent representation of the point-cloud. 

\begin{figure}[hbt]
\centering
\includegraphics[width=\columnwidth]{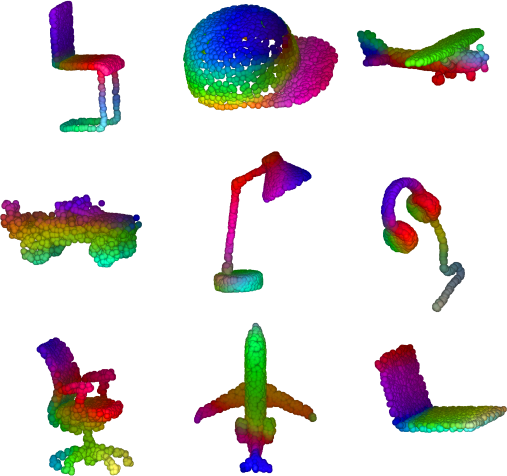}
\caption{Point-clouds colored according to the PCA values of the point-wise features. }
\label{fig:PCA}
\end{figure}

\begin{figure}[hbt]
\centering
\includegraphics[width=\columnwidth]{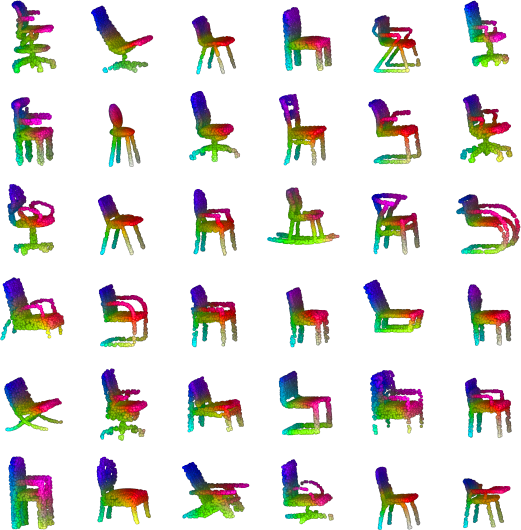}
\caption{Point-clouds colored according to the joint PCA values of the point-wise features of the entire chairs class in ShapeNet-part.}
\label{fig:PCA_chairs}
\end{figure}

Some visual examples of the embedding achieved by our method can be seen in Figures \ref{fig:PCA} and \ref{fig:PCA_chairs}. The figures show point-clouds colored according to the three dimensional PCA values for each point's feature-vector. Each RGB component is equivalent to the point-descriptor's projections on one PCA component and is normalized separately to values between 0 and 1. As can be seen in the figures, points with similar geometric meaning receive similar point-wise features and are therefore colored in a similar fashion in the image. In Figure \ref{fig:PCA_chairs} we show coloring of point-clouds of the same class according to point-wise features' projections on their jointly-computed PCA components. In this example, semantically similar parts share similar colors in all the different point-clouds, e.g. all the chair-arms are colored in magenta. This illustrates that similar parts are embedded closer together in the latent space.

\textbf{Reconstruction loss:} The key idea behind this loss term is the assumption that some of the semantically meaningful information regarding each point is derived from its local environment. Points of which the surrounding areas share similar geometries are more likely to have a similar semantic meanings. 
Therefore, if the feature representation of each point includes information about its local environment, the point-wise representation is prone to be more semantically meaningful. We include such information by using a reconstruction loss on the surrounding area of each point.

Specifically, from each feature representation in the latent space $(f_{i,j})$ we reconstruct a set of points surrounding the original point $(p_{i,j})$ as can be seen in Figure \ref{fig:PatchReconst}. 
Thus, the feature representation of each point must contain information that is relevant not only to that point's coordinates but to its surrounding area as well. The reconstruction loss is calculated using chamfer distance between the reconstructed environment $Rec\left(f_{i,j}\right)$ and the original environment $Env\left(p_{i,j}\right)$ as derived from the input itself:


\begin{flalign}
\label{eq:reconst loss}
\begin{split}
&L_{reconst}=\\
&\Sigma_{x\in{Rec\left(f_{i,j}\right)}}\min_{y\in{Env\left(p_{i,j}\right)}}\Pnorm{x - y}{2}^2+\Sigma_{y\in{Env\left(p_{i,j}\right)}}\min_{x\in{Rec\left(f_{i,j}\right)}}\Pnorm{x - y}{2}^2
\end{split}
\end{flalign}

The environment $Env\left(p_{i,j}\right)$  can be selected utilizing various methods. In this paper we chose to use kNN with euclidean distance as a metric. The term above is calculated for every reconstructed patch. the overall reconstruction loss is the average over all calculated distances.

\textbf{Smoothness loss:} This term regularizes the reconstruction loss and contributes to the smoothness of the latent space. The underlying assumption is that adjacent points in the original 3D space should have a similar embedding in the latent space, while distant points  in the original 3D space should usually have a different embedding in the latent space. This rule of thumb should be broken when the local geometry surrounding nearby points changes dramatically: in that case the points should have a different embedding in the latent space due to the reconstruction loss. Another exception to this assumption occurs for distant points for which the surroundings are very similar. This is solved in the same manner as the previous requirement.
To achieve an embedding which sustains these qualities we use a triplet margin loss to bring representations of adjacent points in the original 3D point-cloud closer together and drive representations of distant points farther apart. 
We use a variation of kNN - common-kNN: two points are considered a positive pair if the number of  overlapping members in both points' k nearest neighbors passes some pre-defined threshold, $m$. More formally, if $KNN_{i,j}$ is the set of k nearest points to point $p_{i,j}$, then $p_{i,j},p_{i,l}$ are common-kNN if:
$$
{kNN_{i,j}}\cap{kNN_{i,l}}\geq m
$$
The common-kNN requirement is stricter than the regular kNN and therefore prevents some of the outlier connections between points. 	
For each point in the point-cloud we randomly choose a set of $N$ triplets to be used in the triplet loss. We form each triplet by choosing a positive point out of the common-kNN and a negative point out of the $k_f$ farthest points randomly.
The positive pairs and the negative pairs for a point $r_i$ are defined as \(G=\{g_i^n : n=1,...,N\}\) and \(B=\{b_i^n : n=1,...,N\}\).
 The triplet loss is defined as follows:

\begin{align}
\label{eq:triplet loss}
\begin{split}
&L_{triplet}=\frac{1}{M} \Sigma_{i=1}^{M}{max\left\{d\left(r_i,g_i^n\right)-d\left(r_i,b_i^n\right)+margin,0\right\}} , \\
\end{split}
\end{align}
where:  $d\left(x,y\right)=\left\|x-y\right\|_p$ and $M$ is the number of points in the point-cloud.

As well as creating a smoother latent space representation along the shape, the smoothness loss also contributes to the convergence of the reconstruction loss. By requiring close points to have a similar embedding, we also encourage close points to yield similar reconstructed patches. Furthermore, by demanding distant points to have a different representation we encourage greater divergence in the latent space and therefore also in the reconstructed environment of each point. These two attributes prevent the reconstruction loss from converging into some local minimum and reconstructing the same generic patch for all points.

For an overview of our method's different phases, see Figure \ref{fig:life_point_cloud}. The figure shows the different outputs of the network - the extracted features, examples of reconstructed patches and part-segmentation results.
\section{Results and Evaluation}

\chapter{Results and Evaluation}
\label{chap:chapter 4}
\subsection{Implementation Details}
Our architecture is based heavily on the segmentation network proposed in \cite{qi2017pointnet} and follows the main principals and properties demonstrated in it in order to extract point-wise features for a given input model. One main difference lies in the global pooling operation, which, in our network, is comprised of maximum, mean an variance operations on the point-wise feature tensor, as opposed to the max-pooling applied in the original network. We found this modification leads to a smoother convergence of the training objectives. We also employ an additional MLP for the purpose of reconstructing a local patch from every point-descriptor.


We use a reconstruction loss weight of 100, and a smoothness loss weight of 0.1. The number of kNN used to create the ground truth patches for the reconstruction is set to 100. For each point we construct three triplets to be processed by the smoothness loss. For each triplet we randomly choose a positive example out of the corresponding point's 20 nearest neighbors and a negative example out of the 200 most distant points using euclidean distance as a metric. The smoothness loss' margin is set to 2. We trained the network for 3 epochs, with a learning rate of 0.001 and a decay of 0.7. We produce 100 dimensional descriptors. 
\subsection{Reconstruction Results}

\begin{figure}[hbt]
\centering
\includegraphics[width=\columnwidth]{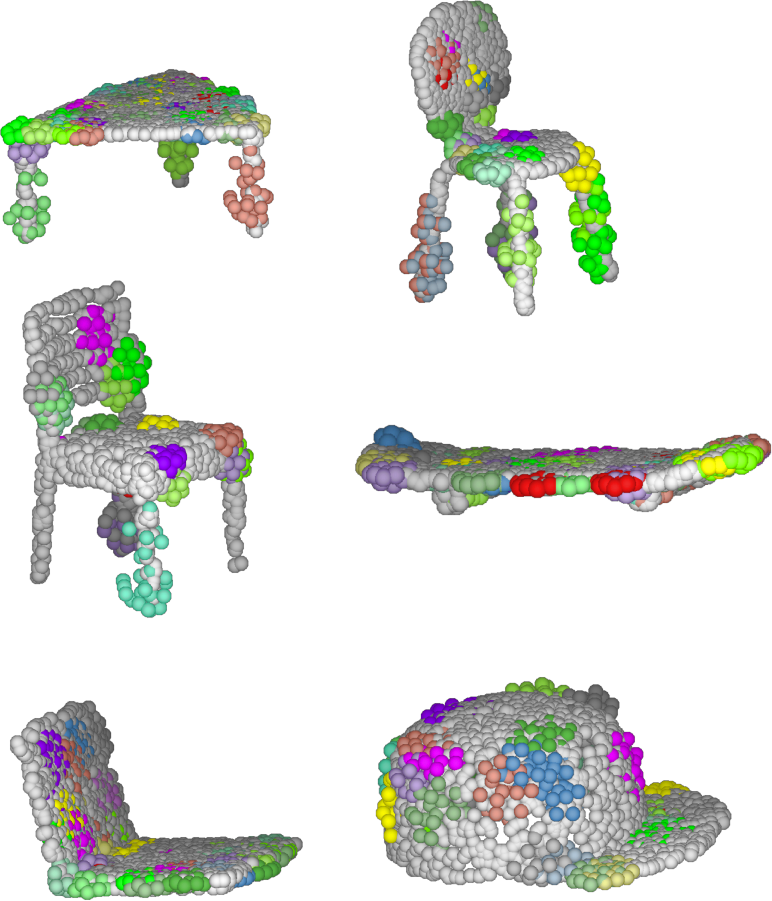}
\caption{Patch reconstruction results of our method.}
\label{fig:PatchReconst}
\end{figure}

In Figure \ref{fig:PatchReconst} we show some of the patches reconstructed by our network with the original point-cloud in the background for comparison. As can be seen in the figure, the reconstructed patches form various local structures existing in the original point-clouds.

\subsection{Segmentation Results}
To demonstrate the quality of the learned point-wise features, we utilize them to present unsupervised part-segmentation. The goal of this demonstration is to divide given shapes into meaningful parts.
Supervised methods formulate the problem of part-segmentation as a point-wise classification problem where the classes correspond to the different segments.
Since our method is completely unsupervised, instead of formulating the part-segmentation problem as a classification problem, we formulate it as a clustering problem where the number of segments in each object is the number of clusters.

We train our network on the entire ShapeNet-part dataset \cite{yi2016scalable}. We use the entire dataset for training rather than only one class at a time in order to allow the network to generalize better to the different parts of the objects. 

There are two possible approaches to performing the segmentation. The first is co-segmentation, where the segmentation is performed on an entire class of objects simultaneously. The second is segmenting each object individually. We chose to follow the latter. For a given number of segments, we use a grid search of parameters for spectral clustering and choose the final segmentation according to the clustering that achieves the highest silhouette indexing score:
\begin{equation}
S = \frac{1}{N}\Sigma_{n=1}^{N}{\frac{b_i-a_i}{max\left(a_i, b_i\right)}}
\end{equation}
where $a_i$ is the average distance from all other point features in the same segment and $b_i$ is the smallest average distance from all point features in each of the other segments. We compute this score twice, using distances in the latent space and in the Cartesian coordinate space. We average the two scores to produce our chosen segmentation. 


As we are not aware of any other methods which extract point-wise features without any supervision, we compare our results to a naive approach - performing the segmentation using each points' coordinates as features.

\subsubsection{Qualitative Evaluation}

\begin{figure}[h]
\centering
\includegraphics[width=\columnwidth]{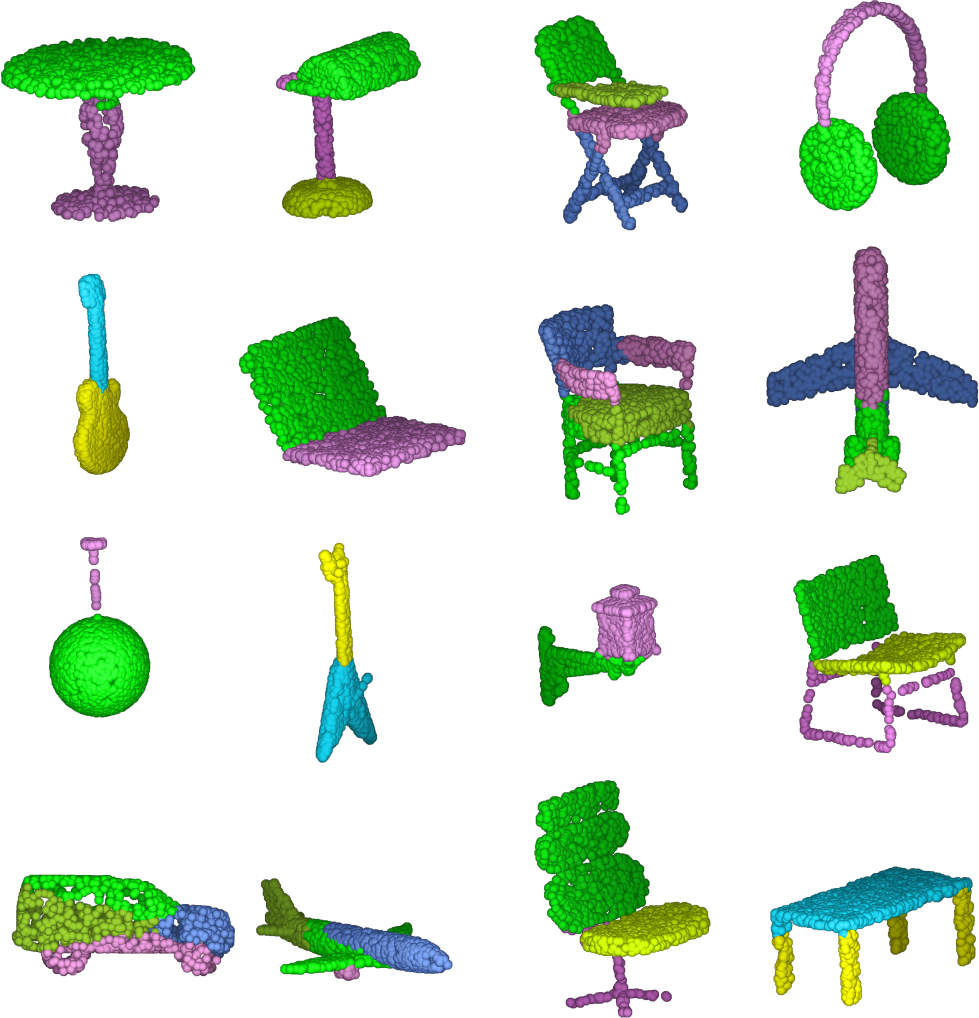}
\caption{Part segmentation results using spectral clustering on our point-wise features}
\label{fig:SegCollection}
\end{figure}

\begin{figure*}[h]
\centering
\includegraphics[width=0.9\textwidth]{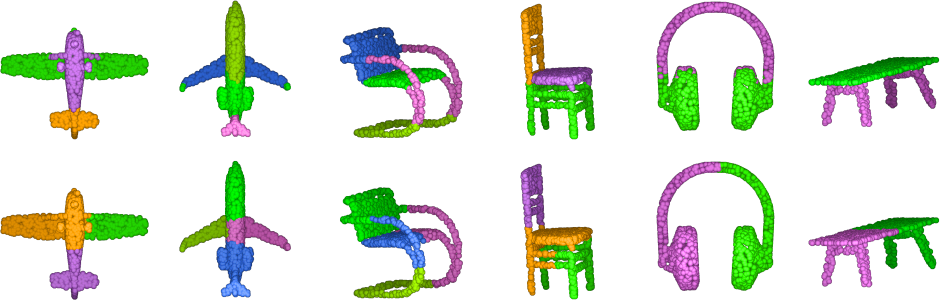}
\caption{Segmentation results on the point-wise features learned by our network (top row) and on point coordinates (bottom row).}
\label{fig:SegResult}
\end{figure*}

Some qualitative examples of our segmentation results can be seen in Figure \ref{fig:SegCollection}.
Appendix \ref{seg_appendix} further demonstrates some of our unsupervised segmentation results, this time performed on models taken from the training data.
In Figure \ref{fig:SegResult} we present a qualitative comparison between segmentation results achieved by clustering over our learned point-wise features and over the points coordinates. 
The objects shown in the figure have not been seen by the network during training.
It is clearly shown that the segmentation using our method achieves far better results than the segmentation applied on the points coordinates alone. The network, for example, learns the symmetry for each object: the wings of an airplane or the armrests of a chair are segmented together unlike in the segmentation performed on the coordinates alone. 

\subsubsection{Quantitative Evaluation}

The average accuracy for part segmentation using the features extracted by our method is 70.1\% compared to the average accuracy using the Cartesian cooardinates which is 61.2\%.
Most of the resulting improvement in our method is reflected by changes in the segmentation of small parts such as armrests or headphone speakers. Therefore the improvement is seen more clearly when viewing the objects than through the accuracy results. 

\begin{figure}[hbt]
\centering
\includegraphics[width=0.6\columnwidth]{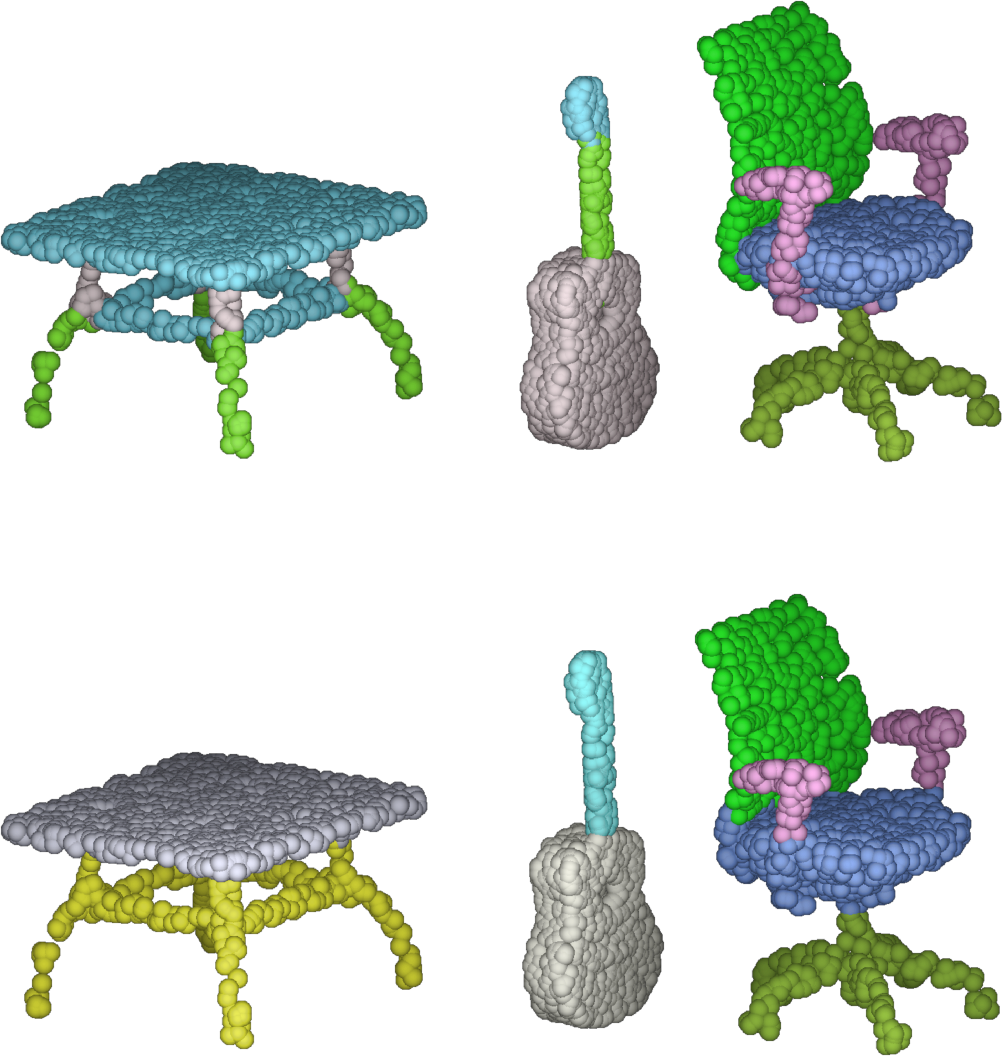}
\caption{The first row shows ground truth part-segmentation and the second row shows part-segmentation using our point-wise features. Even though the part-segmentation using the features and the ground truth are different, they are both reasonable manners of segmenting of the objects.}
\label{fig:SegFewWays}
\end{figure}

Some of the classes in ShapeNet can be semantically segmented in several different fashions. For example all four legs of a chair can be segmented into one part or they could be segmented into two parts - front and back legs. Both divisions make sense. Some of the degradation in accuracy results stems from such cases where the match between the ground truth segmentation and the point-wise features segmentation is poor, however, it does not necessarily mean the resulting segmentation is incorrect. A few examples where this occurs are shown in Figure \ref{fig:SegFewWays}.

For the part-segmentation results we used the number of segments each object contains according to its ground-truth division as an input to the clustering algorithm. For two of the classes in ShapeNet-part we used a different number of segments from that of the ground truth - tables and guitars, where we used two segments instead of three. The reason is that the third segment in each of them can be considered as part of another segment. Especially when the training is performed without supervision. Examples of objects from these two classes with three segments and our segmentation into two segments can be seen in Figure \ref{fig:SegFewWays}.

\subsection{Point Analogies}
\begin{figure}[hbt]
\centering
\includegraphics[width=\columnwidth]{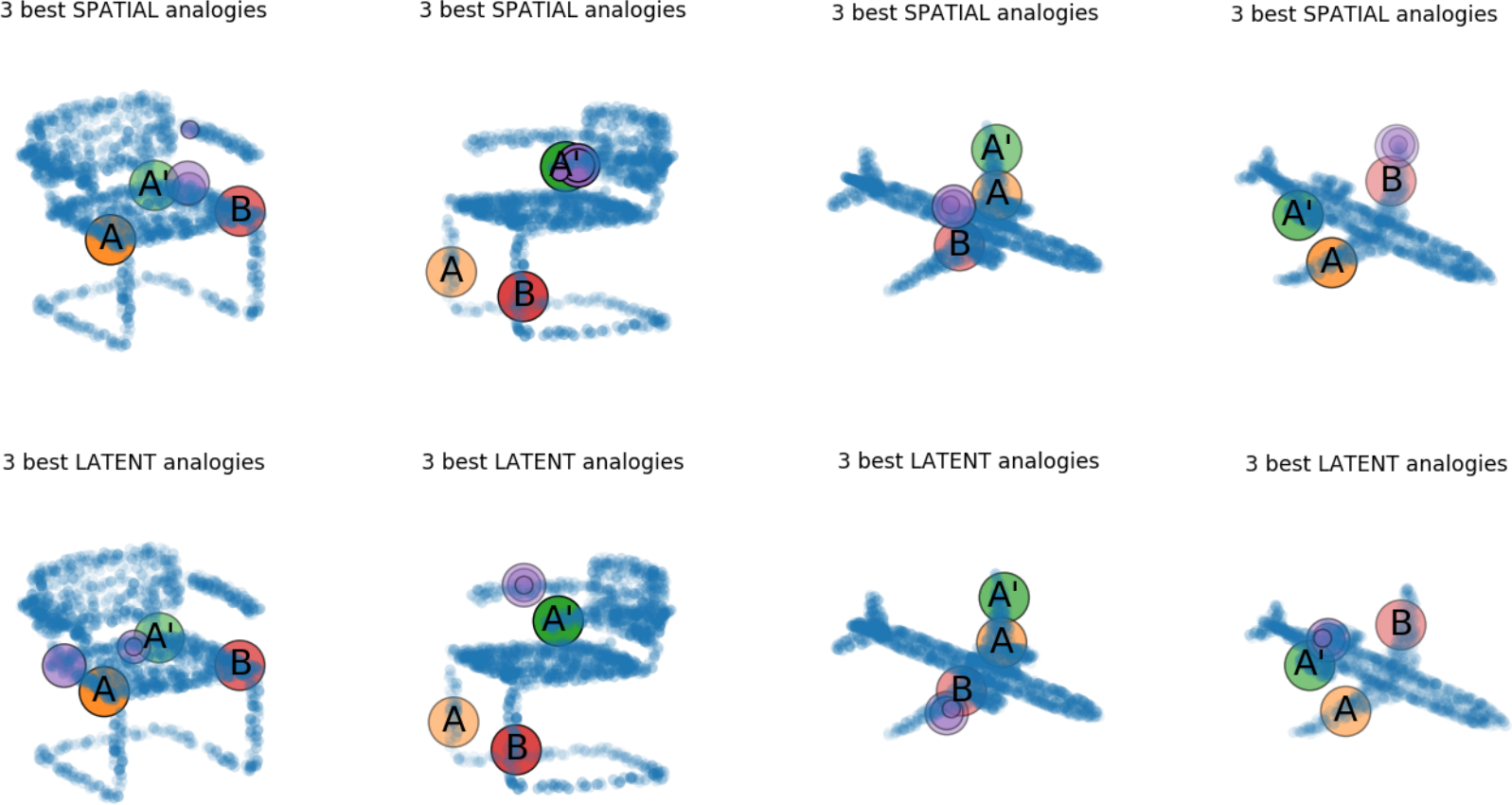}
\caption{We demonstrate the semantic nature of our learned latent-space via finding point analogies. We find the points that relate to $B$ the same way in which $A'$ relates to $A$. The best three matches are presented (in purple). The marker size indicates the respective point's rank among the calculated descriptor's nearest neighbors (bigger is closer). The analogies calculated using our learned features (bottom) are compared to those calculated using point coordinates (top).}
\label{fig:PointAnalogies}
\end{figure}

In order to demonstrate the semantic nature of our learned latent-space, we find point analogies via the performance of arithmetic operations directly on point descriptors. Consider a given point-cloud $C$. For a given pair of points: $A,A'$, and a reference point: $B$, we produce a fourth point: $B'$, that corresponds to $B$ in the same manner $A'$ corresponds to $A$. Under the assumption that our embedding indeed encapsulates semantic properties, this analogy can be retrieved via the performance of simple arithmetic operations. The semantic relationship between points in the spatial domain translates to the euclidean difference between their descriptors in the latent-space. We therefore add the difference between the descriptors of $A'$ and $A$ to the discriptor of $B$ in  order to find the latent representation of $B'$, i.e.
\begin{equation}
h\left(B',C,\theta\right)=h\left(B,C,\theta\right)+\left(h\left(A',C,\theta\right)-h\left(A,C,\theta\right)\right).
\end{equation}
In practice, we calculate the theoretical descriptor and find its nearest neighbors among the existing set of point-wise descriptors of all the points in $C$. For comparison, we conduct the same procedure using the spatial coordinates only.
Some examples of our retrieved analogies are presented in Figure \ref{fig:PointAnalogies}. Note how the spatial relation between $B'$ and $B$ may be completely different than that of $A'$ and $A'$, and yet the semantic relation is preserved. For example, we find points that answer to definitions such as "a point on the armrest on the opposite side from the selected leg" or "a point further toward the edge of the same wing", rather than the definition "a point higher and to the left".

\subsection{Point Correspondence}
\begin{figure}[hbt]
\centering
\includegraphics[width=\columnwidth]{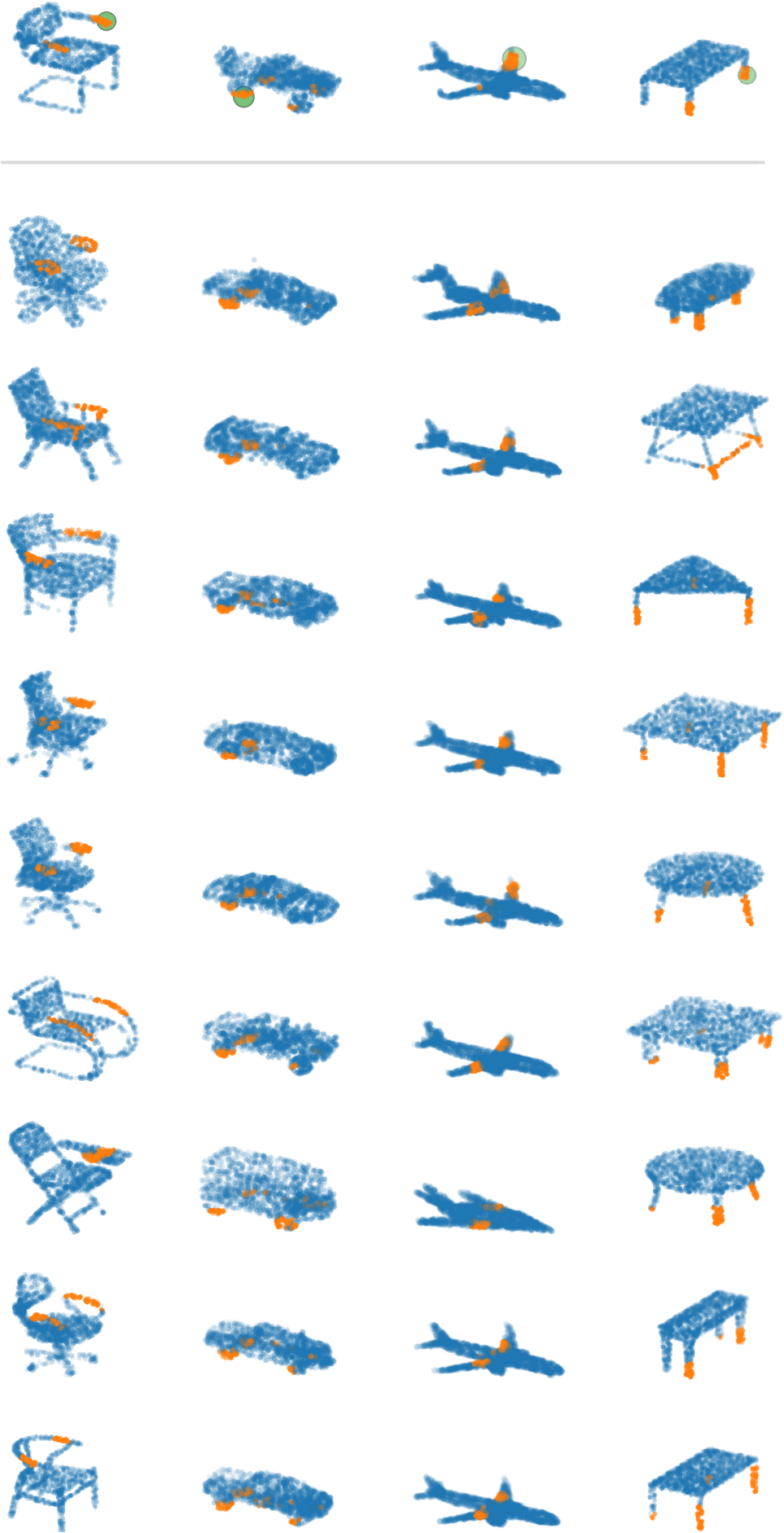}
\centering
\caption{Semantic point-correspondence for a given reference point (marked in green) may be achieved by simply retrieving its K nearest neighbors in our learned latent-space. This process is applicable both within the reference point's point-cloud (top row), and across different point-clouds (bottom rows).}
\label{fig:PointCorrespondence}
\end{figure}

We further demonstrate our network's ability to learn a meaningful point-wise embedding by presenting both inter-set in intra-set point correspondence. For a given reference point within a given point-cloud, we retrieve the subset of its nearest-neighbors in the latent space. We conduct this operation both on the point-cloud containing the reference point and on other point-clouds of objects of the same class. Figure \ref{fig:PointCorrespondence} includes some examples of this application. Note how the reference point's nearest-neighbors in the latent representation share a semantic meaning rather than spatial proximity. This is true when examining the neighbors within the cloud containing the reference point as well as those originating in other clouds. 

\section{Conclusion}
We present a method for extracting a meaningful point-wise representation of point-clouds.
Unlike previous works, our framework requires no access to labeling of either the individual points or the entire point-clouds.
The key idea is defining training-objectives that allow deep-neural networks to learn descriptive features using only self-supervision derived directly from the raw data. We use two types of loss terms which, when combined, enable the network to learn a semantically meaningful point-wise representation.
The first is a local patch reconstruction loss which is guided by the notion that the structure of the local environment surrounding each point plays an important role in characterizing it.
The second loss term relies on the idea that adjacent points in the 3D point-cloud should have a similar representation while distant points will usually have a different representation. By incorporating this triplet loss into the loss term the transition between the embedding of close points becomes smoother while the distinction between distant point's features leads to a more diverse latent space.
We have demonstrated the semantic nature of our learned embedding through three applications: We perform point-cloud part-segmentation by clustering the point-wise descriptors produced by our network; We retrieve semantic point-analogies by performing arithmetic operations on our point-wise descriptors; We perform semantic point-correspondence by finding a given point's nearest neighbors in our learned latent-space.
Through these application we emphasize the fact that euclidean proximity between point-wise descriptors in our learned latent-space tends to imply a semantic  similarity rather than a spatial one.
This framework enables the analysis of three-dimensional point-clouds, utilizing the large and ever-growing available bodies of unlabeled data.

Other applications of our work still remain unexplored. One of the framework's products is a decoder meant to reconstruct local "point patches" from each of the produced point-wise feature vectors. This module opens a window to various applications such as shape-completion and consolidation.

Our framework is demonstrated here using a very basic feature extractor as presented in \cite{qi2017pointnet}, but can be implemented using any architecture aimed at extracting point-wise features from point-clouds. Our feature extractor encompasses the inherent limitation of very limited local-awareness. It is dichotomous by design, and does not take into consideration local structures of any scale but a single point. This property confines the network's ability to produce and aggregate relevant information to be embedded in the point-wise features.
We predict that incorporating a more advanced neural-network with local-environment convolution ability - such as presented in \cite{li2018pointcnn} or \cite{wang2018dynamic} - into out framework will produce an even richer, more descriptive embedding.

Future development of the presented concept may include adaptation of our framework to the domain of two-dimensional images. By utilizing an architecture fitting for pixel-wise feature extraction (e.g. any CNN designed for pixel-wise image segmentation) and reconstructing image patches rather than point patches, one may be able to learn a rich and descriptive pixel-wise embedding in an unsupervised manner. Such an embedding may open a window to interesting application as unsupervised image segmentation, pixel-wise correspondence and more.

\newpage
\bibliographystyle{ACM-Reference-Format}
\bibliography{point_clouds}
\newpage

\appendix
\section{Appendix - Segmentation results on samples from the training data}
\label{seg_appendix}

We present various examples of unsupervised part segmentation performed on models taken from the training data.



\begin{figure*}
\begin{center}
	\jsubfig{\includegraphics[height=3.49cm]{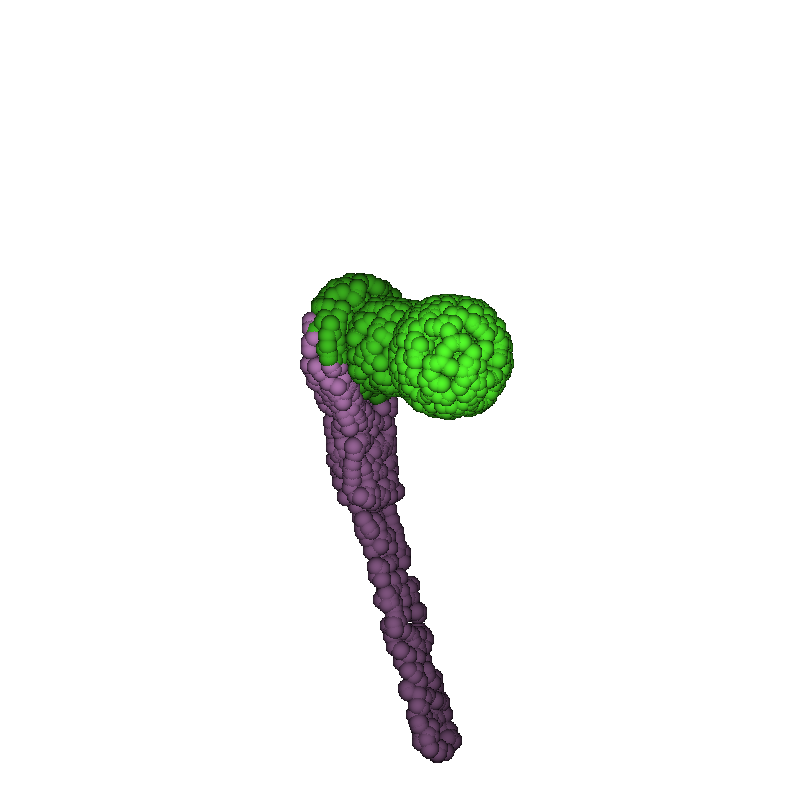}}
	{}%
	\hfill%
	\jsubfig{\includegraphics[height=3.49cm]{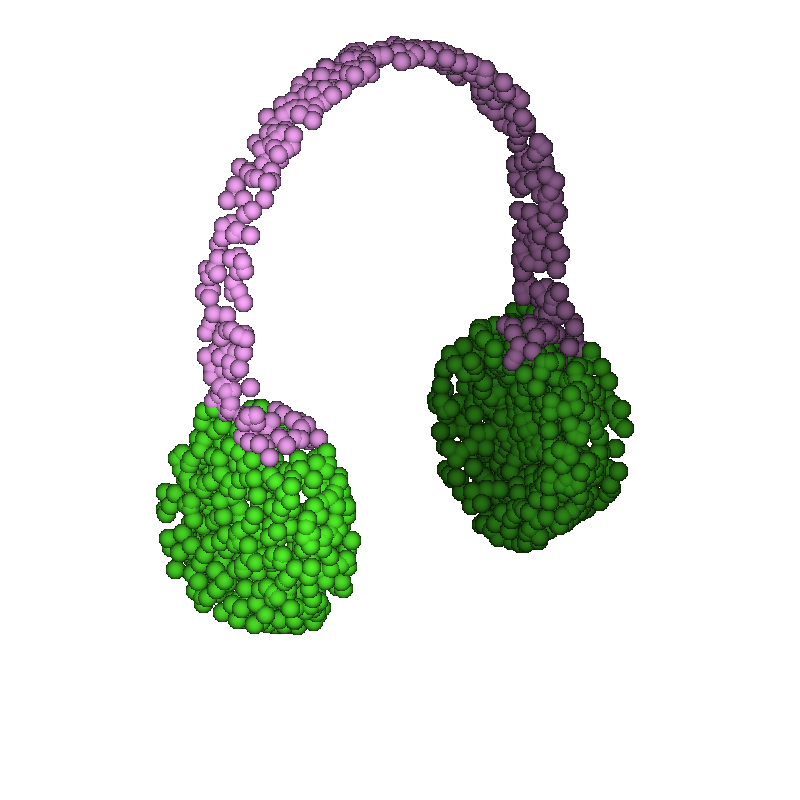}}
	{}%
	\hfill%
	\jsubfig{\includegraphics[height=3.49cm]{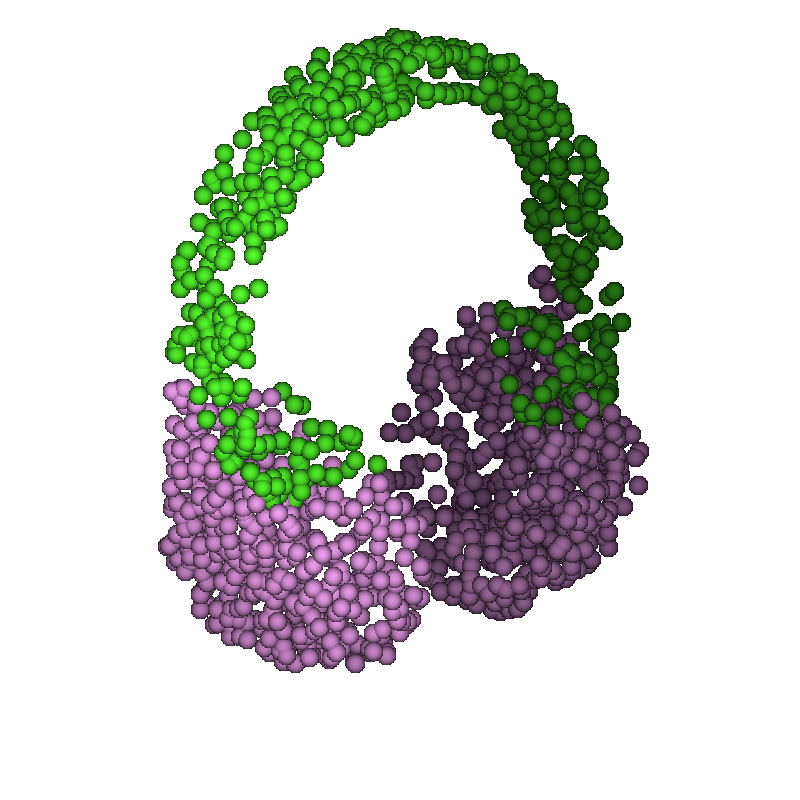}}
	{}%
	\hfill%
	\jsubfig{\includegraphics[height=3.49cm]{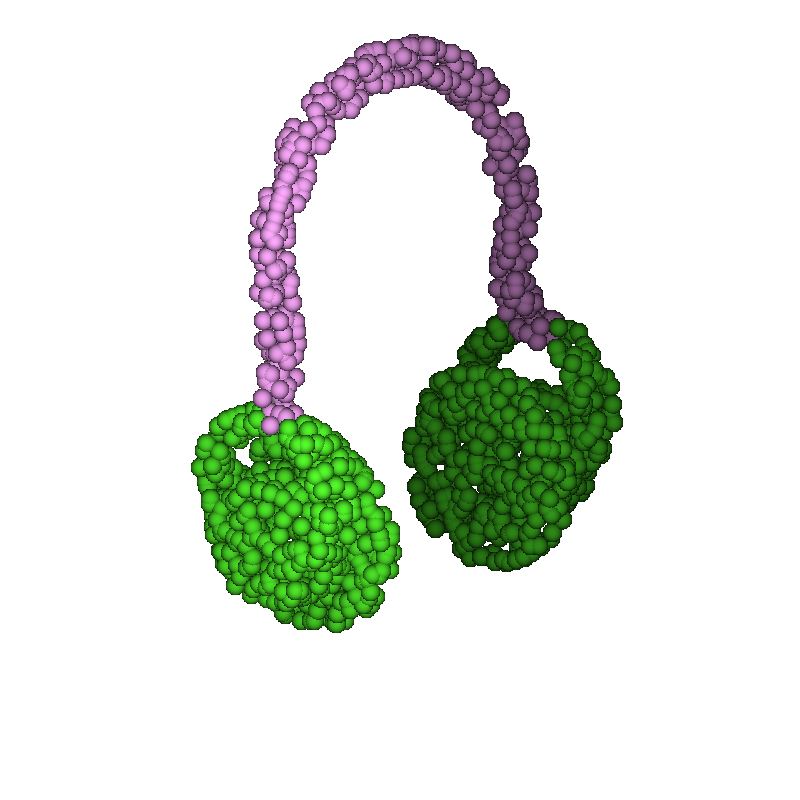}}
	{}%
\\
\vspace{6pt}
	\jsubfig{\includegraphics[height=3.49cm]{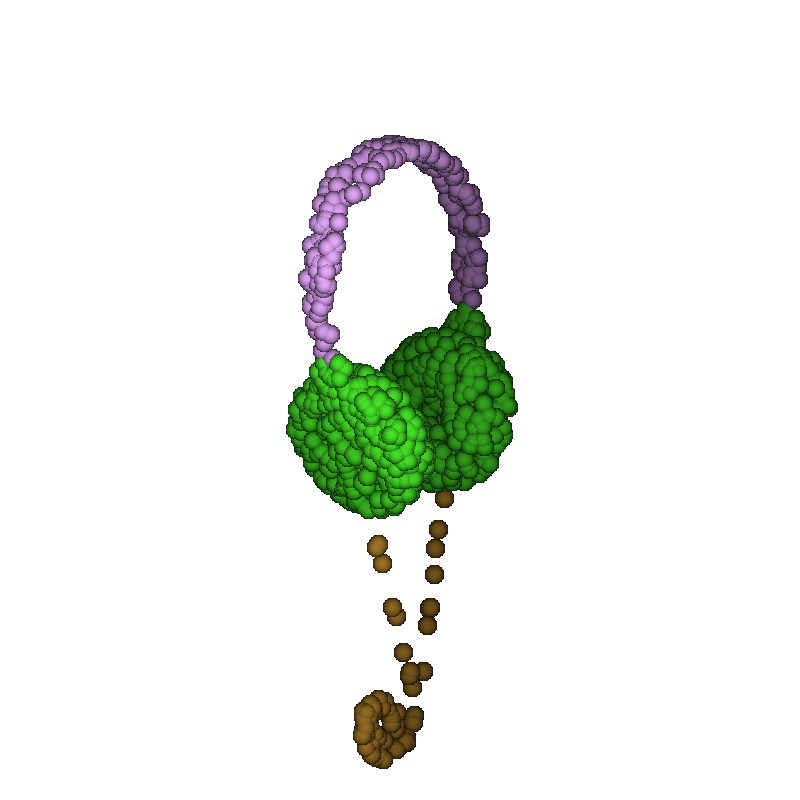}}
	{}%
	\hfill%
	\jsubfig{\includegraphics[height=3.49cm]{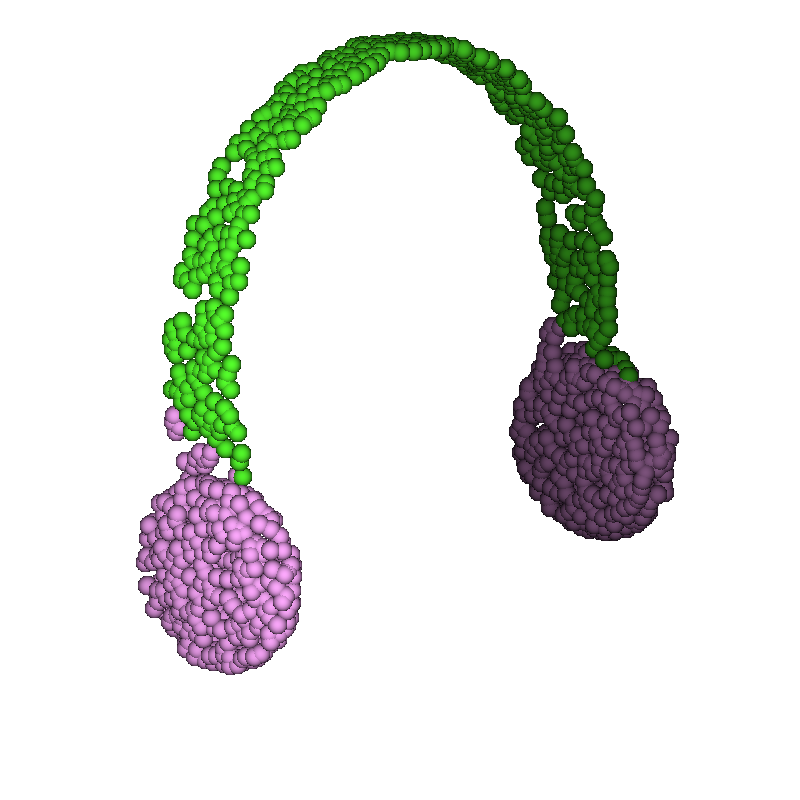}}
	{}%
	\hfill%
	\jsubfig{\includegraphics[height=3.49cm]{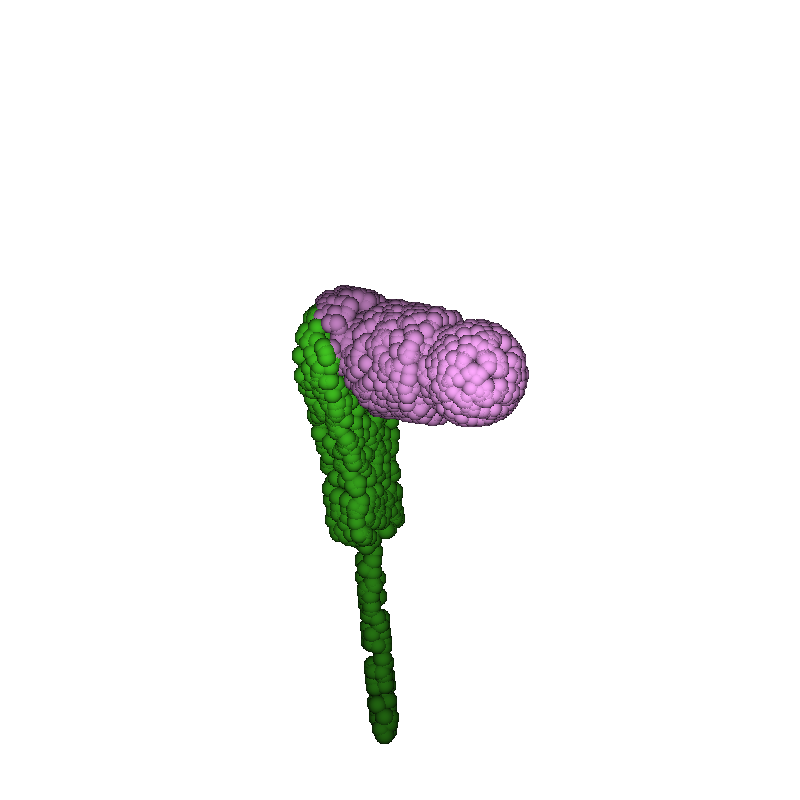}}
	{}%
	\hfill%
	\jsubfig{\includegraphics[height=3.49cm]{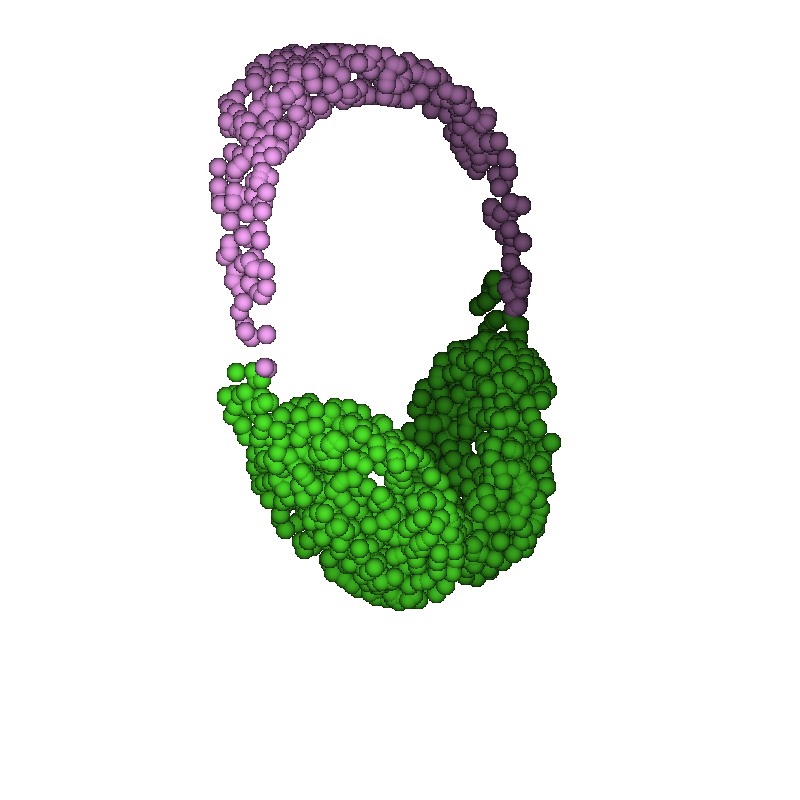}}
	{}%
\end{center}
\caption{Unsupervised segmentation examples on earphones taken from the training data.}

\label{fig:seg_train_earphones}
\end{figure*}

\begin{figure*}[h]
\begin{center}
	\jsubfig{\includegraphics[height=3.49cm]{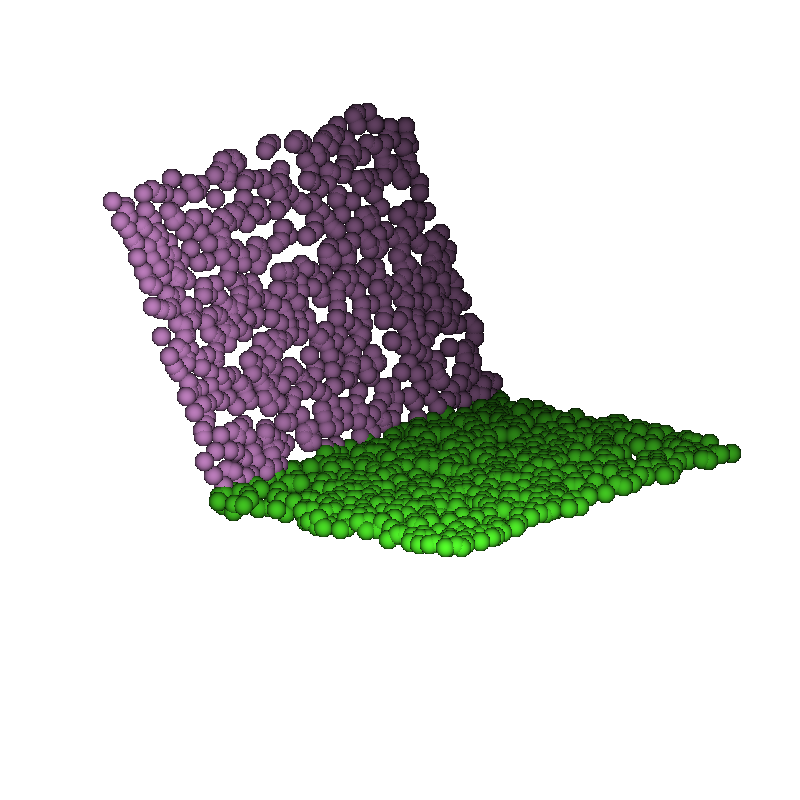}}
	{}%
	\hfill%
	\jsubfig{\includegraphics[height=3.49cm]{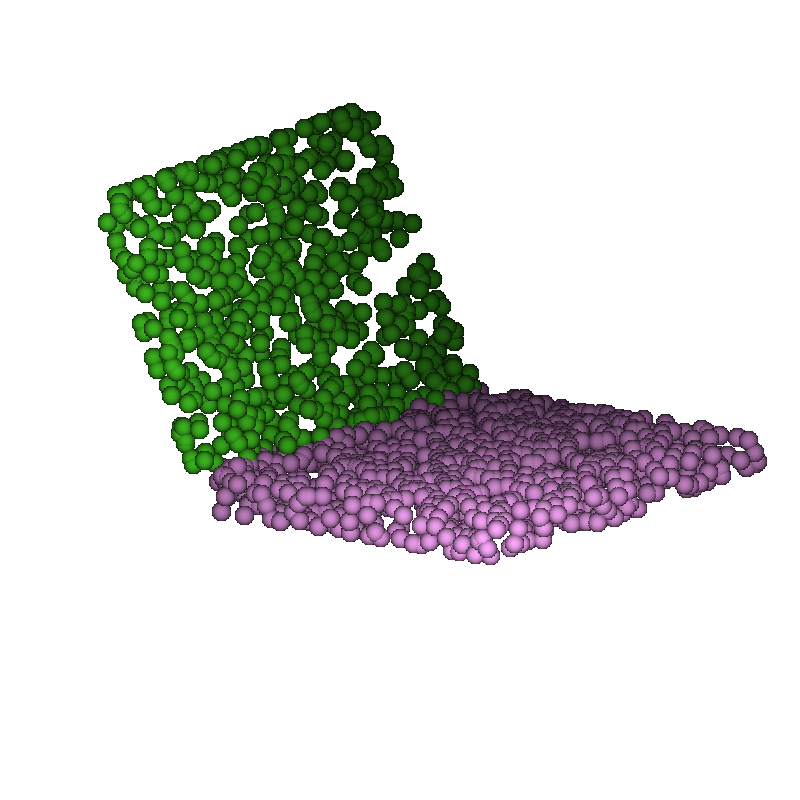}}
	{}%
	\hfill%
	\jsubfig{\includegraphics[height=3.49cm]{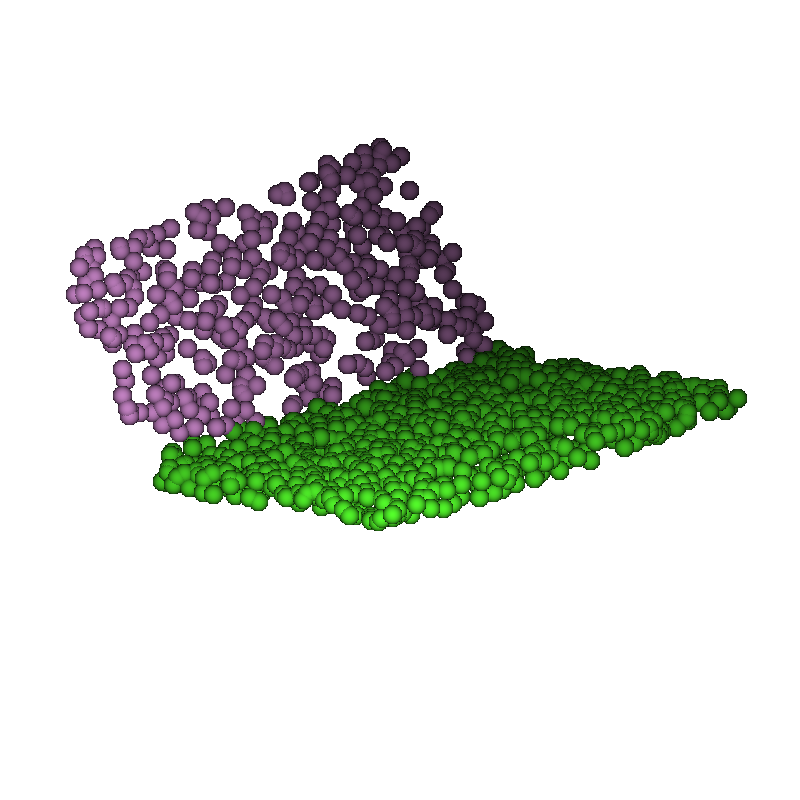}}
	{}%
	\hfill%
	\jsubfig{\includegraphics[height=3.49cm]{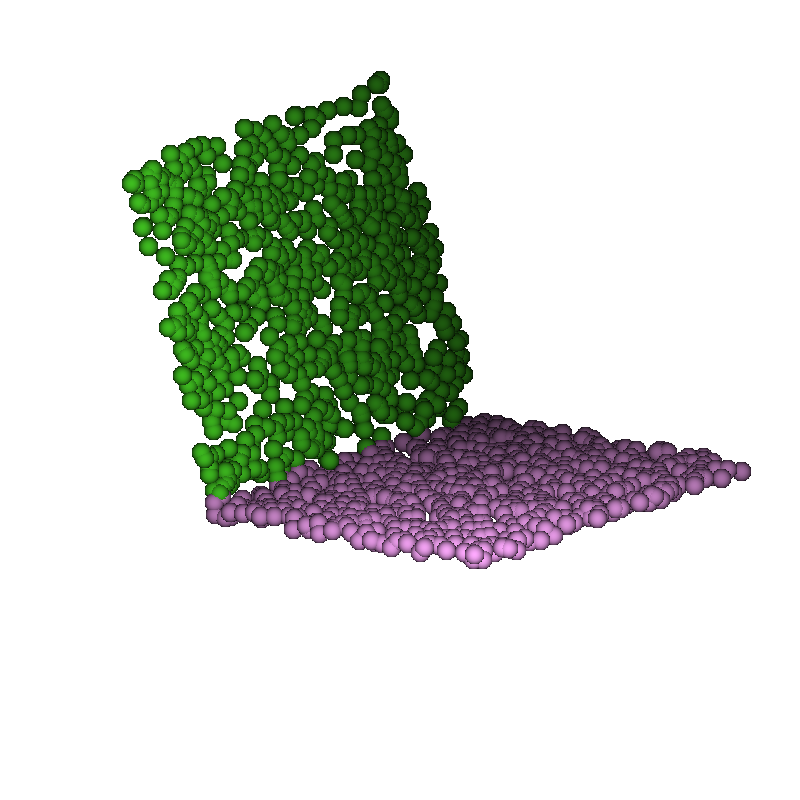}}
	{}%
\\
\vspace{6pt}
	\jsubfig{\includegraphics[height=3.49cm]{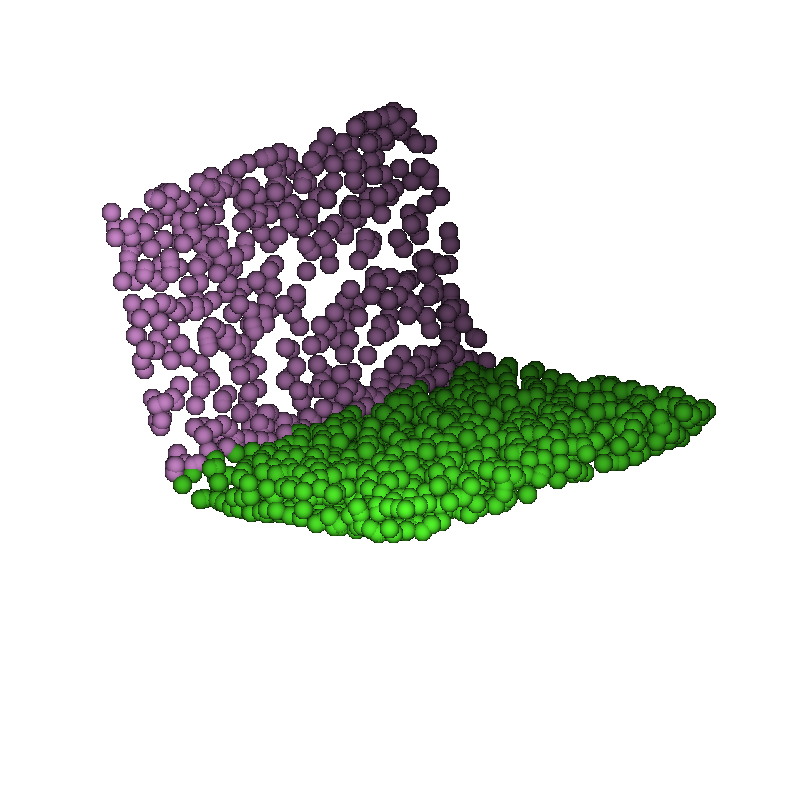}}
	{}%
	\hfill%
	\jsubfig{\includegraphics[height=3.49cm]{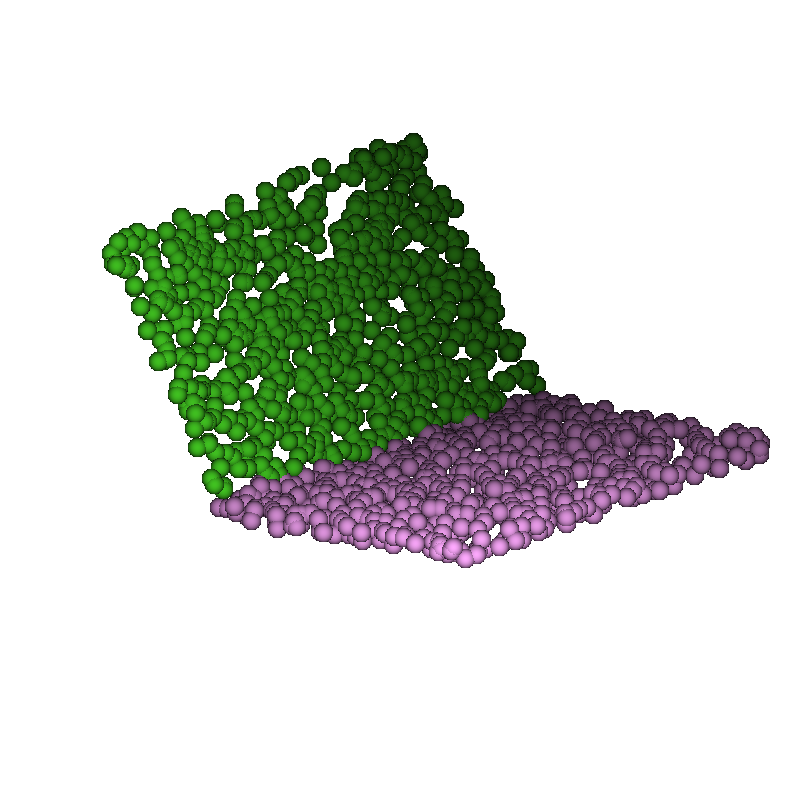}}
	{}%
	\hfill%
	\jsubfig{\includegraphics[height=3.49cm]{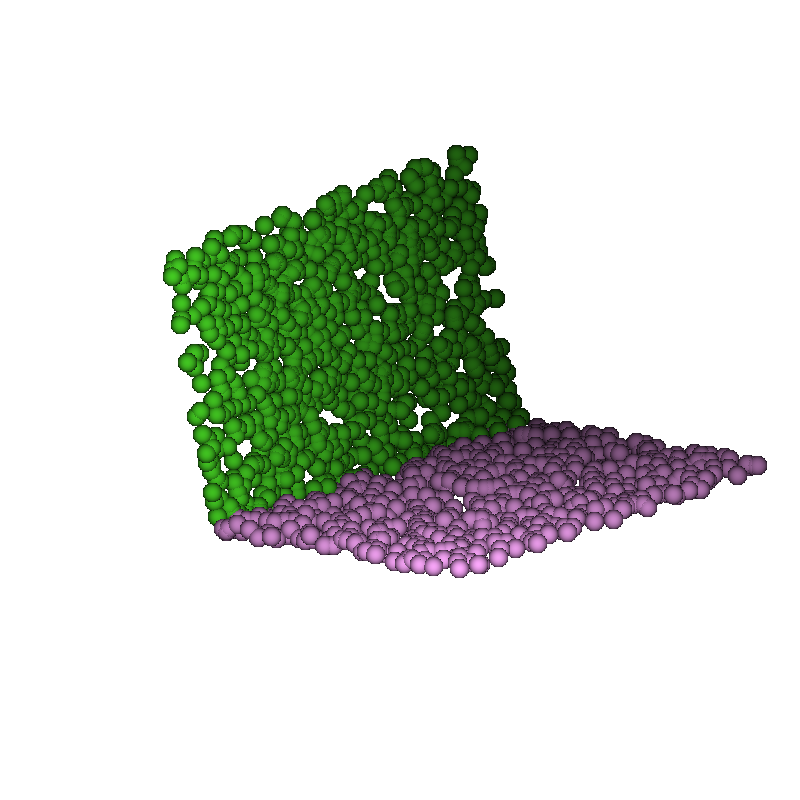}}
	{}%
	\hfill%
	\jsubfig{\includegraphics[height=3.49cm]{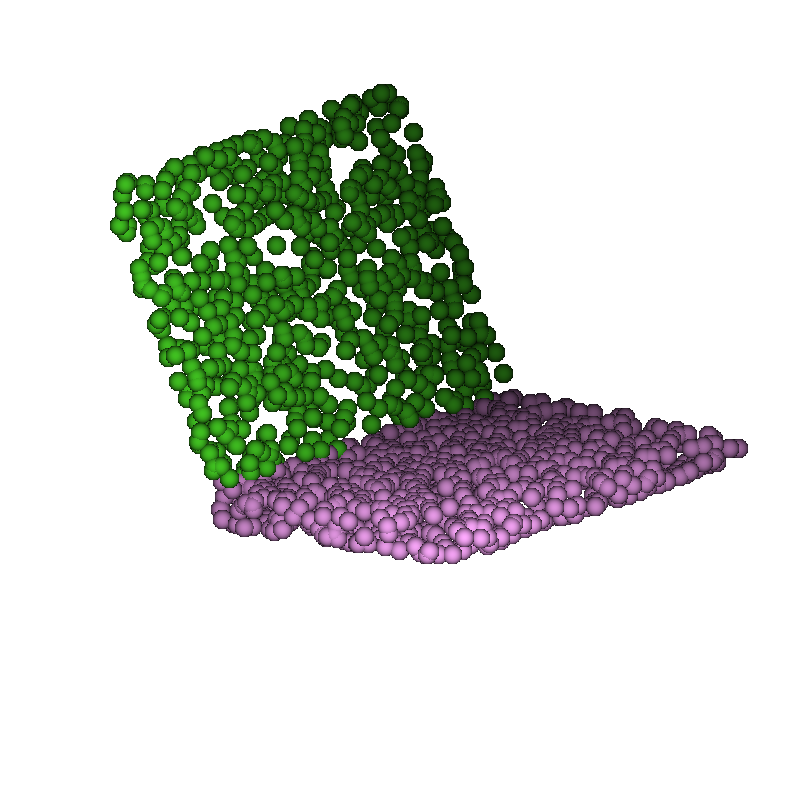}}
	{}%
\end{center}
\caption{Unsupervised segmentation examples on laptops taken from the training data.}

\label{fig:seg_train_laptops}
\end{figure*}

\begin{figure*}
\begin{center}
	\jsubfig{\includegraphics[height=3.49cm]{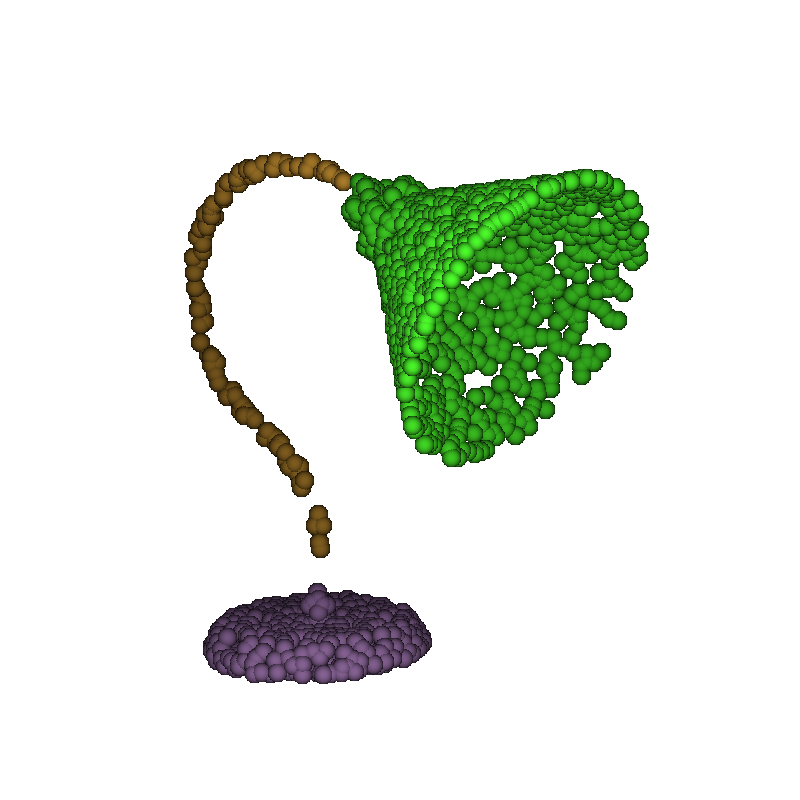}}
	{}%
	\hfill%
	\jsubfig{\includegraphics[height=3.49cm]{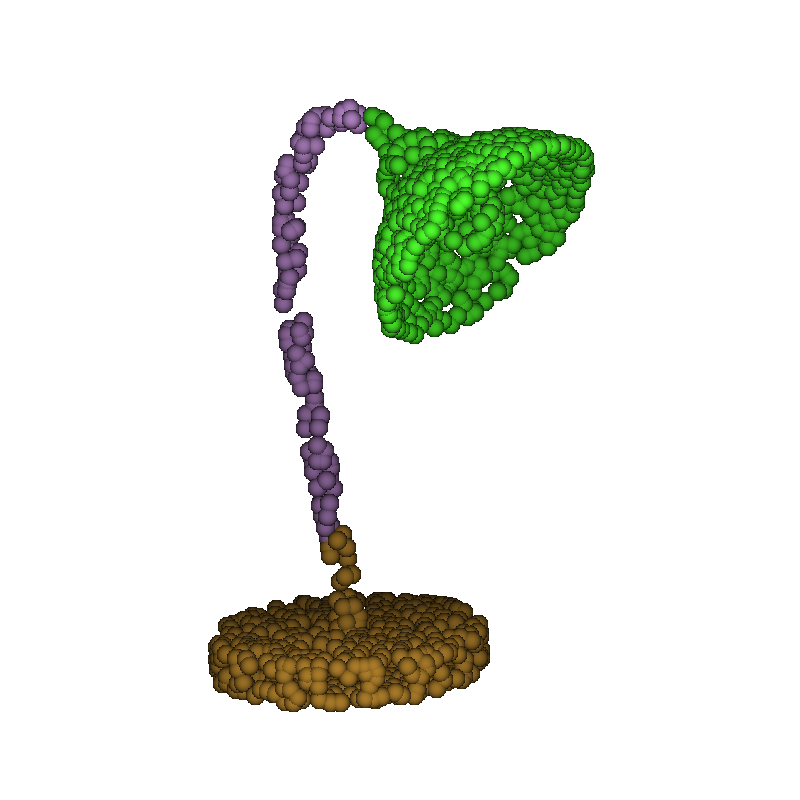}}
	{}%
	\hfill%
	\jsubfig{\includegraphics[height=3.49cm]{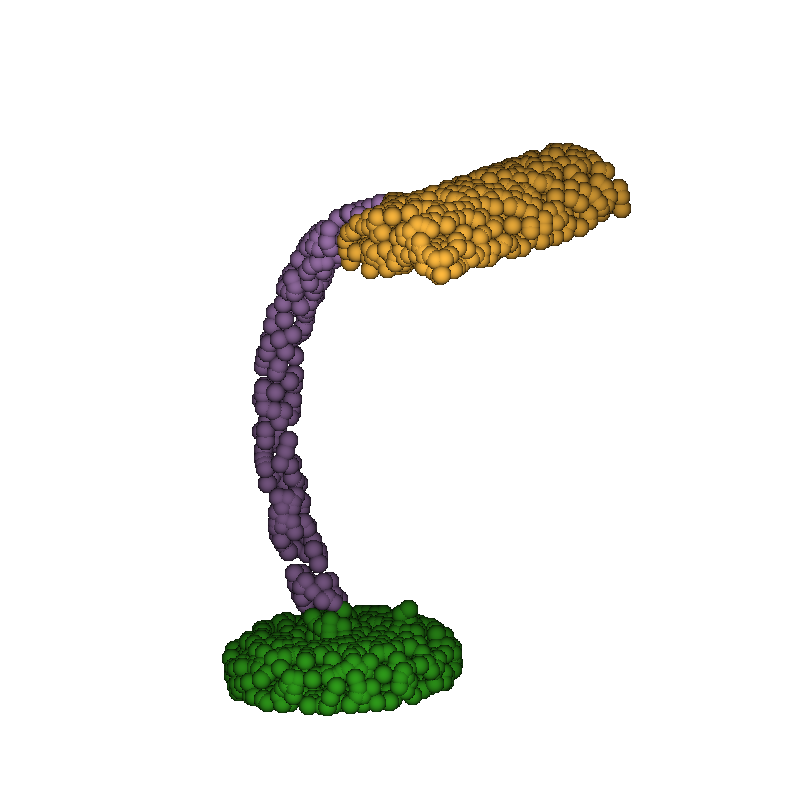}}
	{}%
	\hfill%
	\jsubfig{\includegraphics[height=3.49cm]{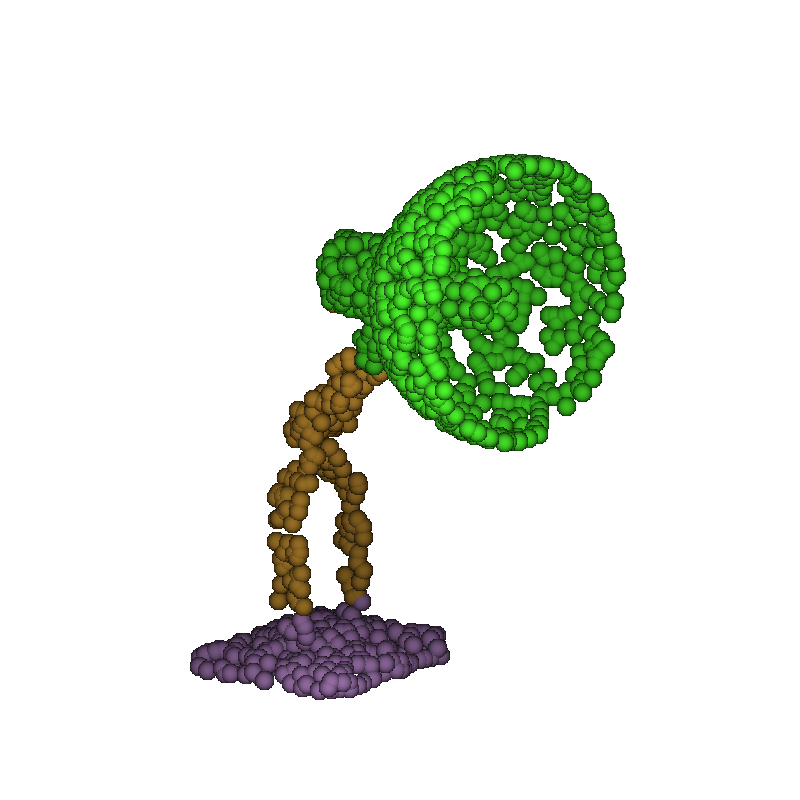}}
	{}%
\\
\vspace{6pt}
	\jsubfig{\includegraphics[height=3.49cm]{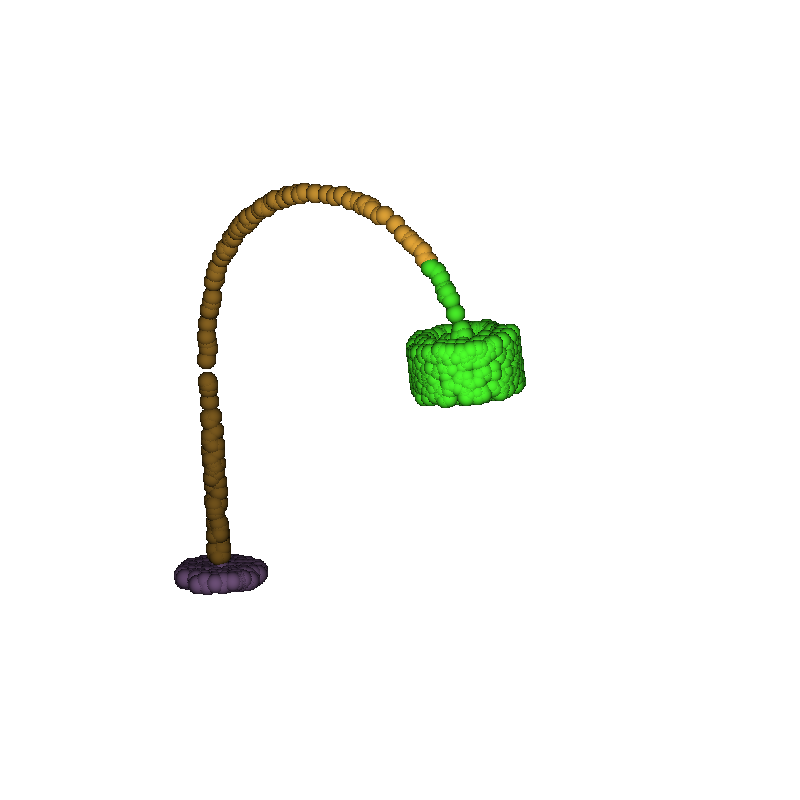}}
	{}%
	\hfill%
	\jsubfig{\includegraphics[height=3.49cm]{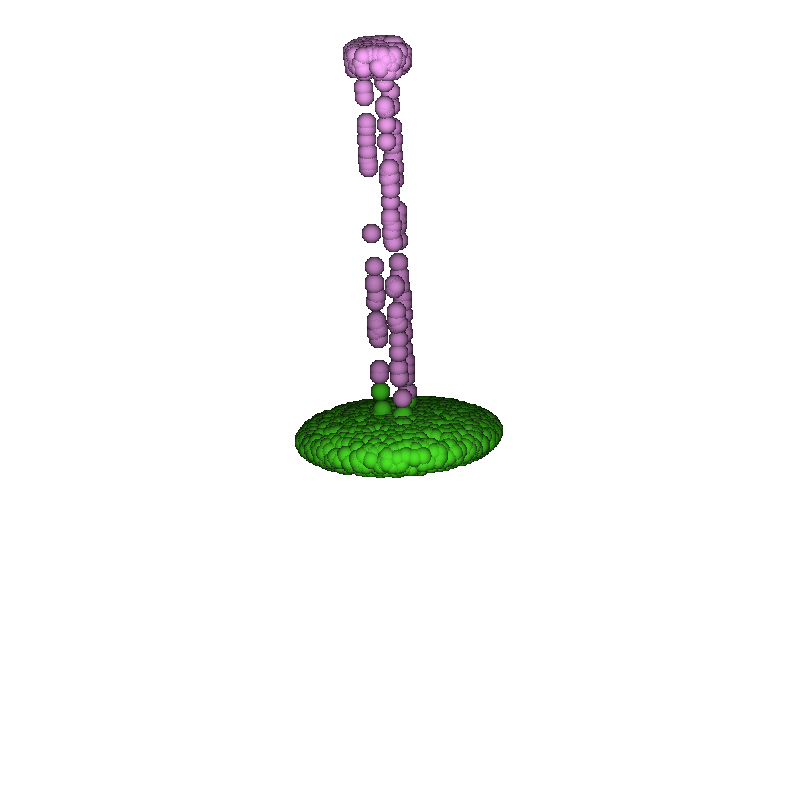}}
	{}%
	\hfill%
	\jsubfig{\includegraphics[height=3.49cm]{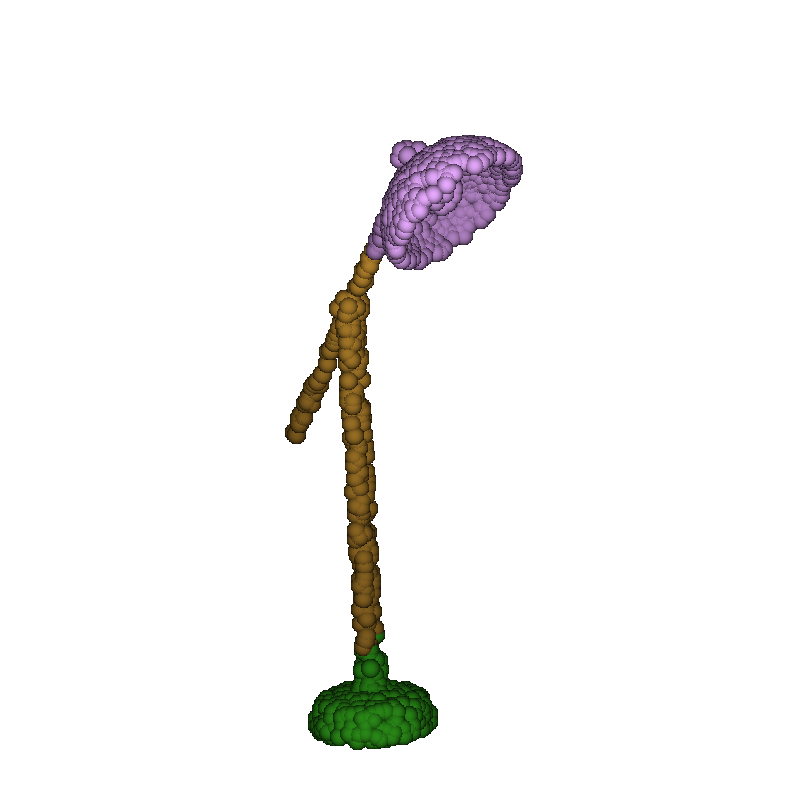}}
	{}%
	\hfill%
	\jsubfig{\includegraphics[height=3.49cm]{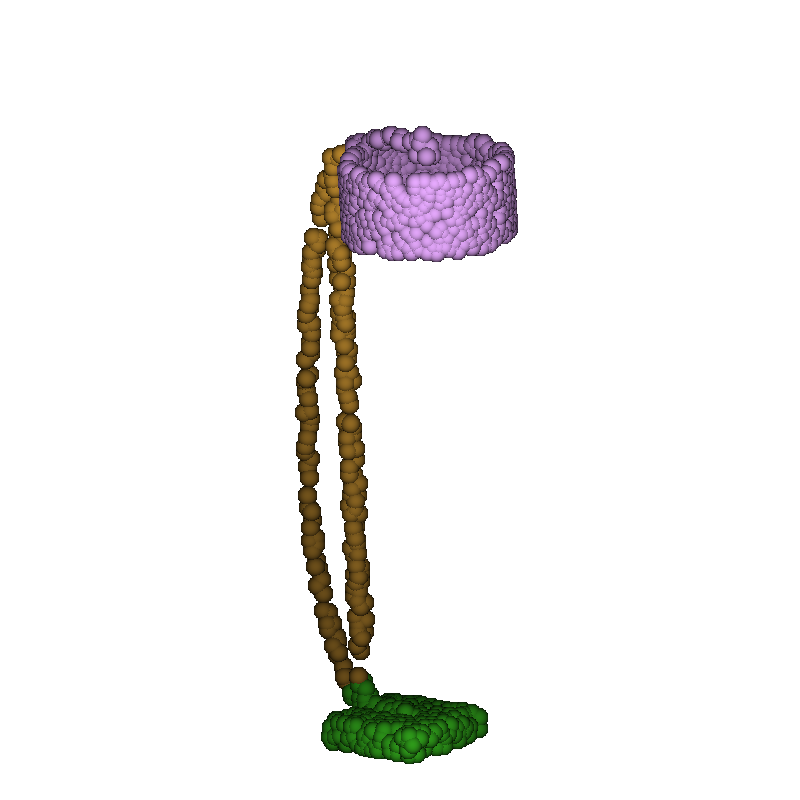}}
	{}%
\\
\vspace{6pt}
	\jsubfig{\includegraphics[height=3.49cm]{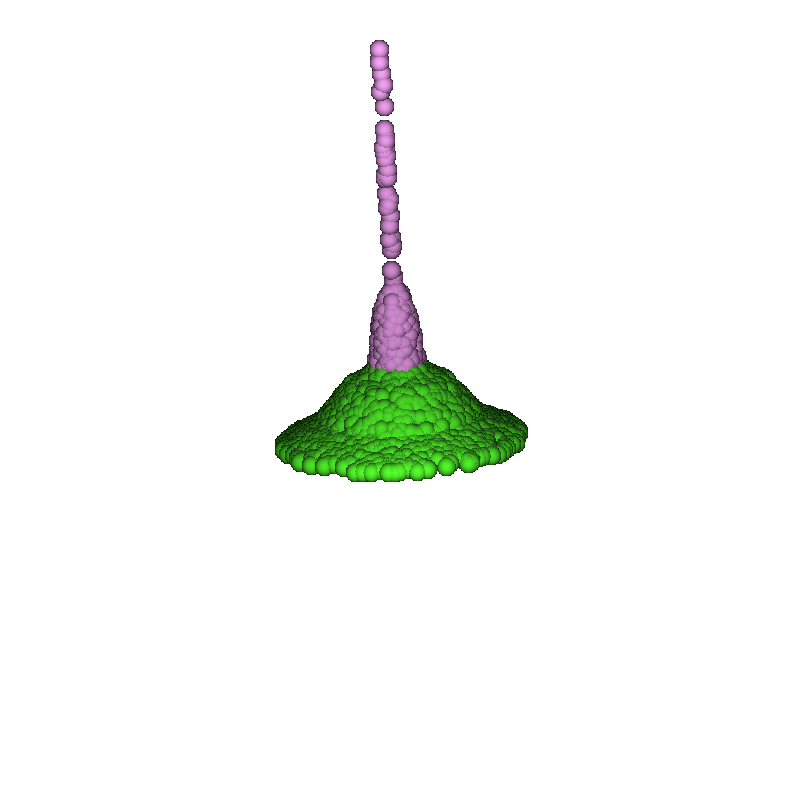}}
	{}%
	\hfill%
	\jsubfig{\includegraphics[height=3.49cm]{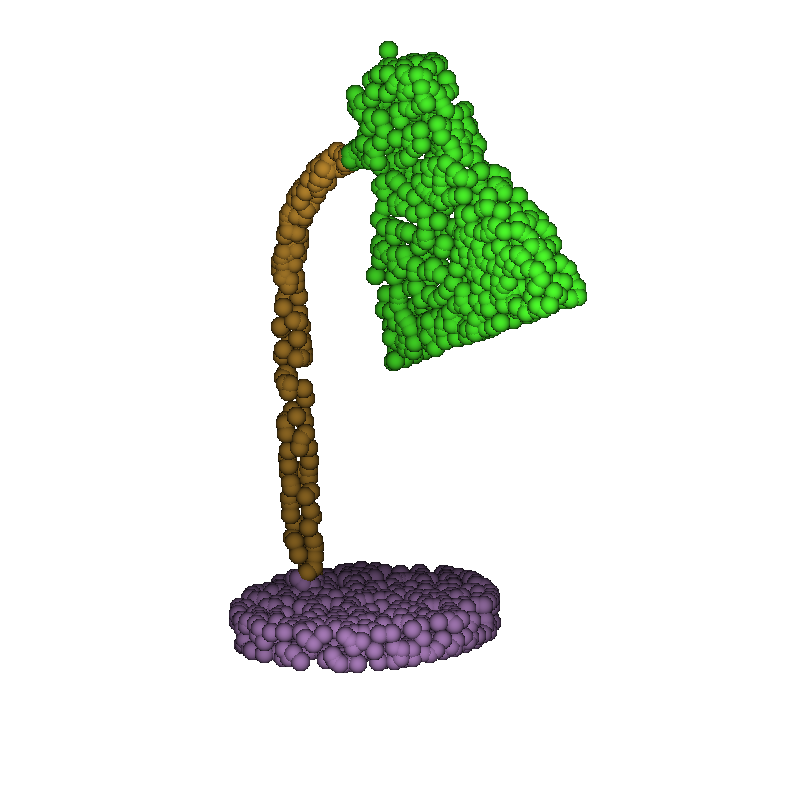}}
	{}%
	\hfill%
	\jsubfig{\includegraphics[height=3.49cm]{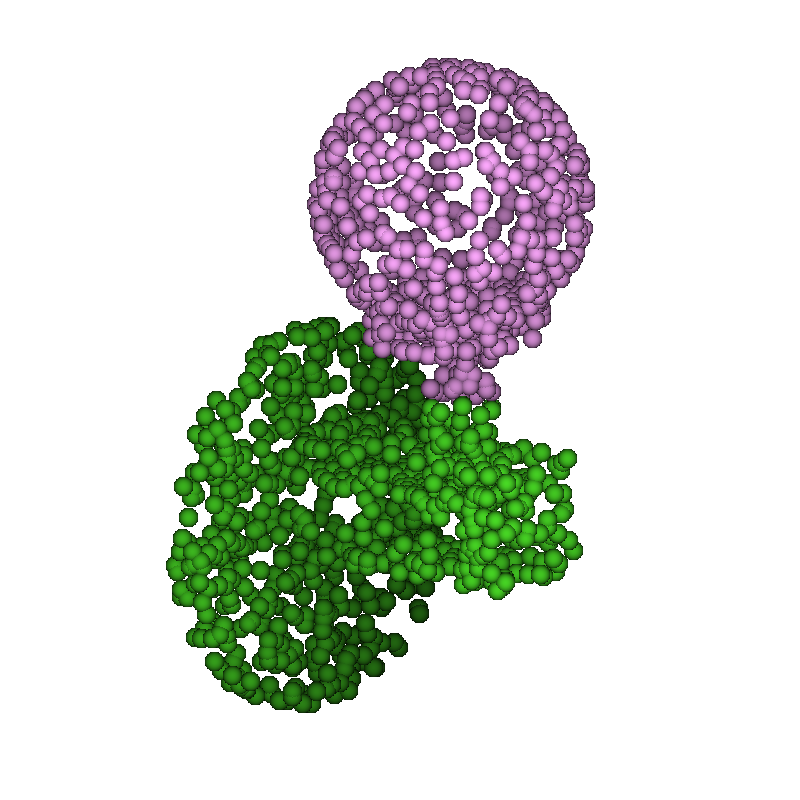}}
	{}%
	\hfill%
	\jsubfig{\includegraphics[height=3.49cm]{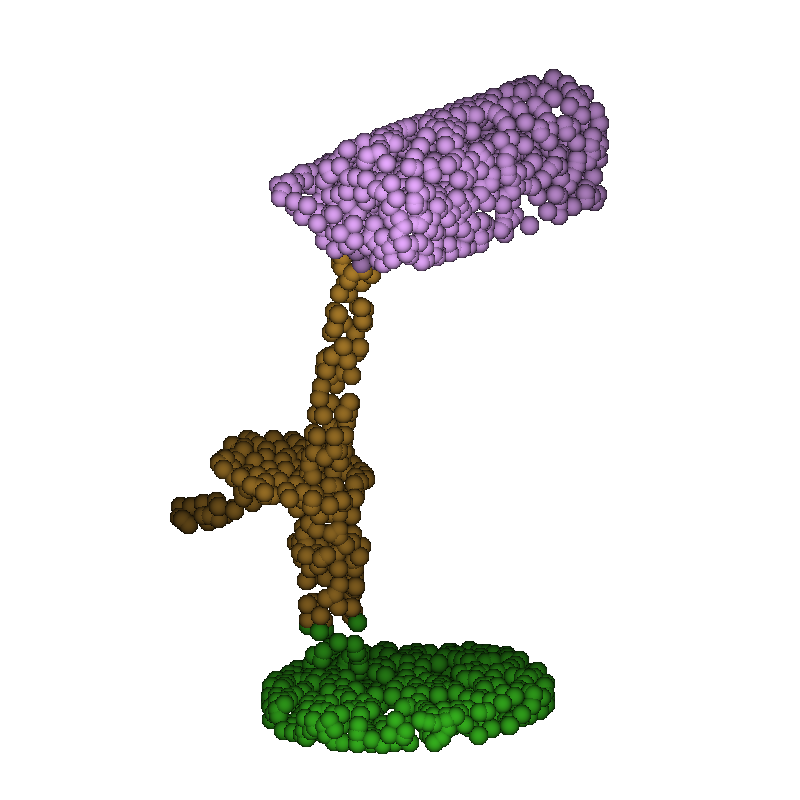}}
	{}%
\\
\vspace{6pt}
	\jsubfig{\includegraphics[height=3.49cm]{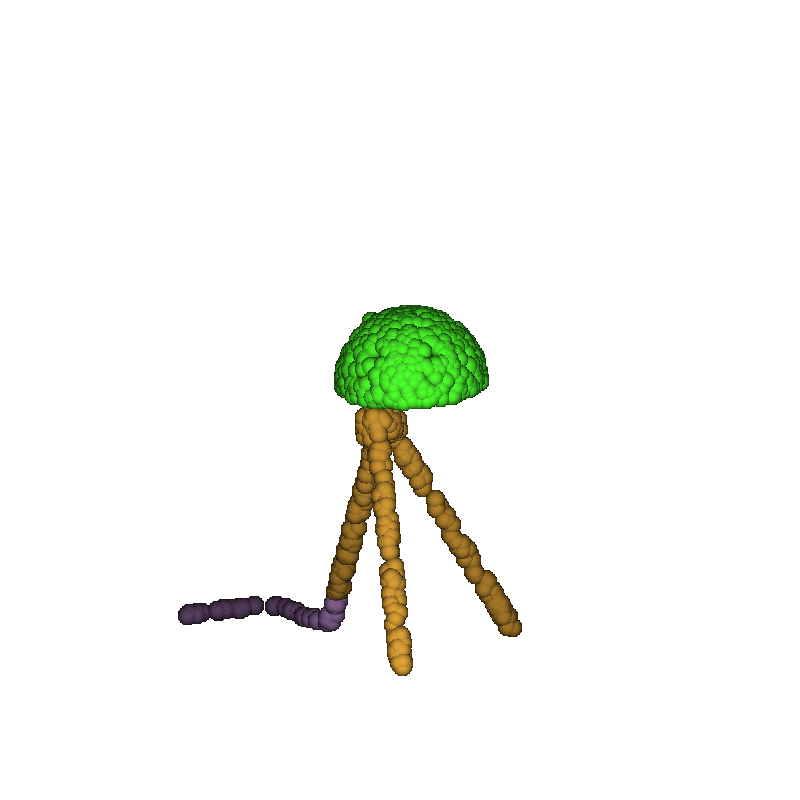}}
	{}%
	\hfill%
	\jsubfig{\includegraphics[height=3.49cm]{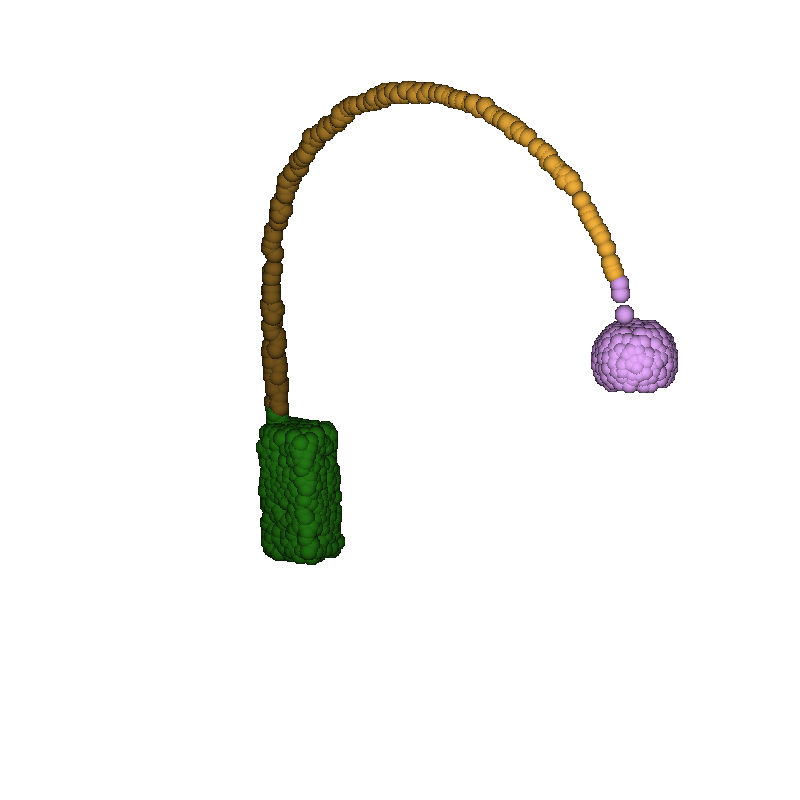}}
	{}%
	\hfill%
	\jsubfig{\includegraphics[height=3.49cm]{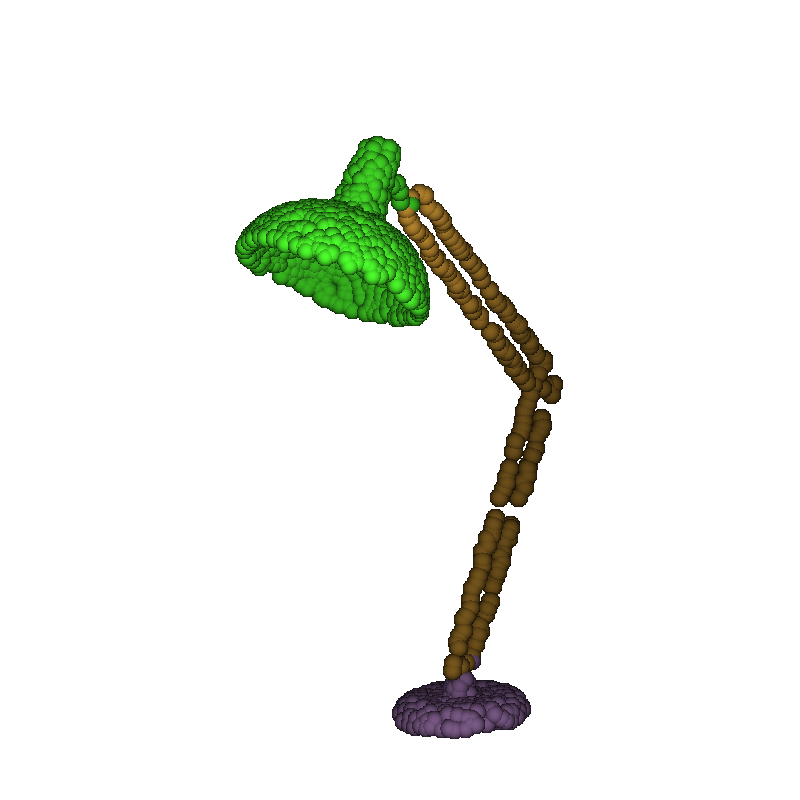}}
	{}%
	\hfill%
	\jsubfig{\includegraphics[height=3.49cm]{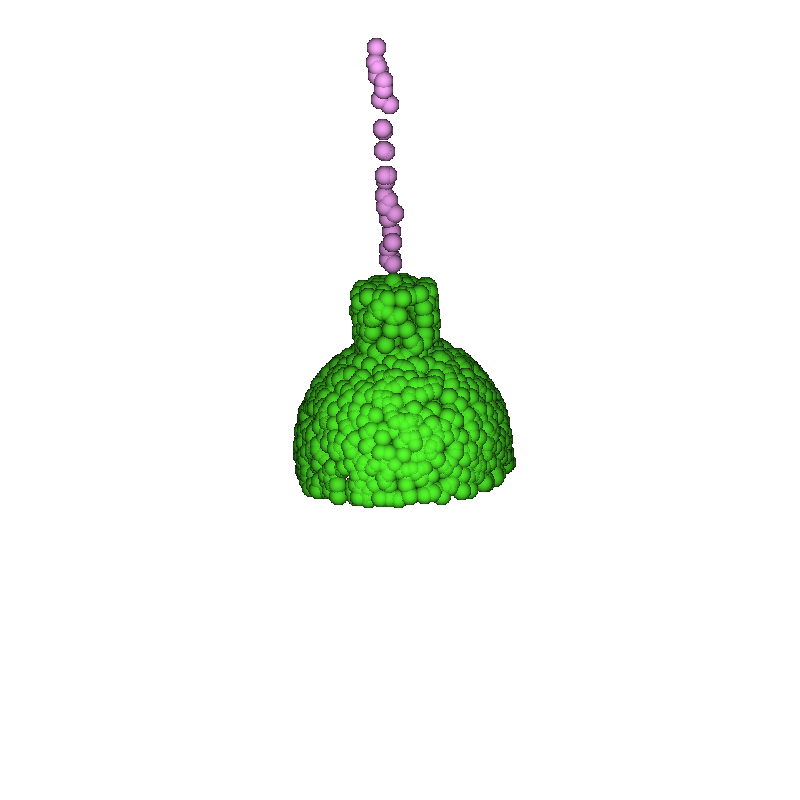}}
	{}%
\\
\vspace{6pt}
	\jsubfig{\includegraphics[height=3.49cm]{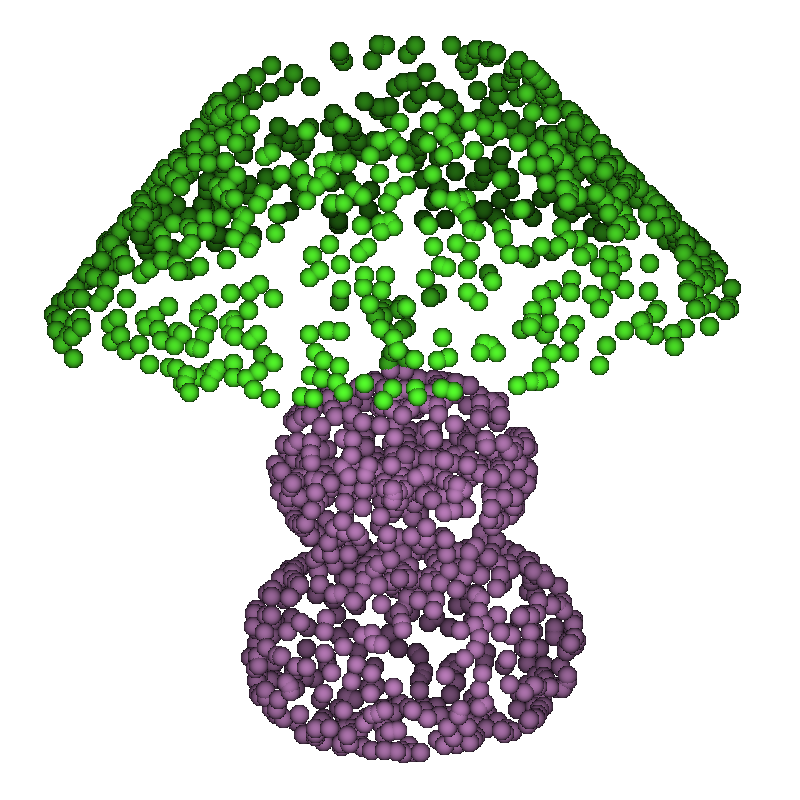}}
	{}%
	\hfill%
	\jsubfig{\includegraphics[height=3.49cm]{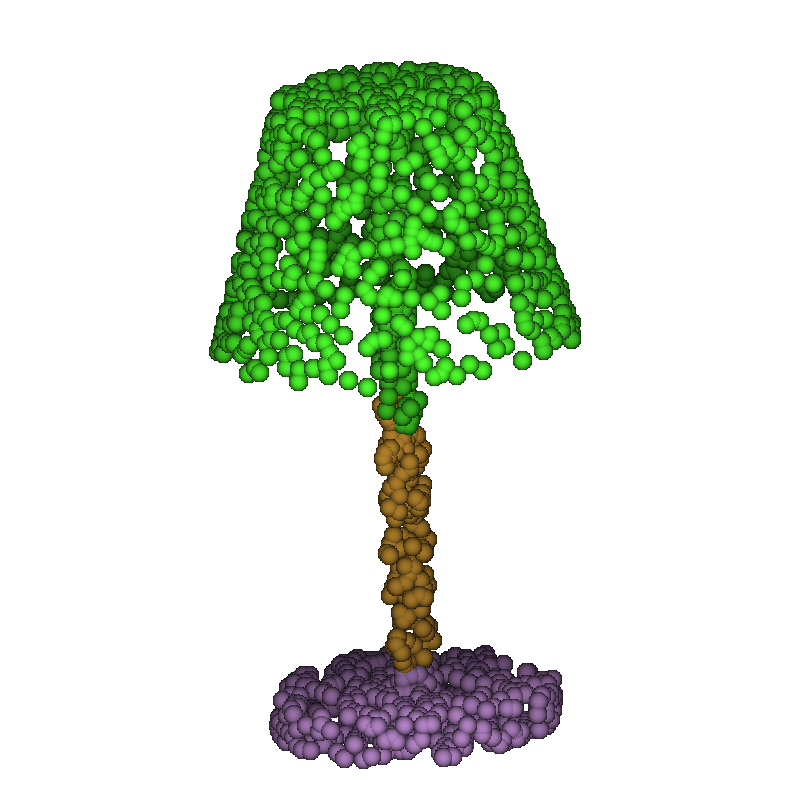}}
	{}%
	\hfill%
	\jsubfig{\includegraphics[height=3.49cm]{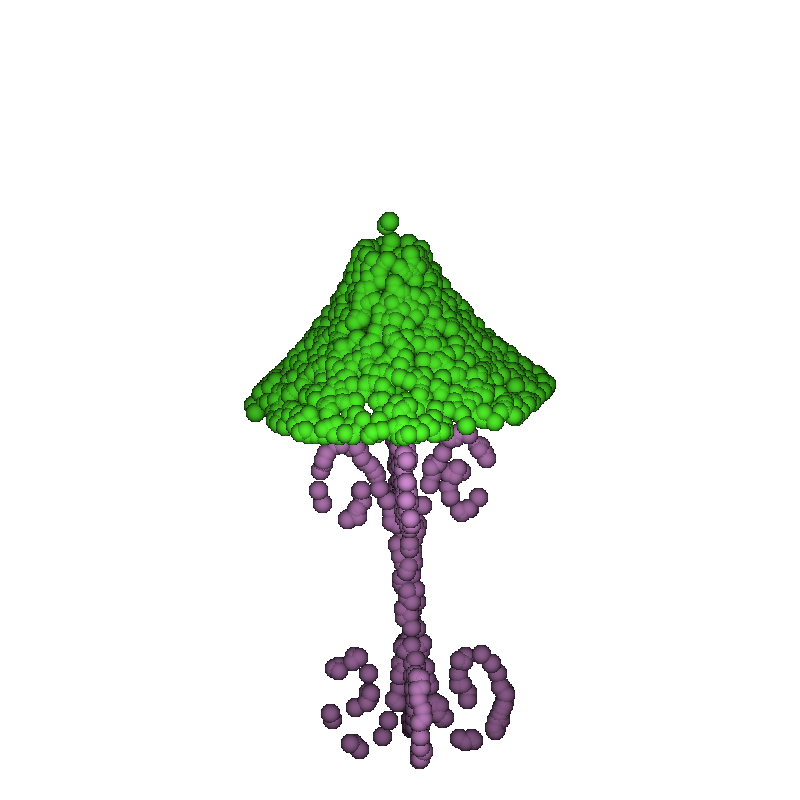}}
	{}%
	\hfill%
	\jsubfig{\includegraphics[height=3.49cm]{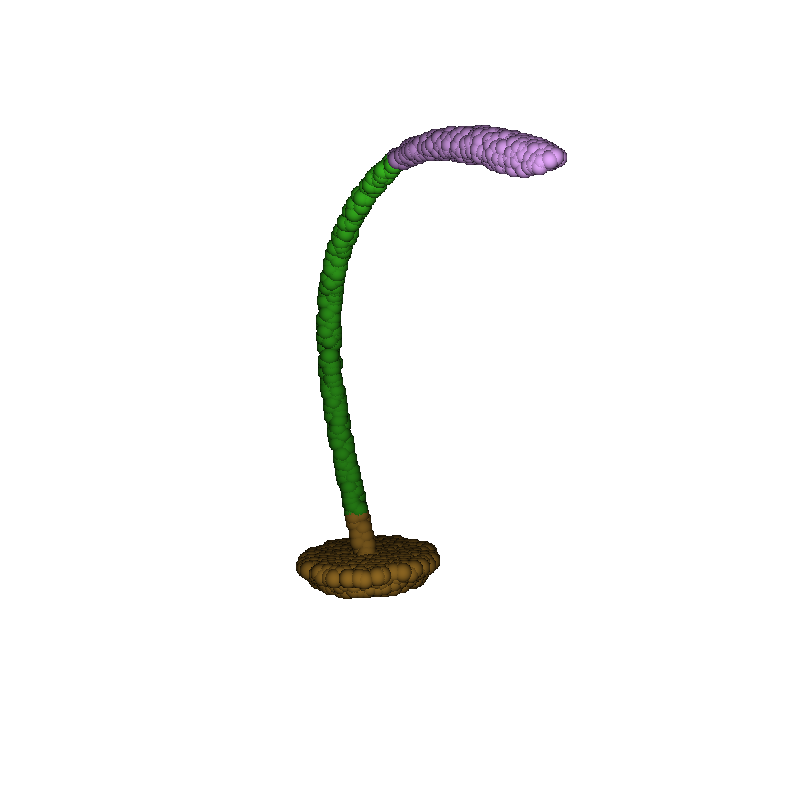}}
	{}%
\\
\vspace{6pt}
	\jsubfig{\includegraphics[height=3.49cm]{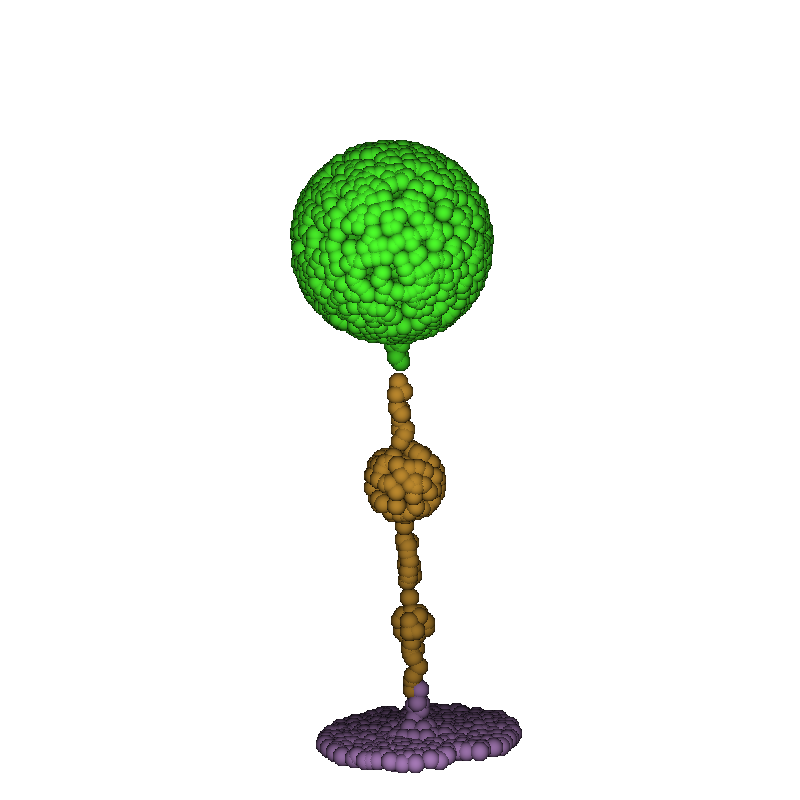}}
	{}%
	\hfill%
	\jsubfig{\includegraphics[height=3.49cm]{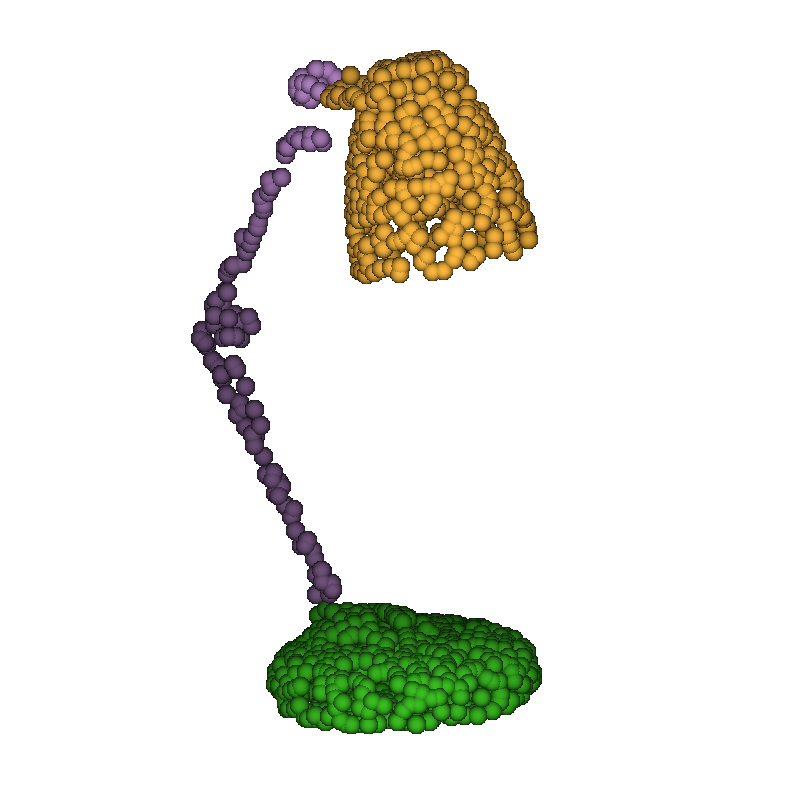}}
	{}%
	\hfill%
	\jsubfig{\includegraphics[height=3.49cm]{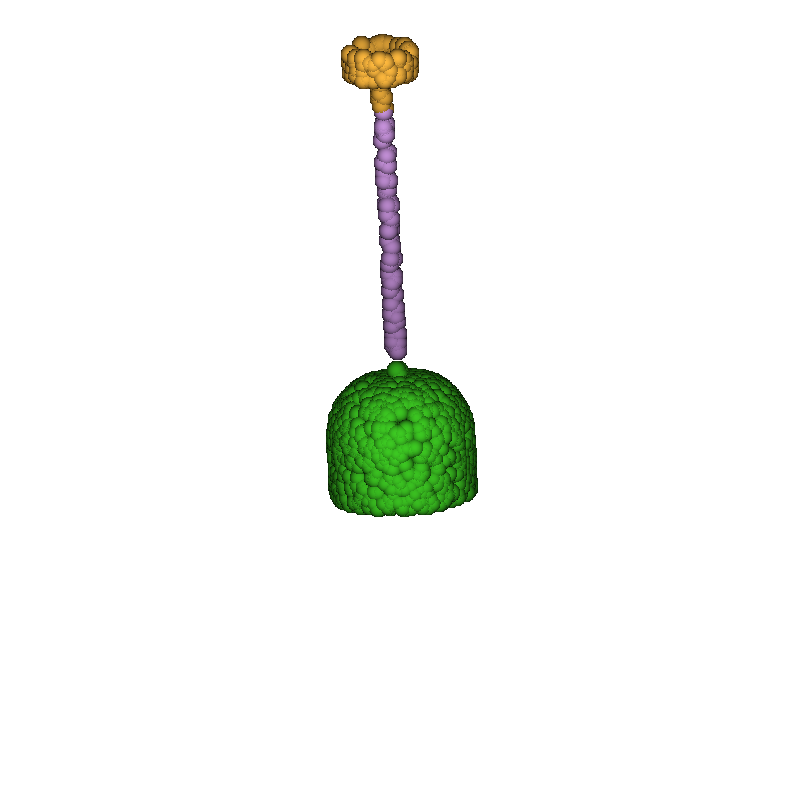}}
	{}%
	\hfill%
	\jsubfig{\includegraphics[height=3.49cm]{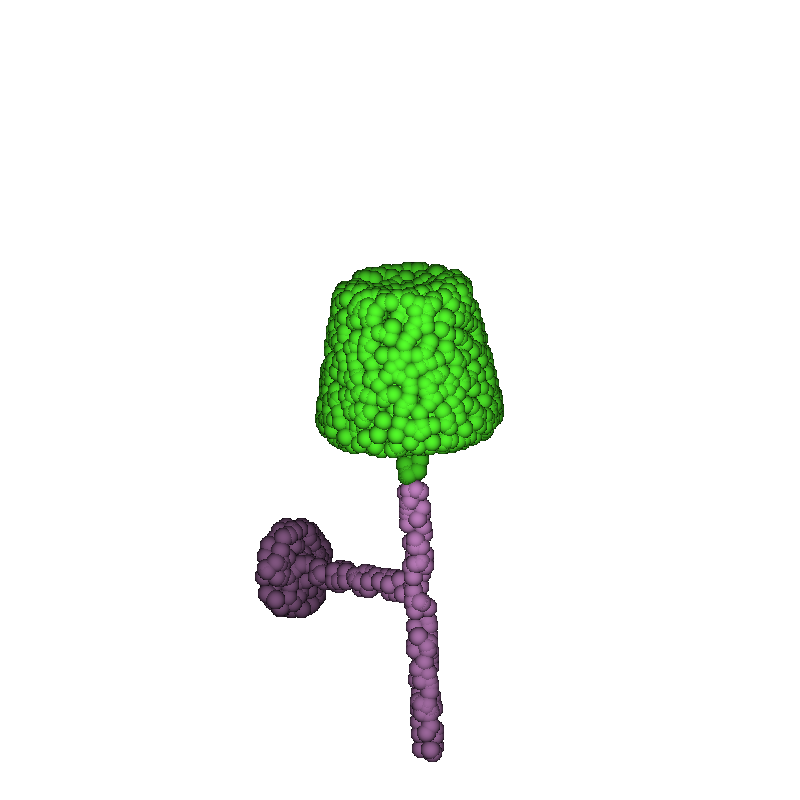}}
	{}%
\end{center}
\caption{Unsupervised segmentation examples on lamps taken from the training data.}

\label{fig:seg_train_lamps}
\end{figure*}

\begin{figure*}[h]
\begin{center}
	\jsubfig{\includegraphics[height=3.49cm]{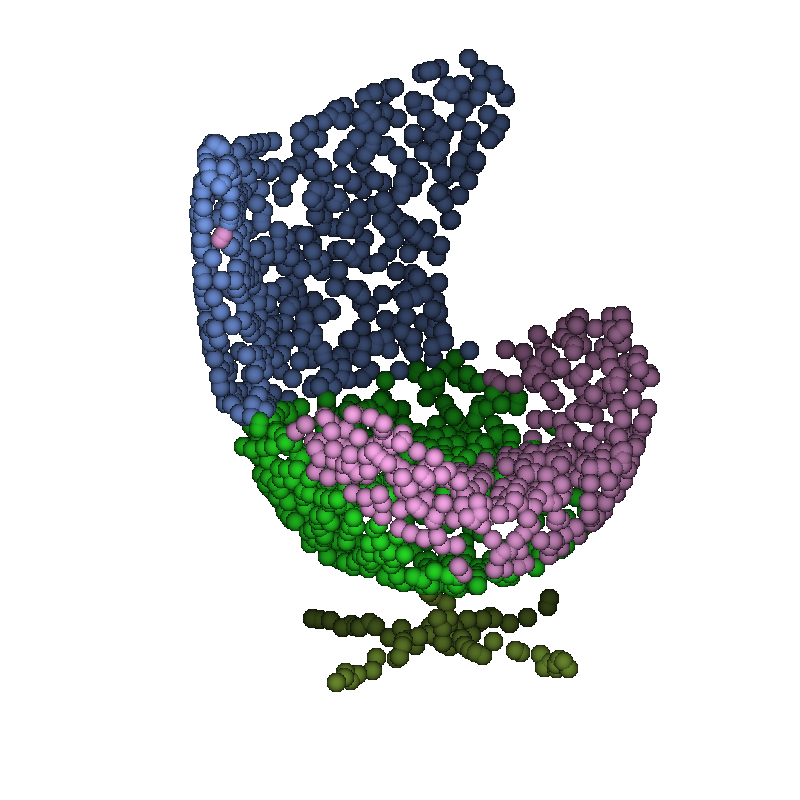}}
	{}%
	\hfill%
	\jsubfig{\includegraphics[height=3.49cm]{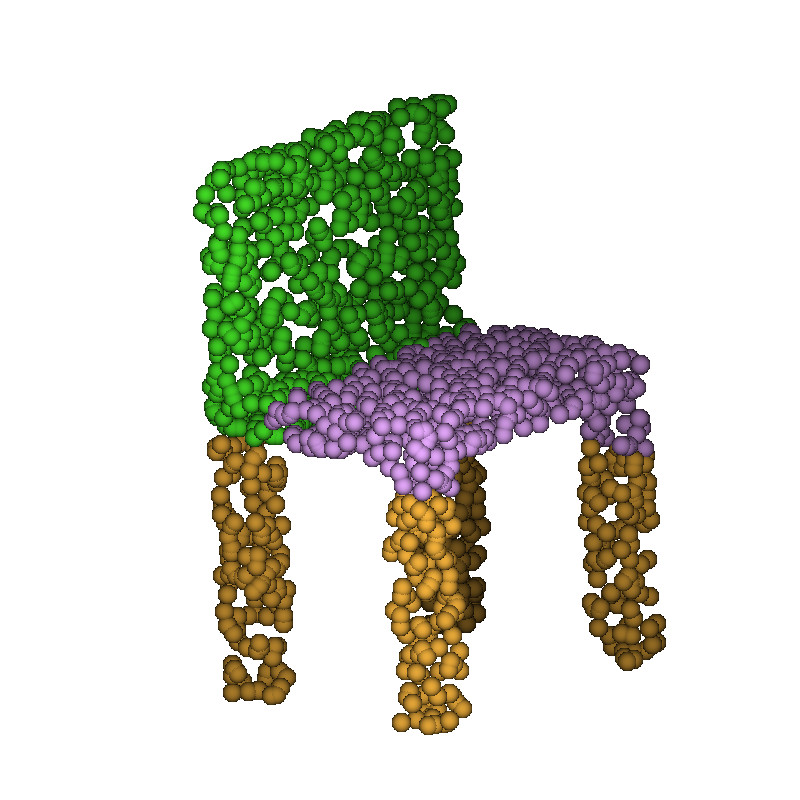}}
	{}%
	\hfill%
	\jsubfig{\includegraphics[height=3.49cm]{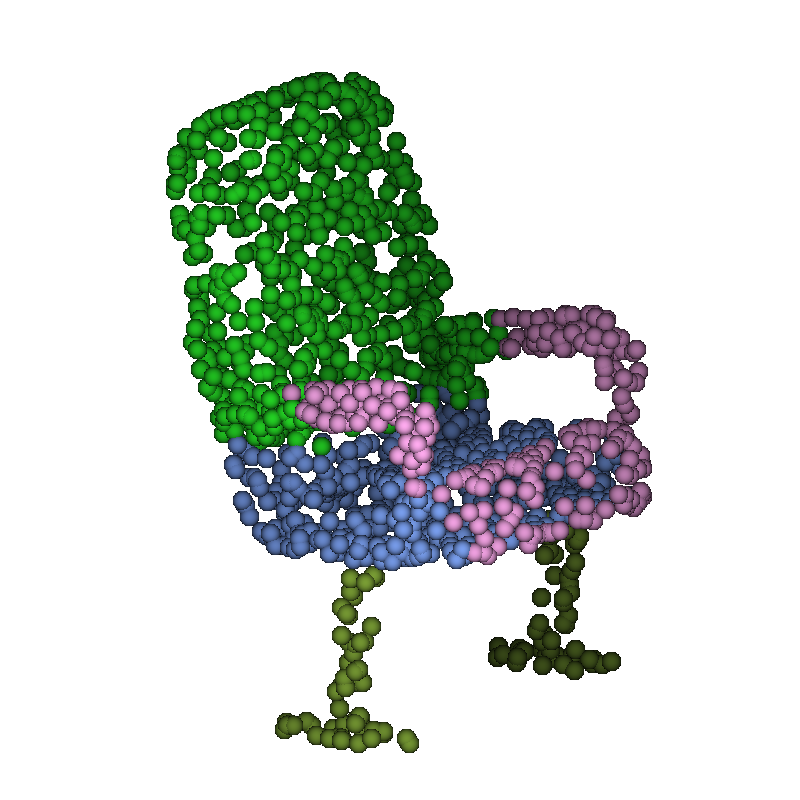}}
	{}%
	\hfill%
	\jsubfig{\includegraphics[height=3.49cm]{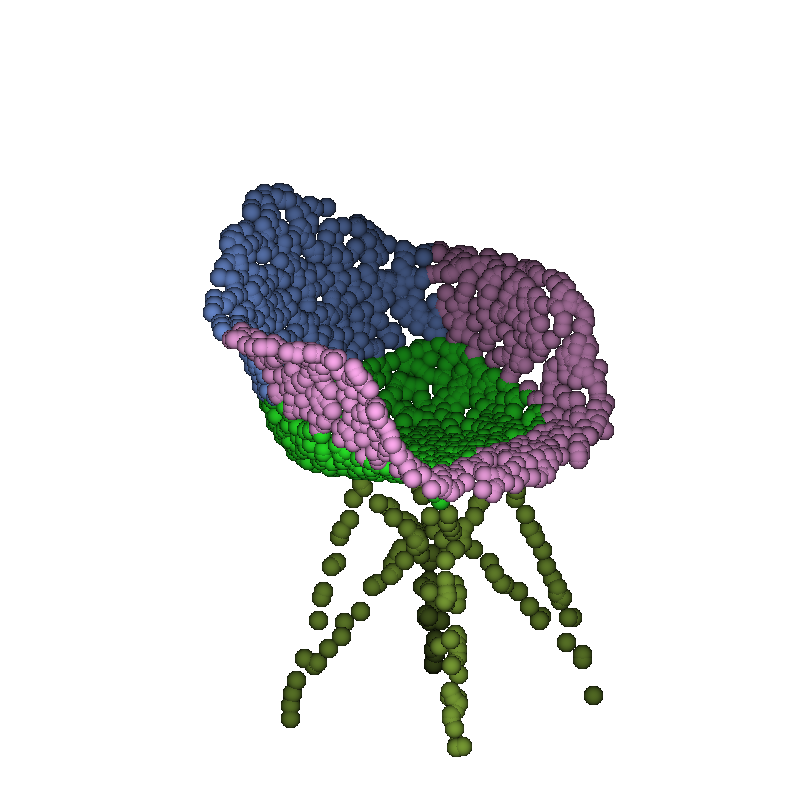}}
	{}%
\\
\vspace{6pt}
	\jsubfig{\includegraphics[height=3.49cm]{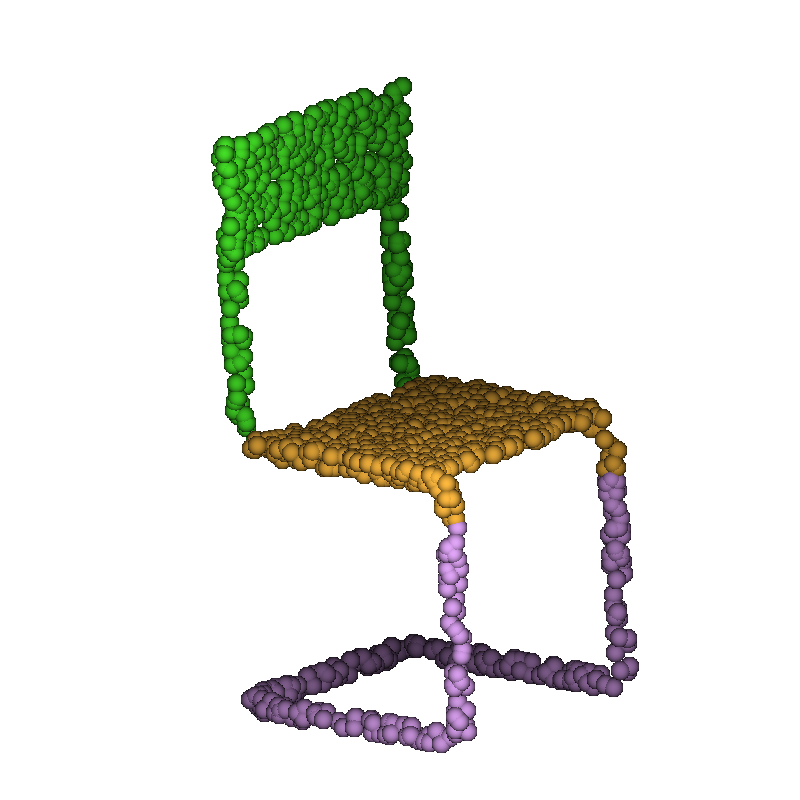}}
	{}%
	\hfill%
	\jsubfig{\includegraphics[height=3.49cm]{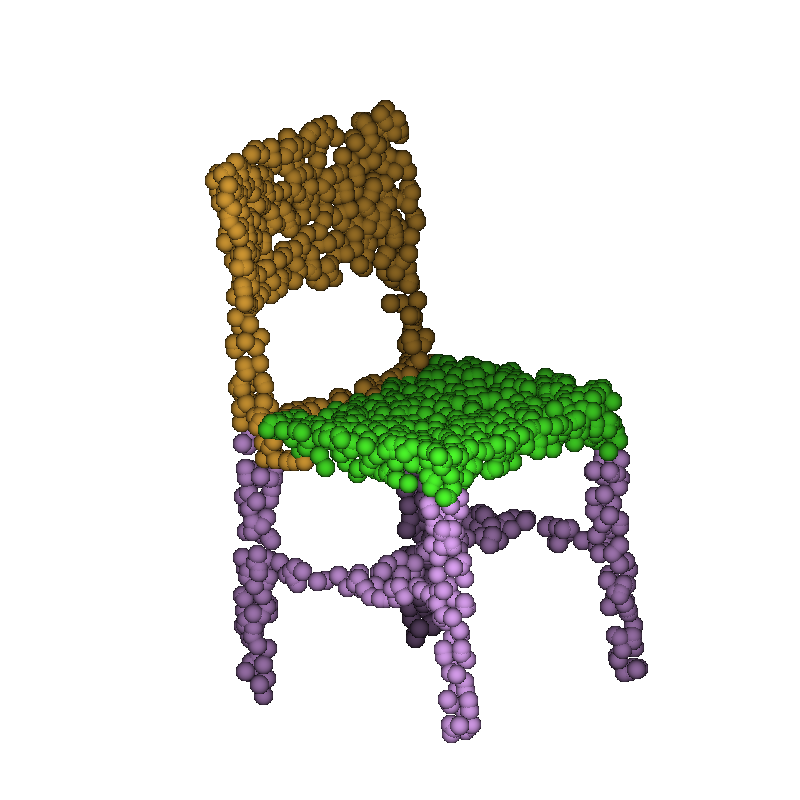}}
	{}%
	\hfill%
	\jsubfig{\includegraphics[height=3.49cm]{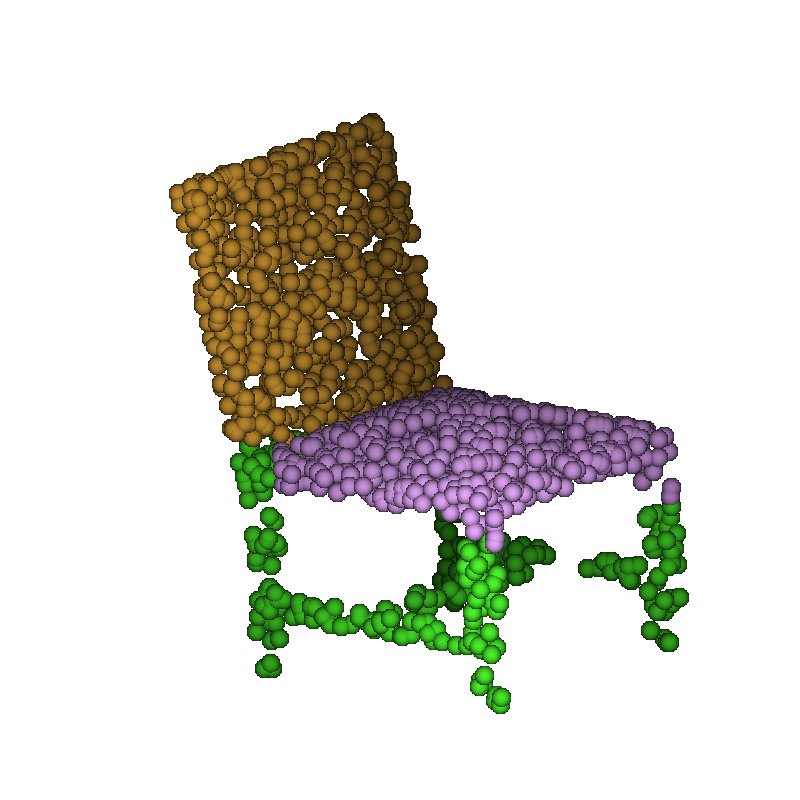}}
	{}%
	\hfill%
	\jsubfig{\includegraphics[height=3.49cm]{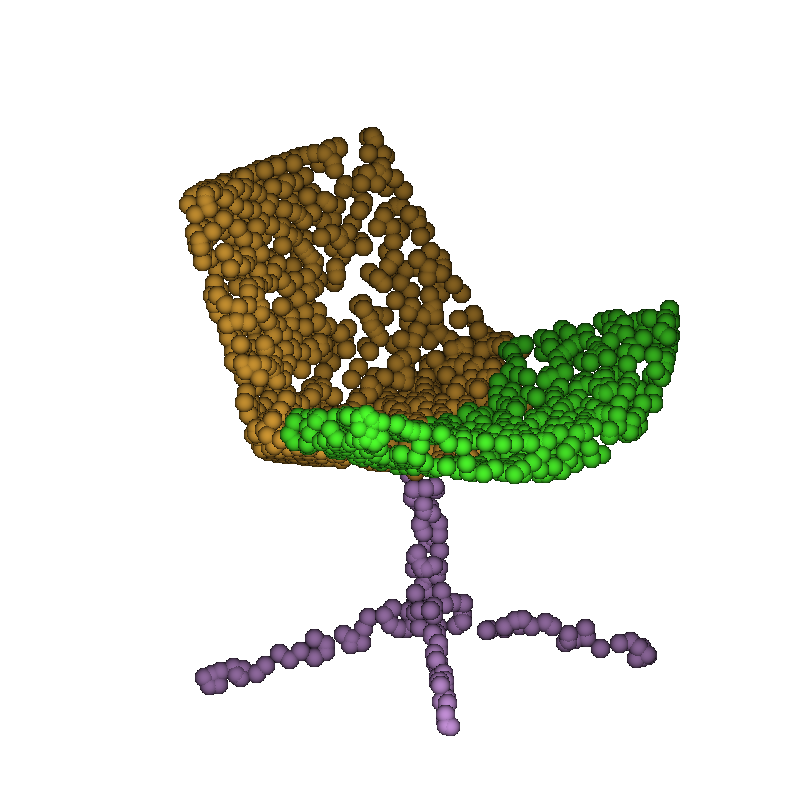}}
	{}%
\\
\vspace{6pt}
	\jsubfig{\includegraphics[height=3.49cm]{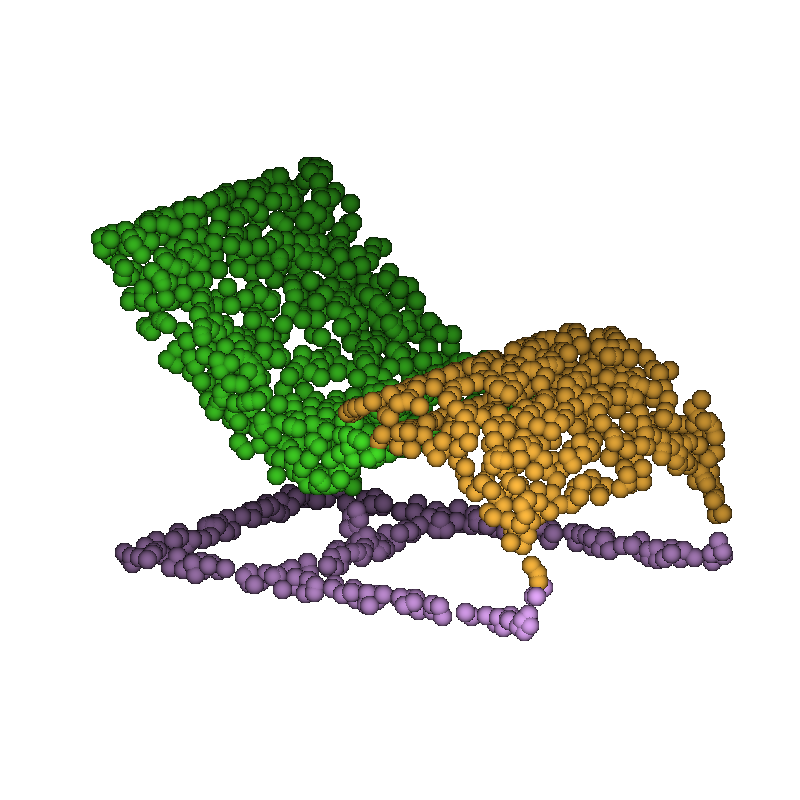}}
	{}%
	\hfill%
	\jsubfig{\includegraphics[height=3.49cm]{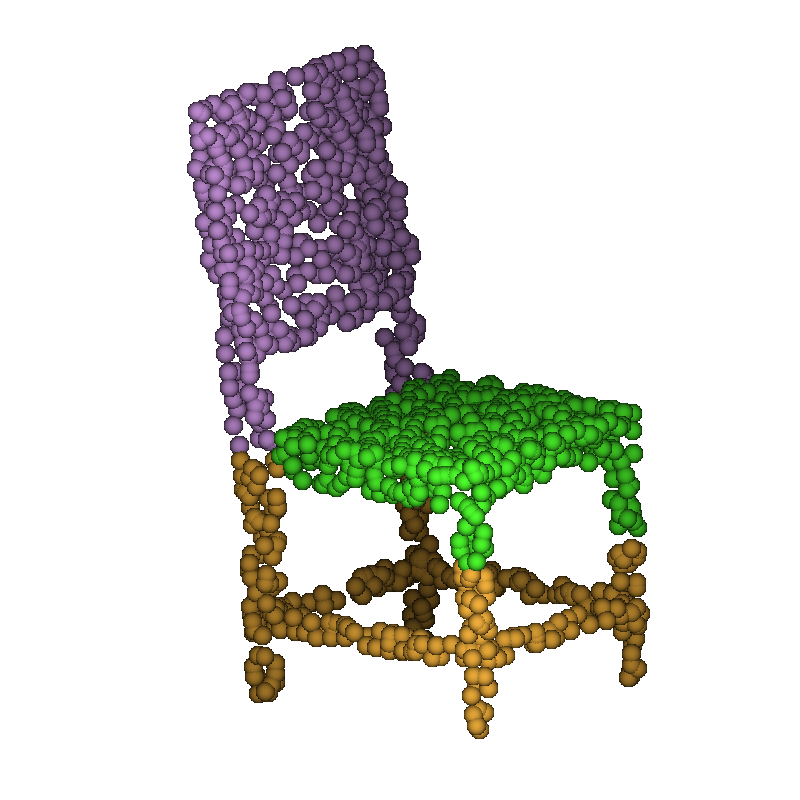}}
	{}%
	\hfill%
	\jsubfig{\includegraphics[height=3.49cm]{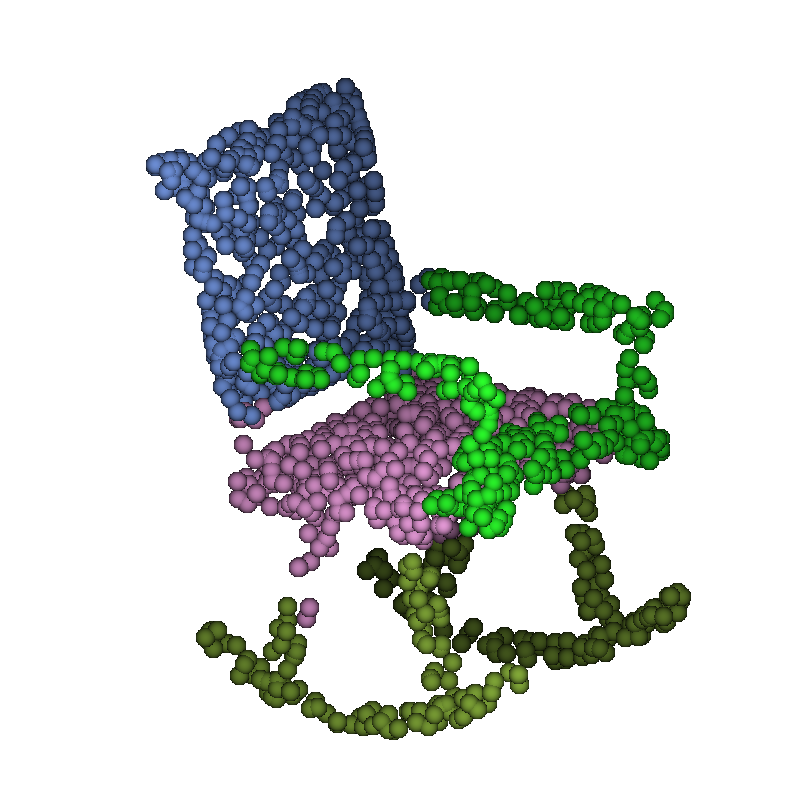}}
	{}%
	\hfill%
	\jsubfig{\includegraphics[height=3.49cm]{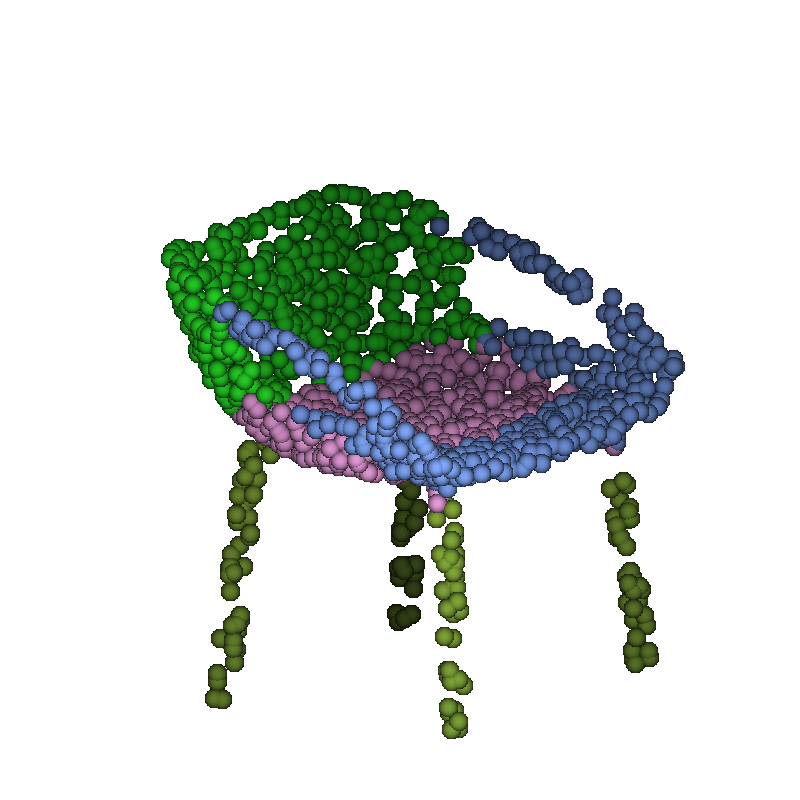}}
	{}%
\\
\vspace{6pt}
	\jsubfig{\includegraphics[height=3.49cm]{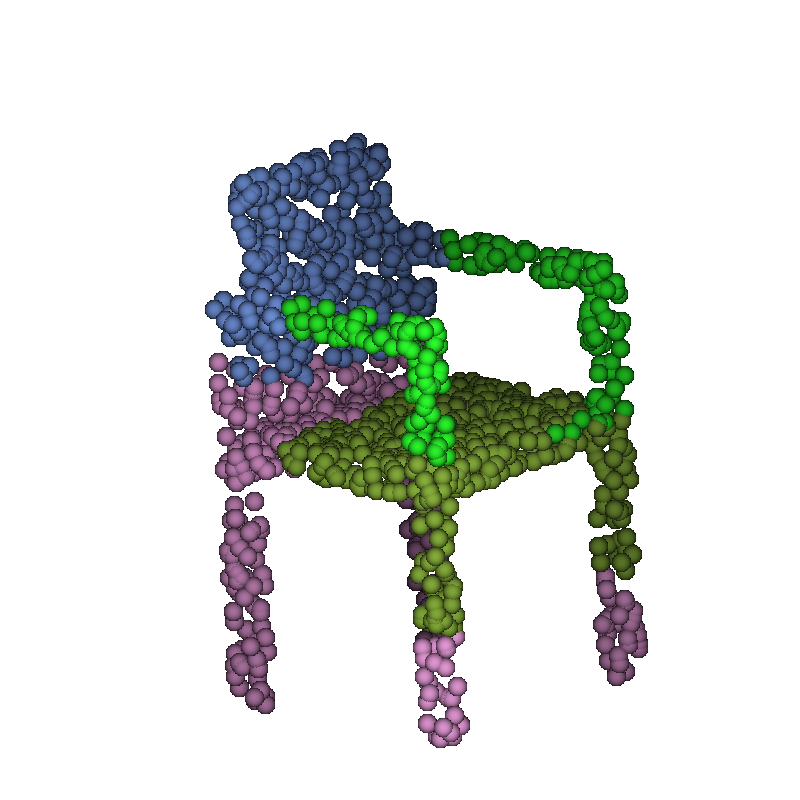}}
	{}%
	\hfill%
	\jsubfig{\includegraphics[height=3.49cm]{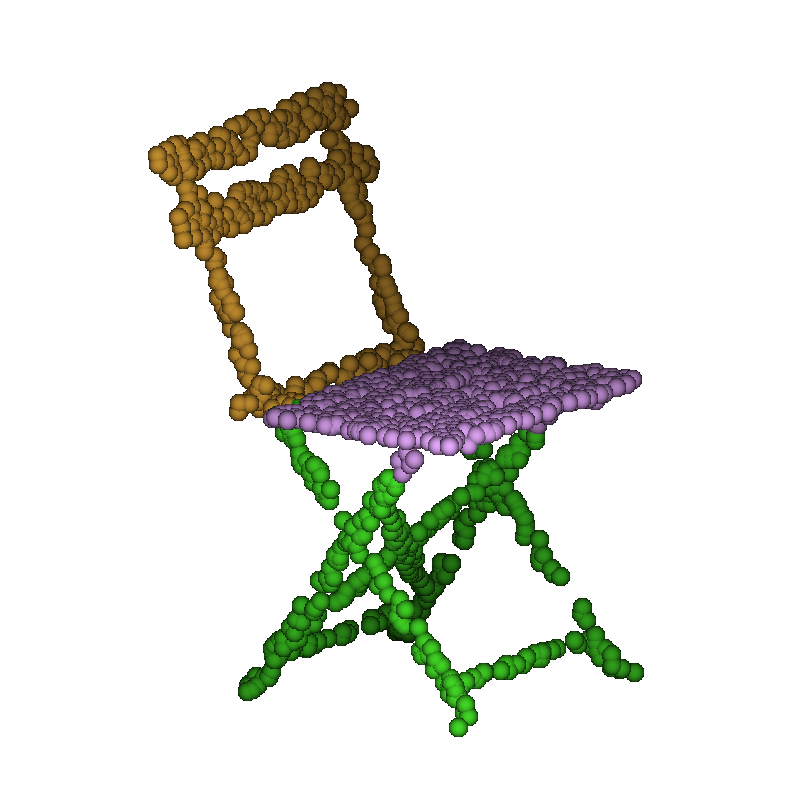}}
	{}%
	\hfill%
	\jsubfig{\includegraphics[height=3.49cm]{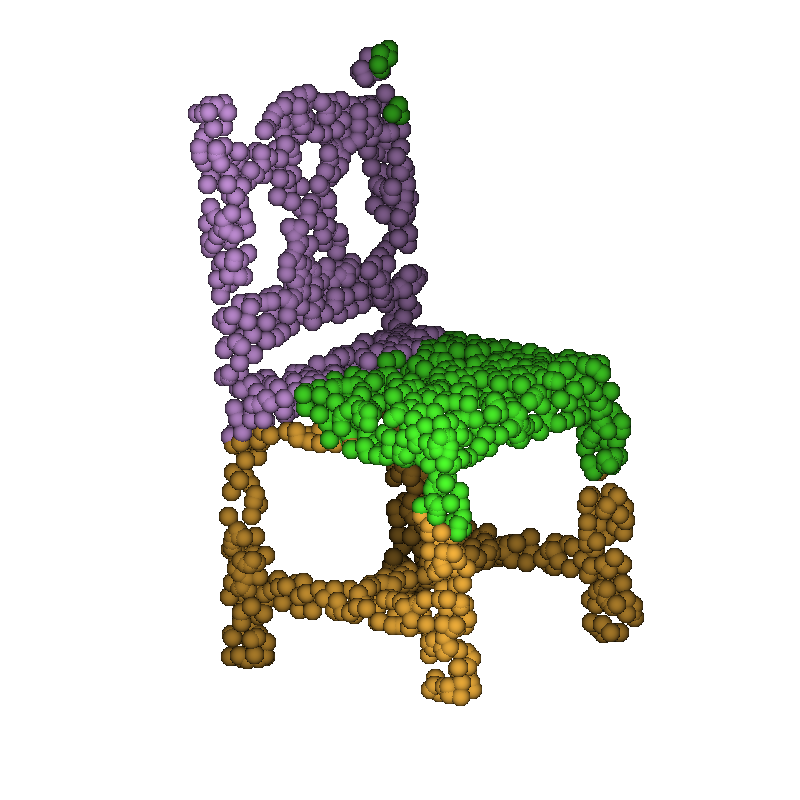}}
	{}%
	\hfill%
	\jsubfig{\includegraphics[height=3.49cm]{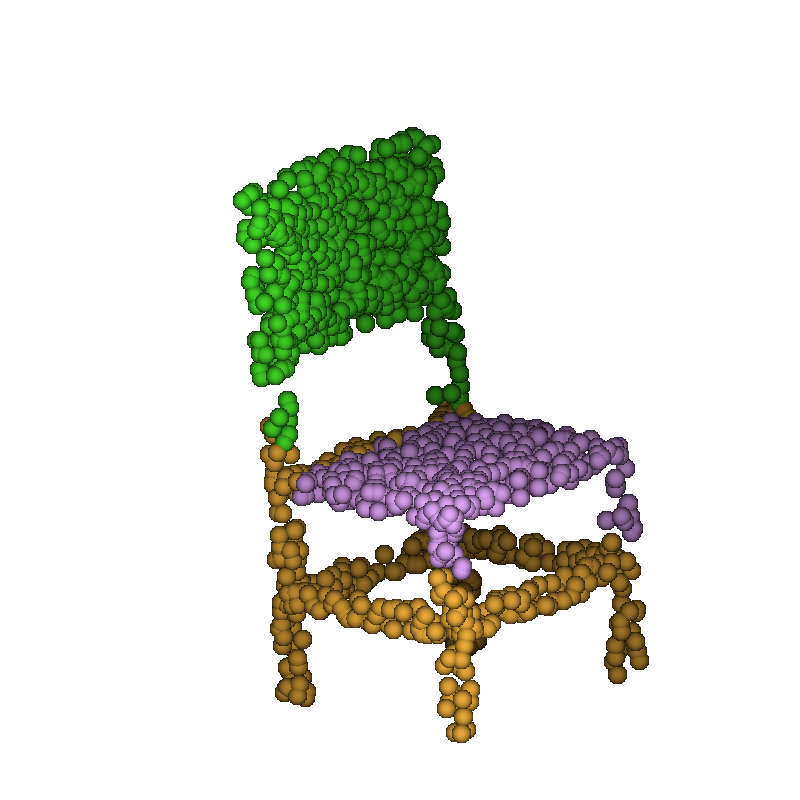}}
	{}%
\\
\vspace{6pt}
	\jsubfig{\includegraphics[height=3.49cm]{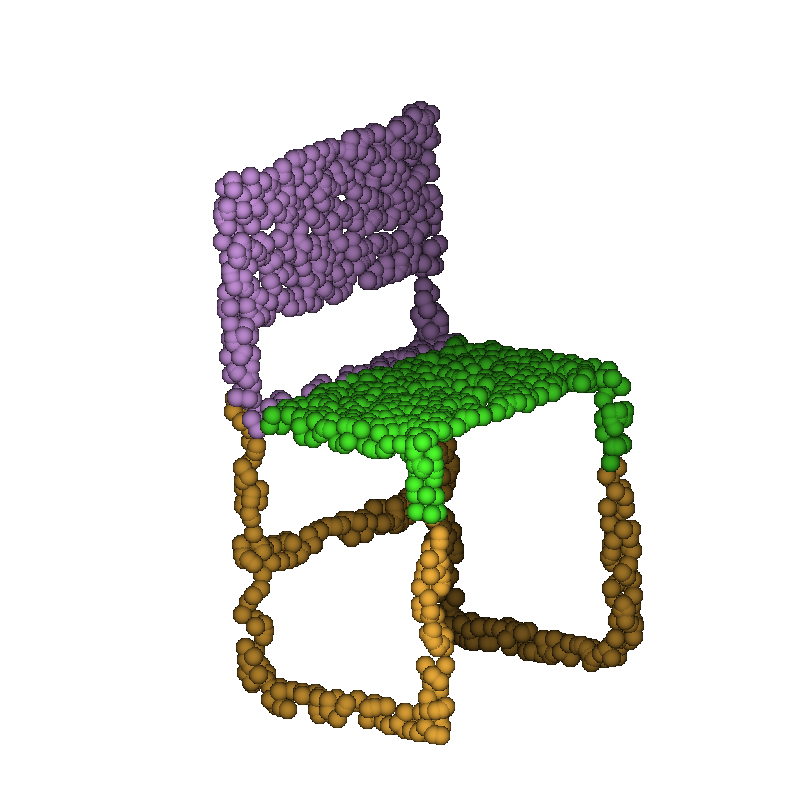}}
	{}%
	\hfill%
	\jsubfig{\includegraphics[height=3.49cm]{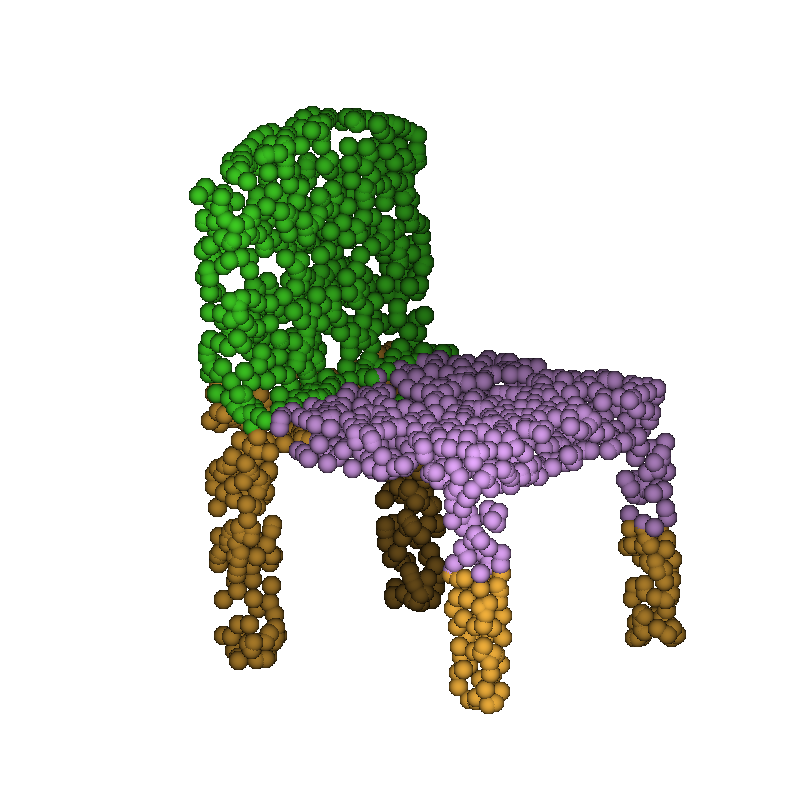}}
	{}%
	\hfill%
	\jsubfig{\includegraphics[height=3.49cm]{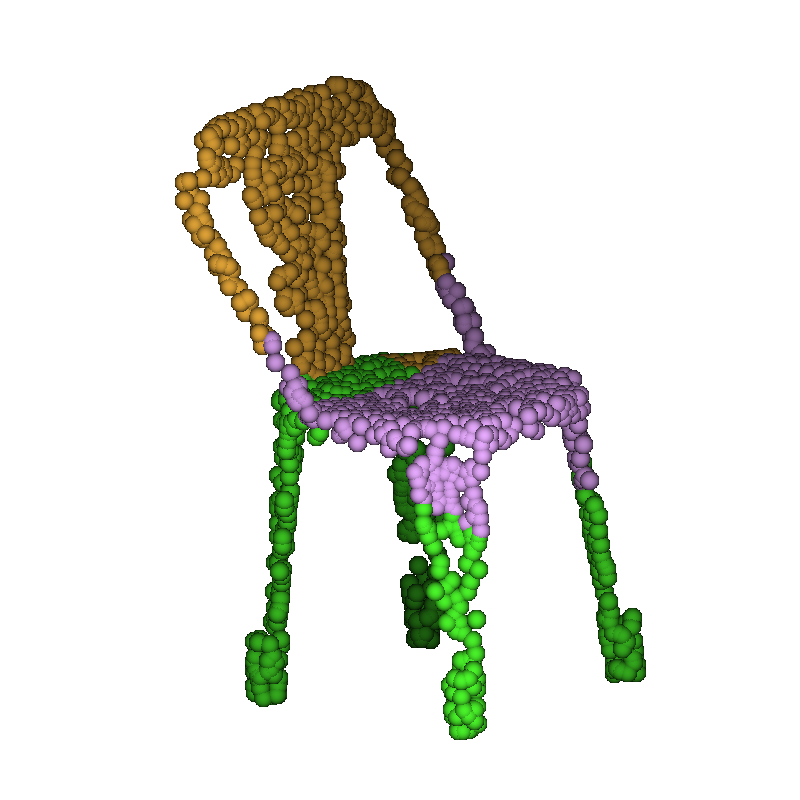}}
	{}%
	\hfill%
	\jsubfig{\includegraphics[height=3.49cm]{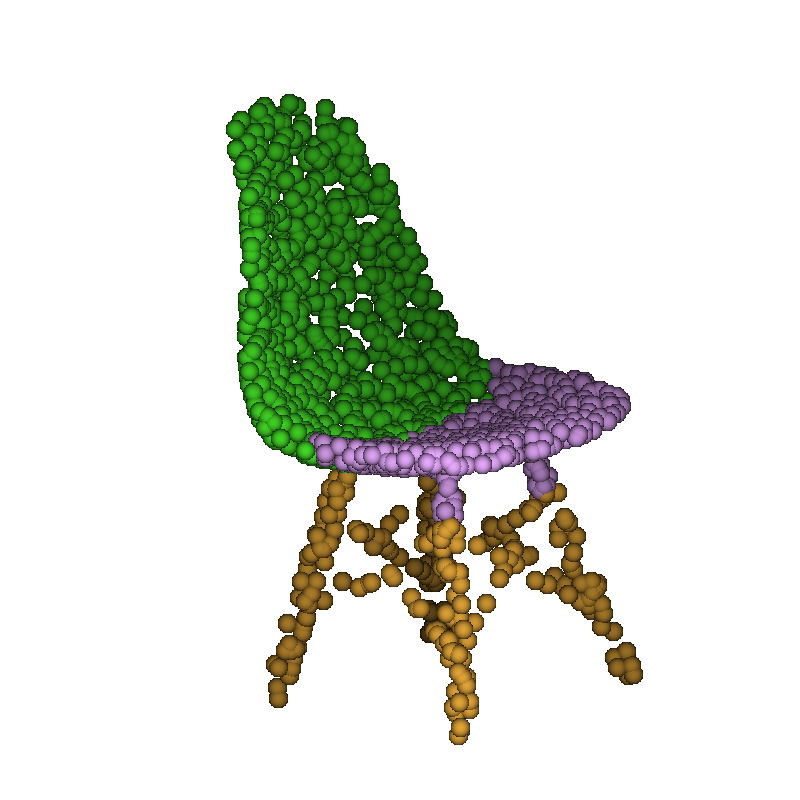}}
	{}%
	
\end{center}
\caption{Unsupervised segmentation examples on chairs taken from the training data.}

\label{fig:seg_train_chairs}
\end{figure*}


\begin{figure*}
\begin{center}
	\jsubfig{\includegraphics[height=3.49cm]{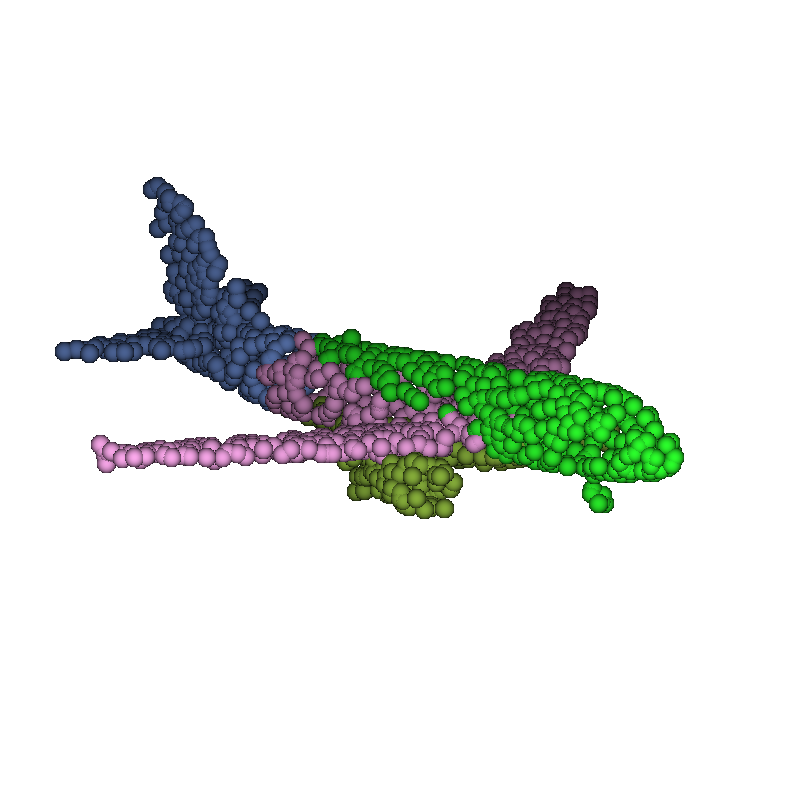}}
	{}%
	\hfill%
	\jsubfig{\includegraphics[height=3.49cm]{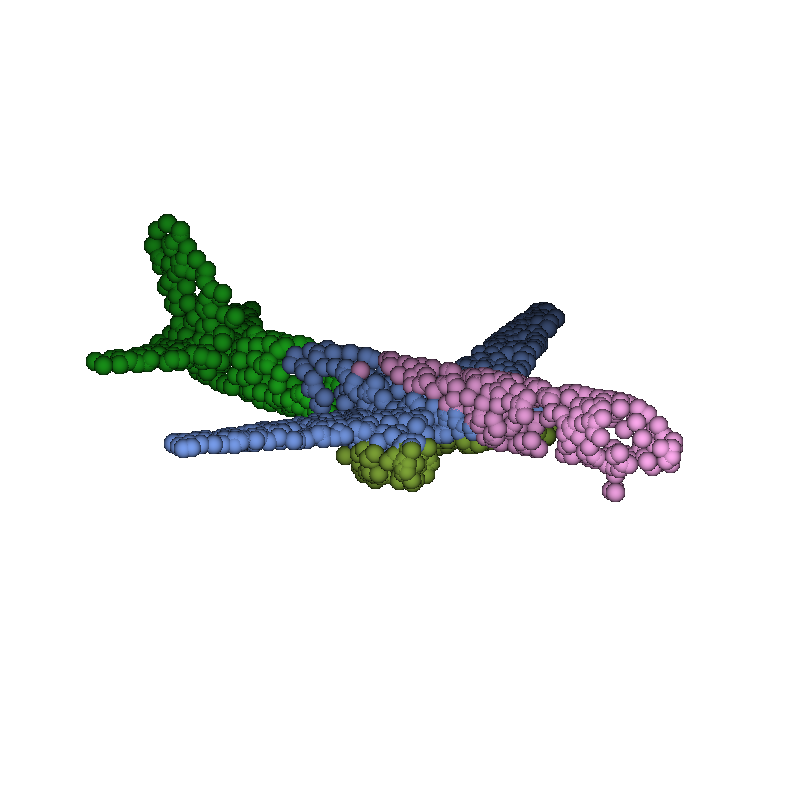}}
	{}%
	\hfill%
	\jsubfig{\includegraphics[height=3.49cm]{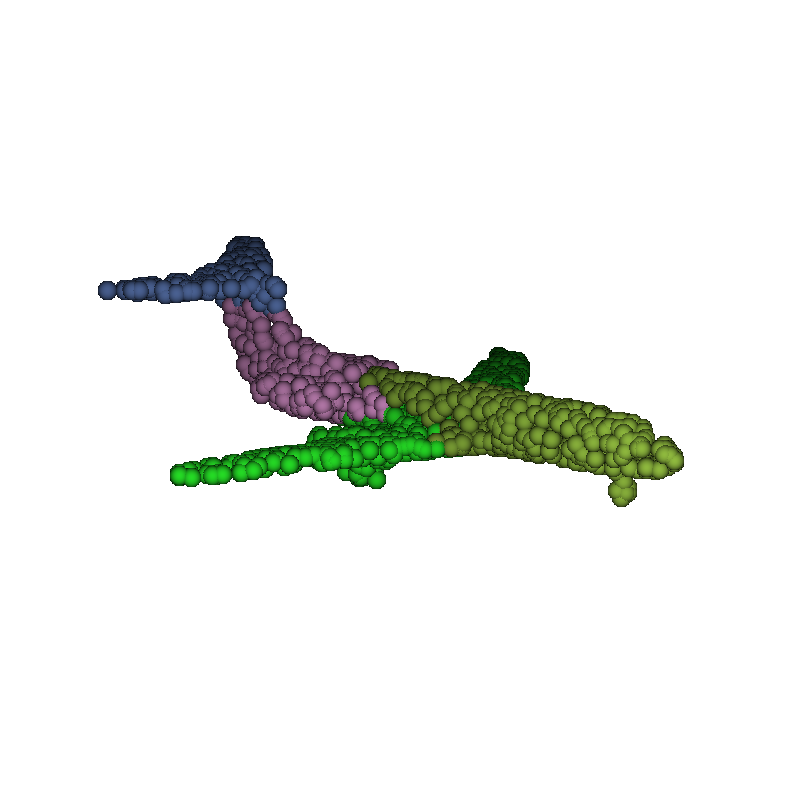}}
	{}%
	\hfill%
	\jsubfig{\includegraphics[height=3.49cm]{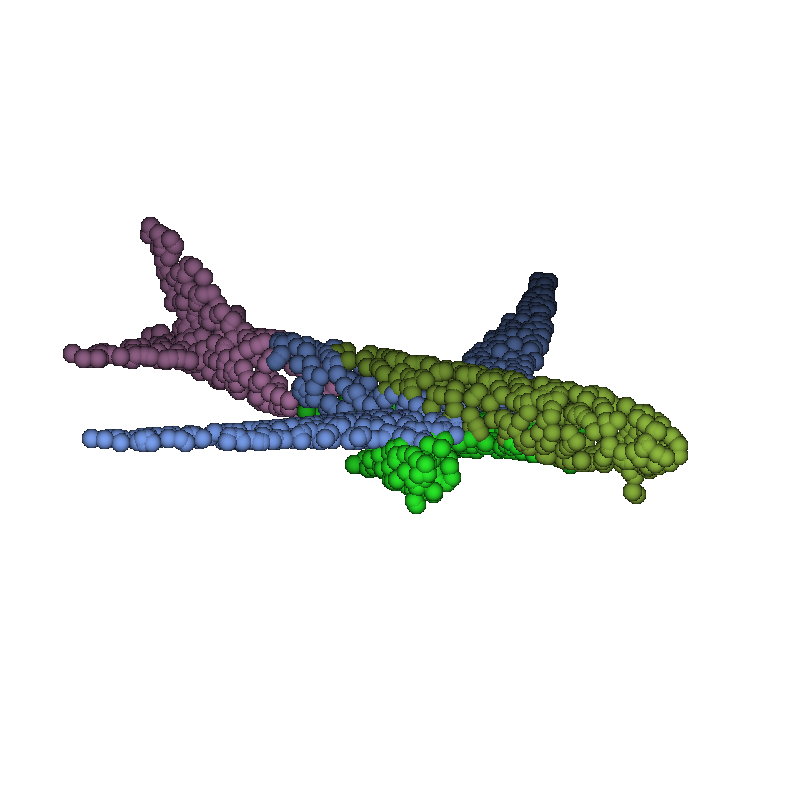}}
	{}%
\\
\vspace{6pt}
	\jsubfig{\includegraphics[height=3.49cm]{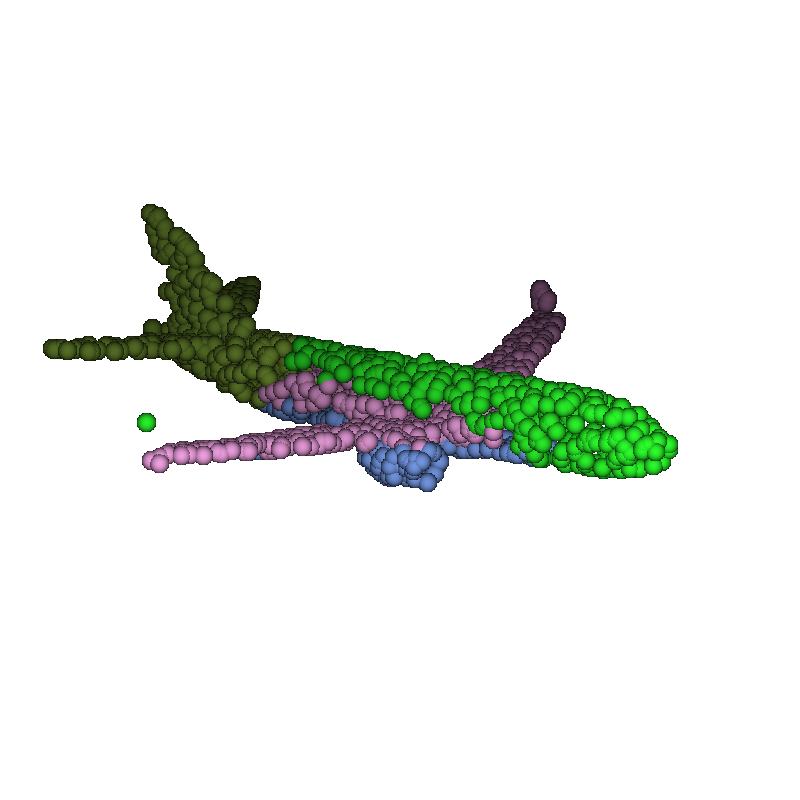}}
	{}%
	\hfill%
	\jsubfig{\includegraphics[height=3.49cm]{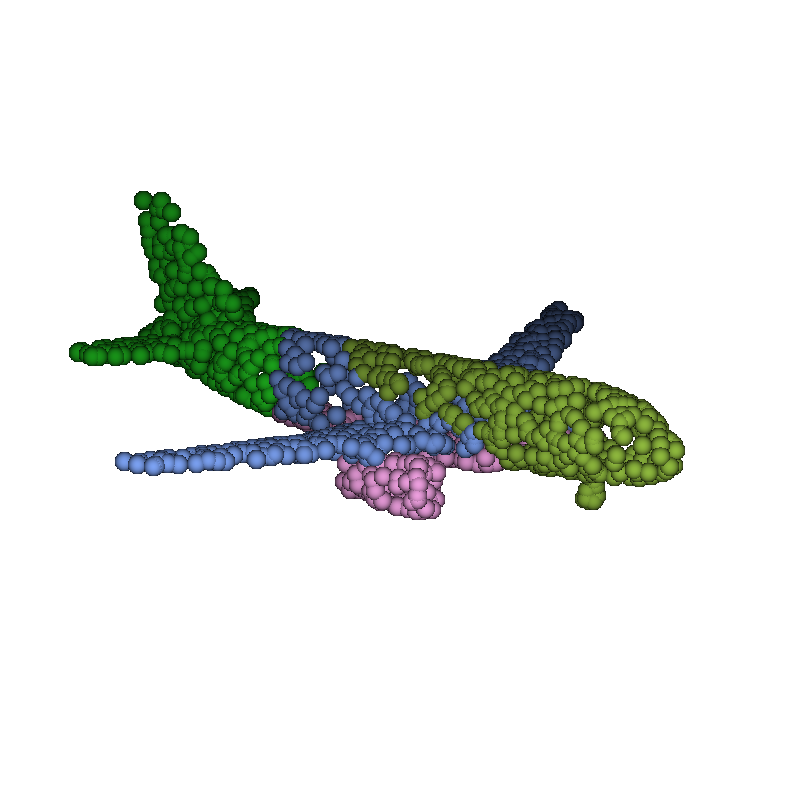}}
	{}%
	\hfill%
	\jsubfig{\includegraphics[height=3.49cm]{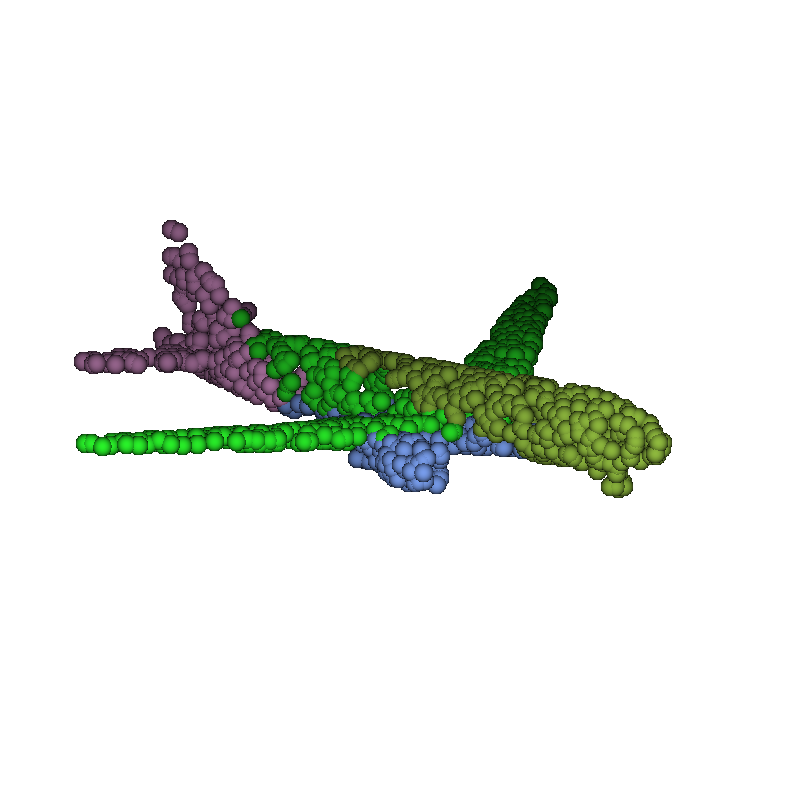}}
	{}%
	\hfill%
	\jsubfig{\includegraphics[height=3.49cm]{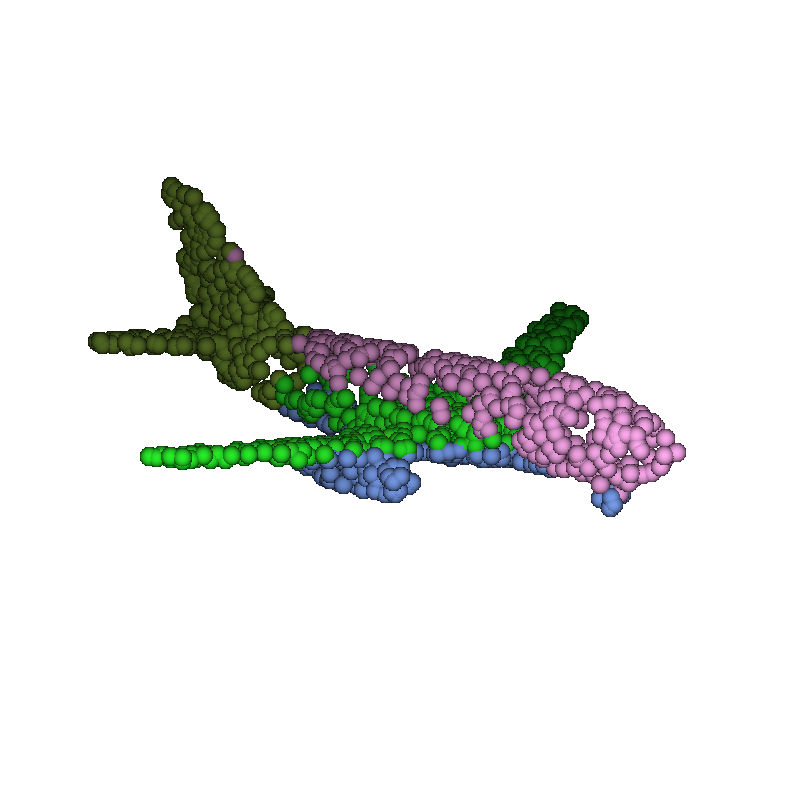}}
	{}%
\\
\vspace{6pt}
	\jsubfig{\includegraphics[height=3.49cm]{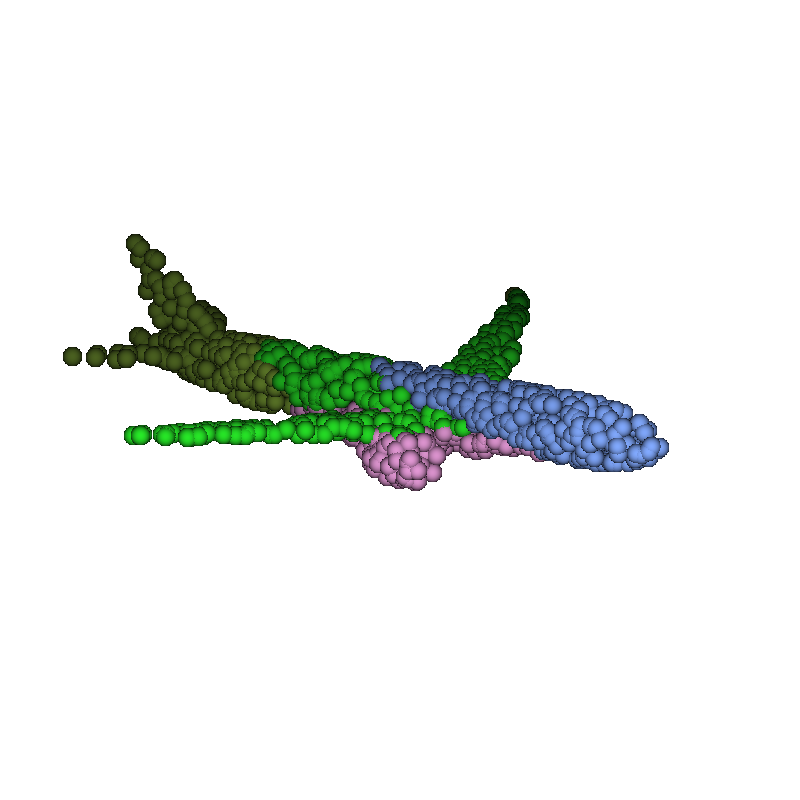}}
	{}%
	\hfill%
	\jsubfig{\includegraphics[height=3.49cm]{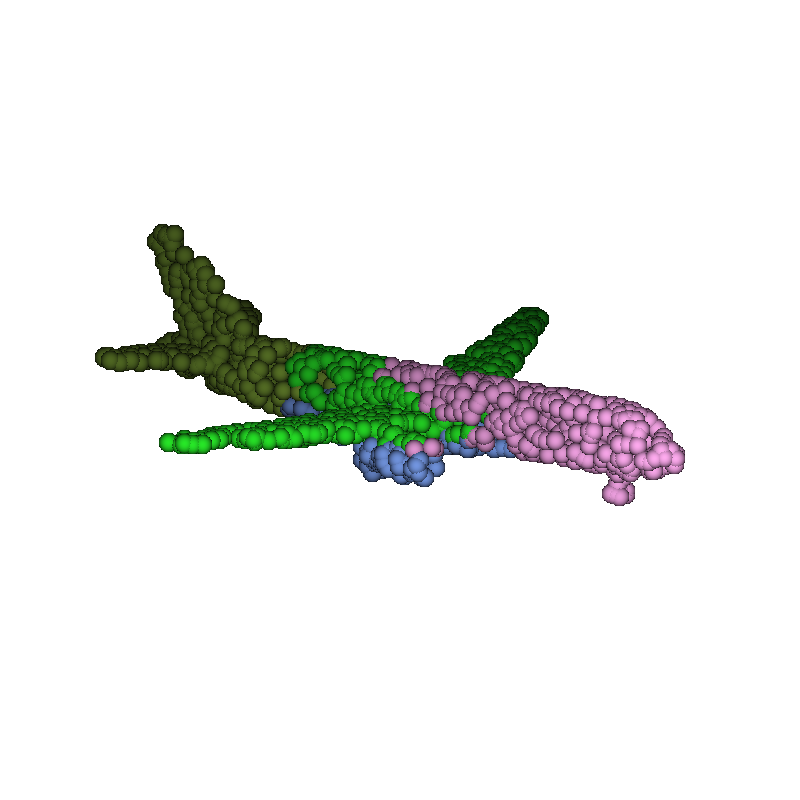}}
	{}%
	\hfill%
	\jsubfig{\includegraphics[height=3.49cm]{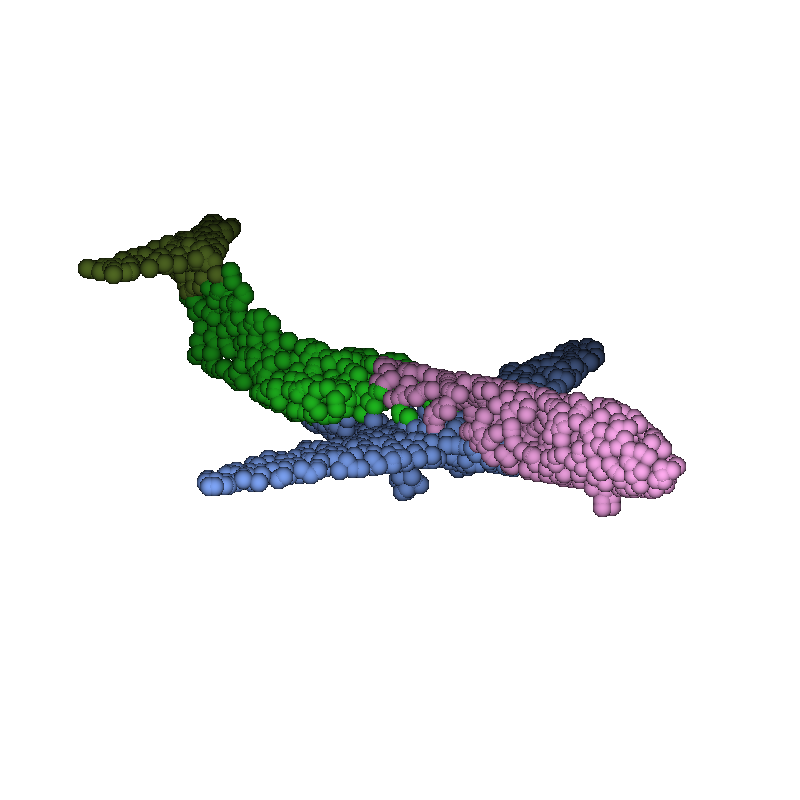}}
	{}%
	\hfill%
	\jsubfig{\includegraphics[height=3.49cm]{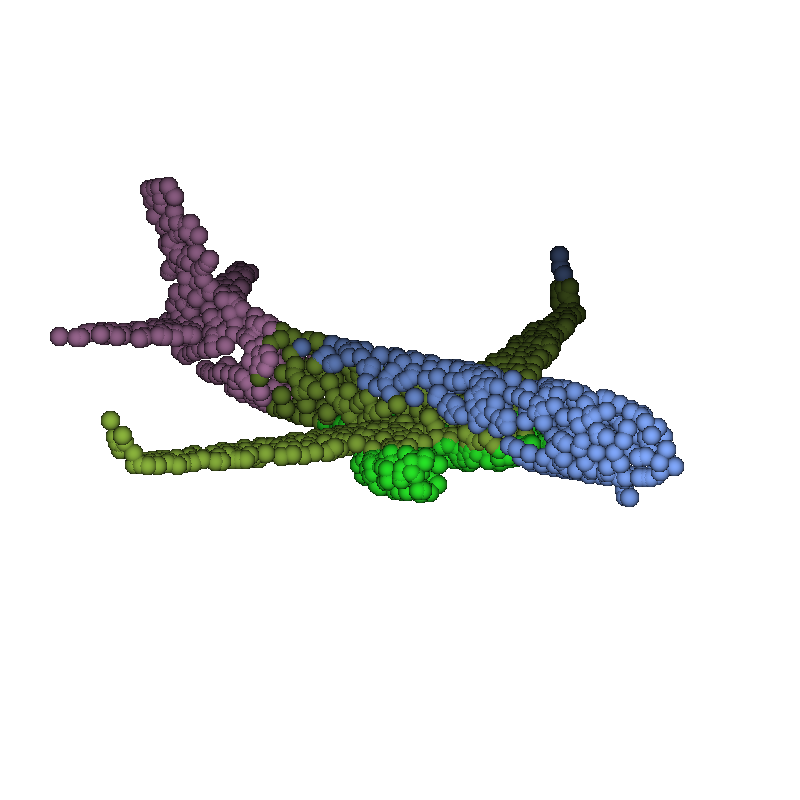}}
	{}%
\\
\vspace{6pt}
	\jsubfig{\includegraphics[height=3.49cm]{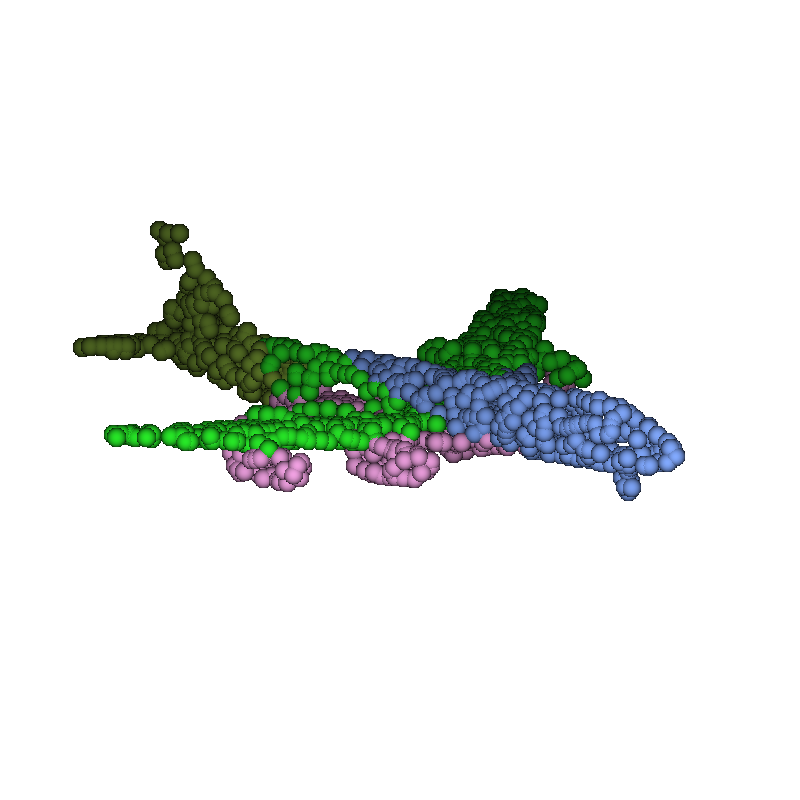}}
	{}%
	\hfill%
	\jsubfig{\includegraphics[height=3.49cm]{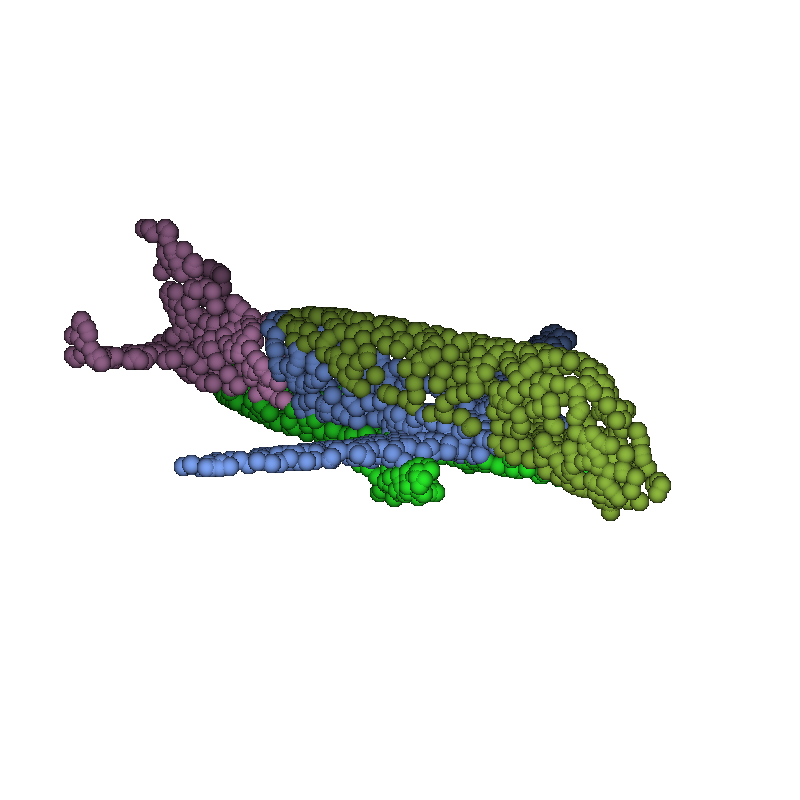}}
	{}%
	\hfill%
	\jsubfig{\includegraphics[height=3.49cm]{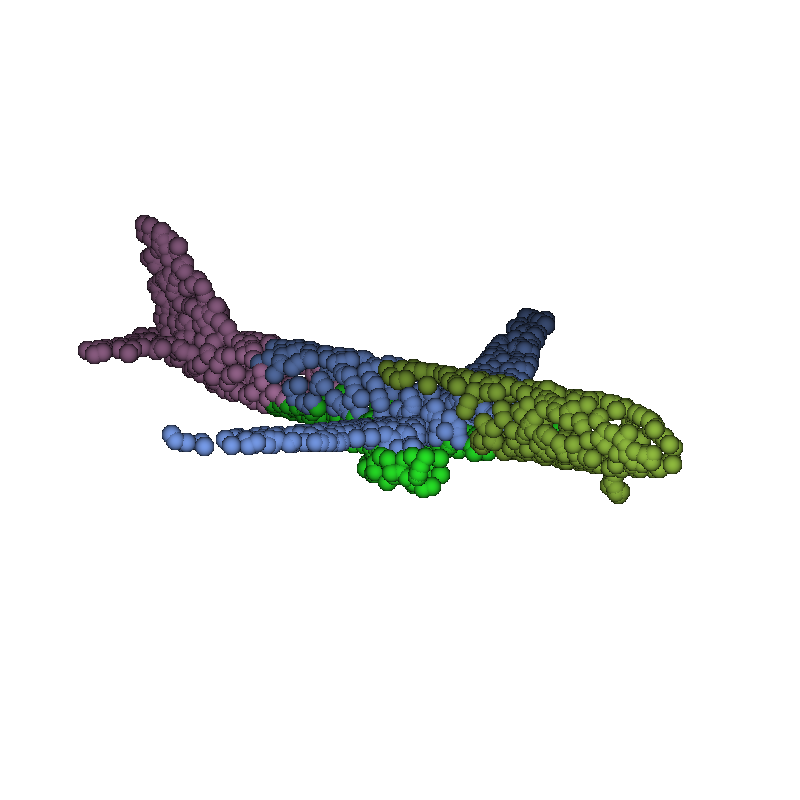}}
	{}%
	\hfill%
	\jsubfig{\includegraphics[height=3.49cm]{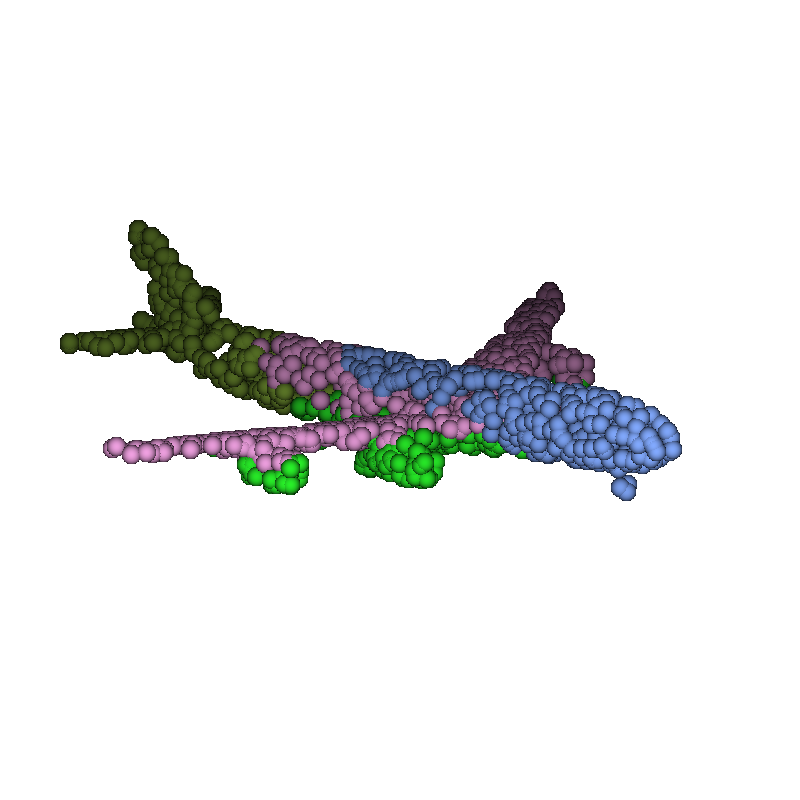}}
	{}%
\\
\vspace{6pt}
	\jsubfig{\includegraphics[height=3.49cm]{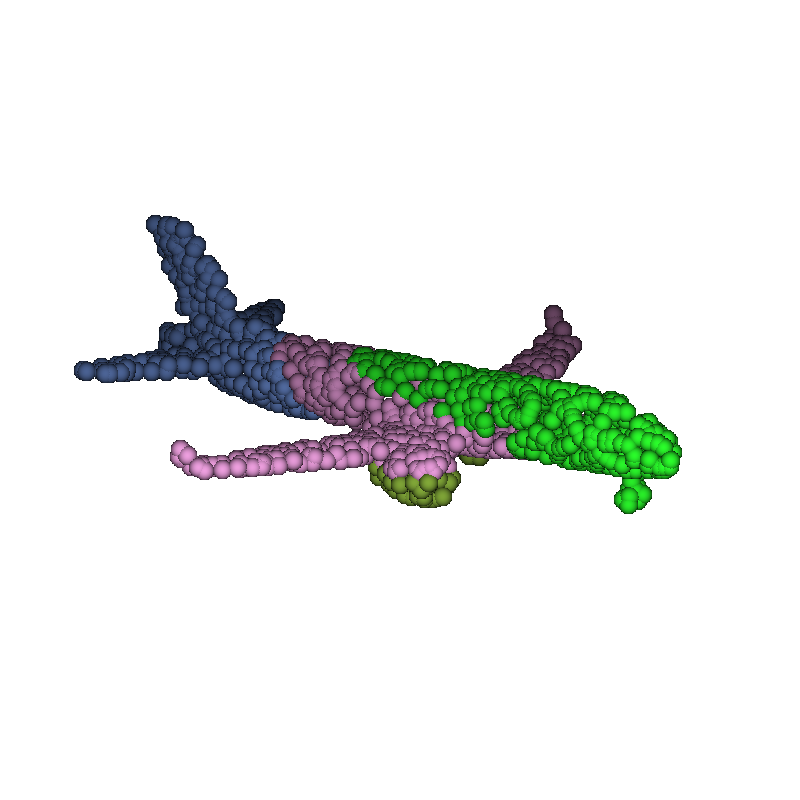}}
	{}%
	\hfill%
	\jsubfig{\includegraphics[height=3.49cm]{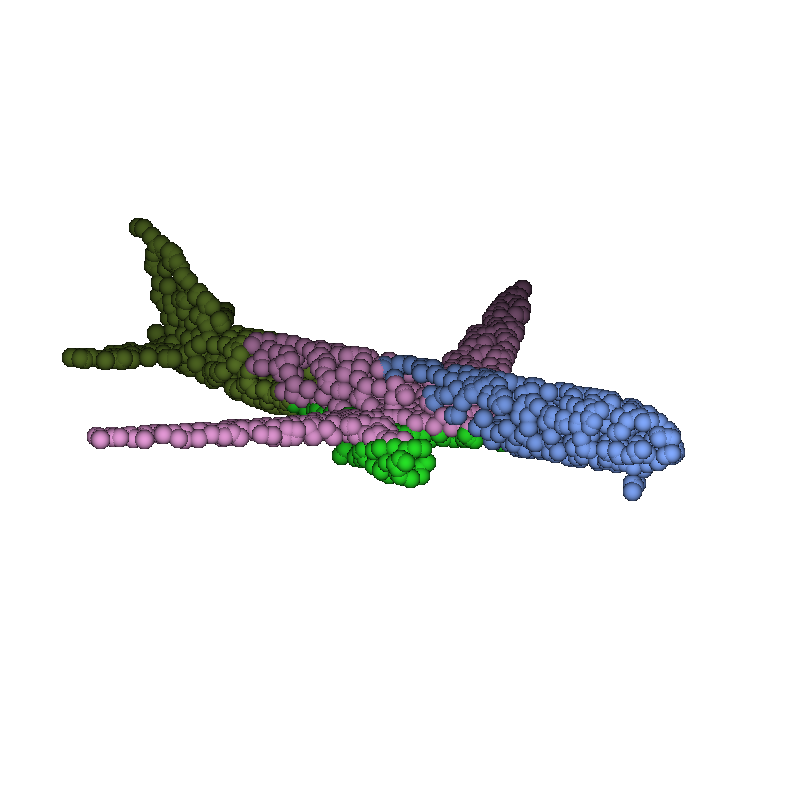}}
	{}%
	\hfill%
	\jsubfig{\includegraphics[height=3.49cm]{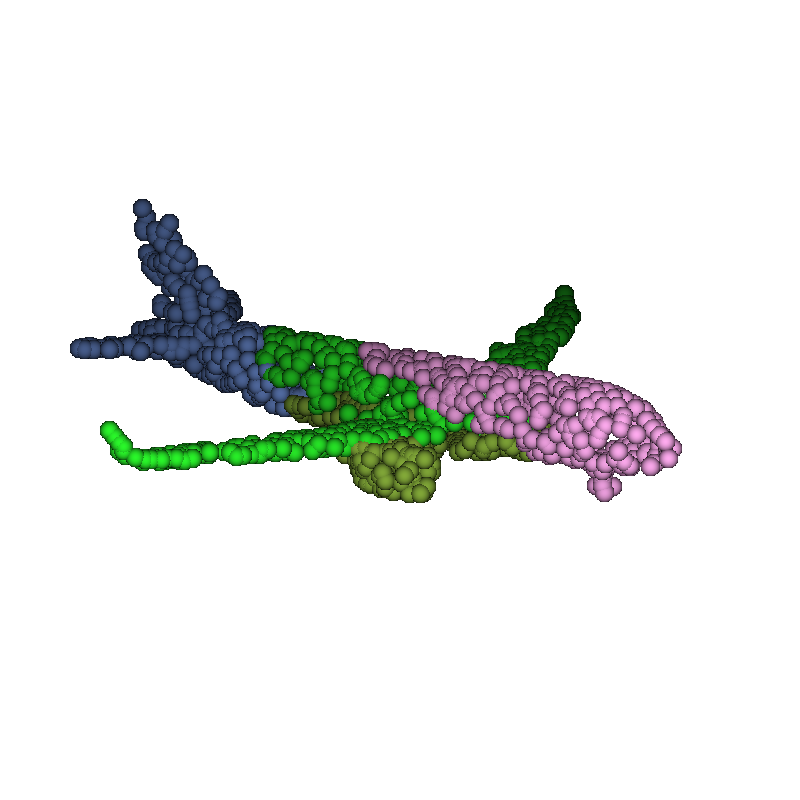}}
	{}%
	\hfill%
	\jsubfig{\includegraphics[height=3.49cm]{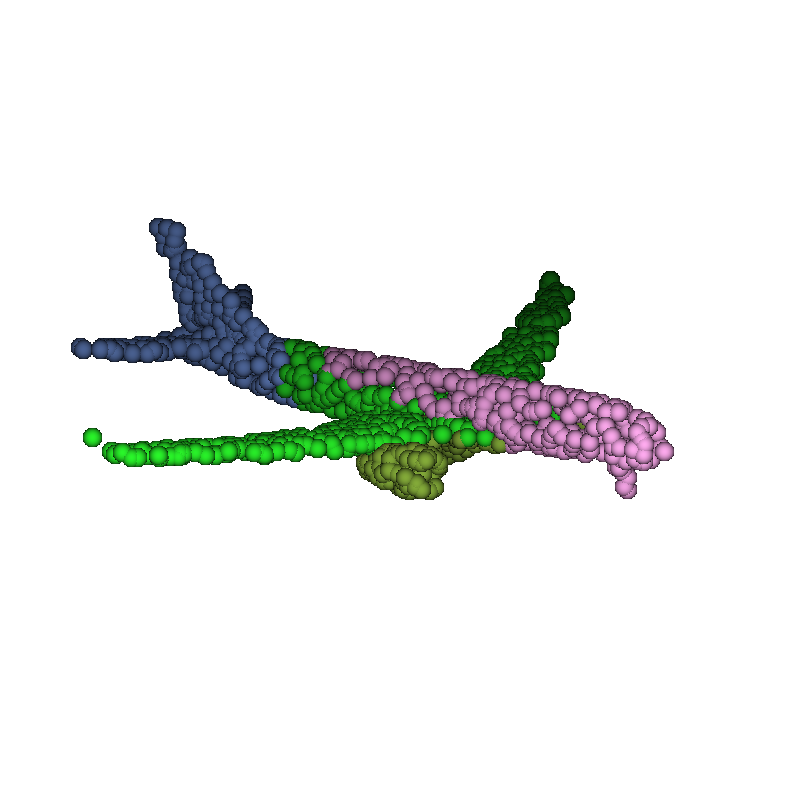}}
	{}%
	
\end{center}
\caption{Unsupervised segmentation examples on airplanes taken from the training data.}

\label{fig:seg_train_airplanes}
\end{figure*}


\begin{figure*}
\begin{center}
	\jsubfig{\includegraphics[height=3.49cm]{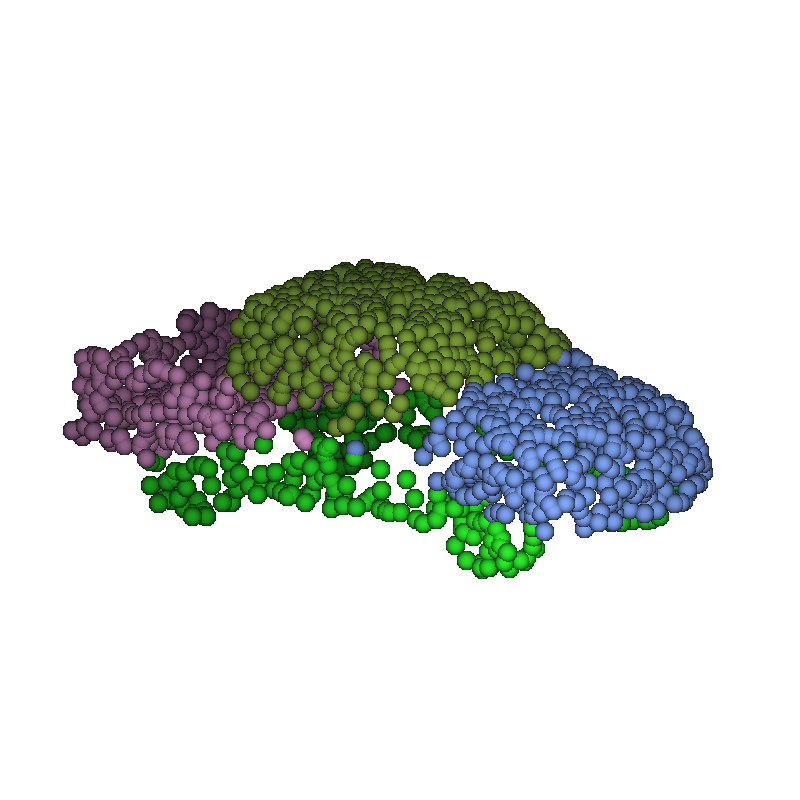}}
	{}%
	\hfill%
	\jsubfig{\includegraphics[height=3.49cm]{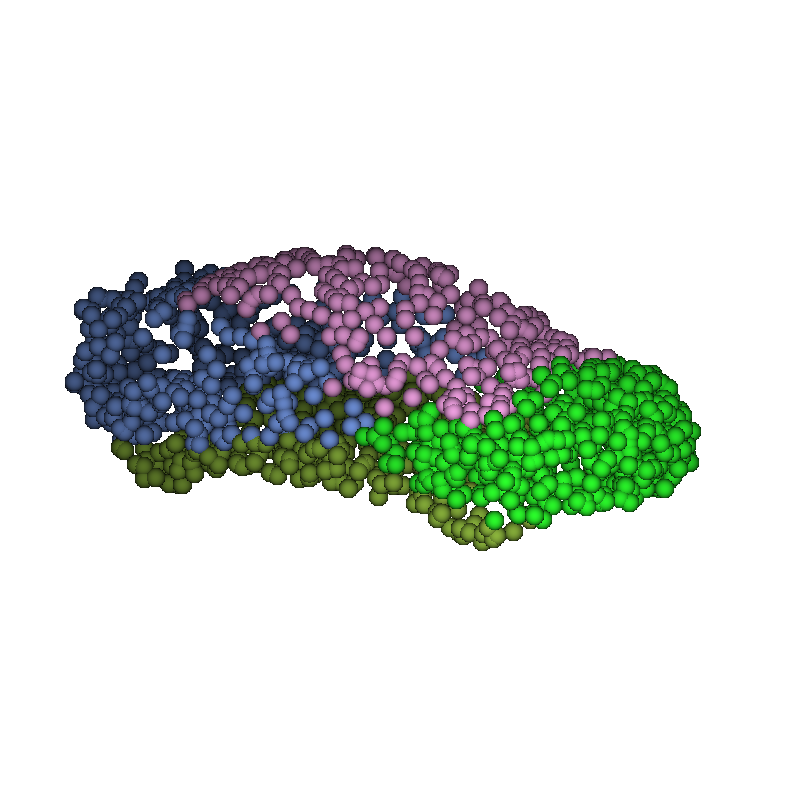}}
	{}%
	\hfill%
	\jsubfig{\includegraphics[height=3.49cm]{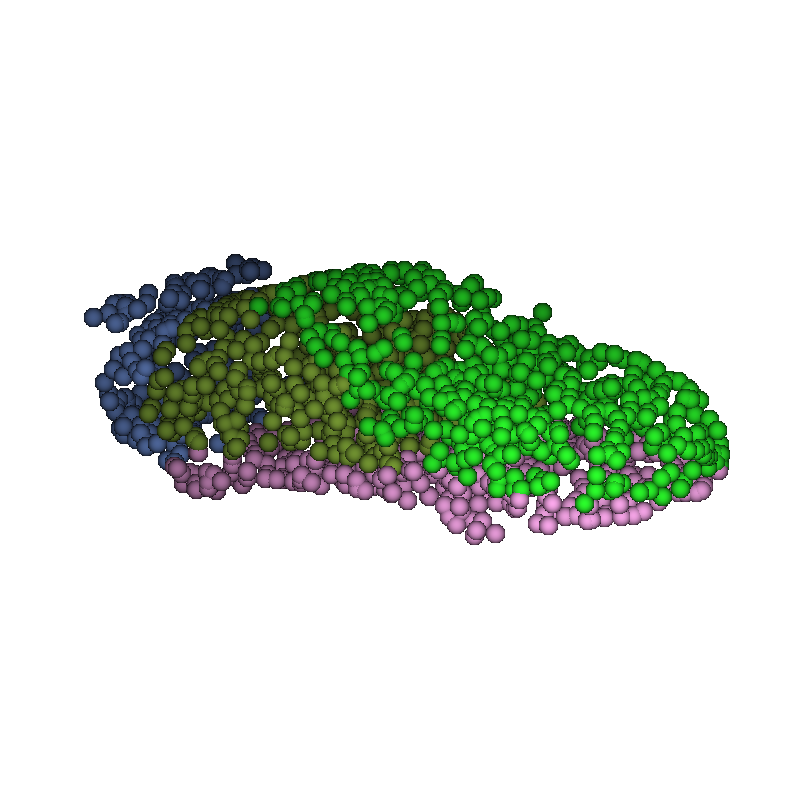}}
	{}%
	\hfill%
	\jsubfig{\includegraphics[height=3.49cm]{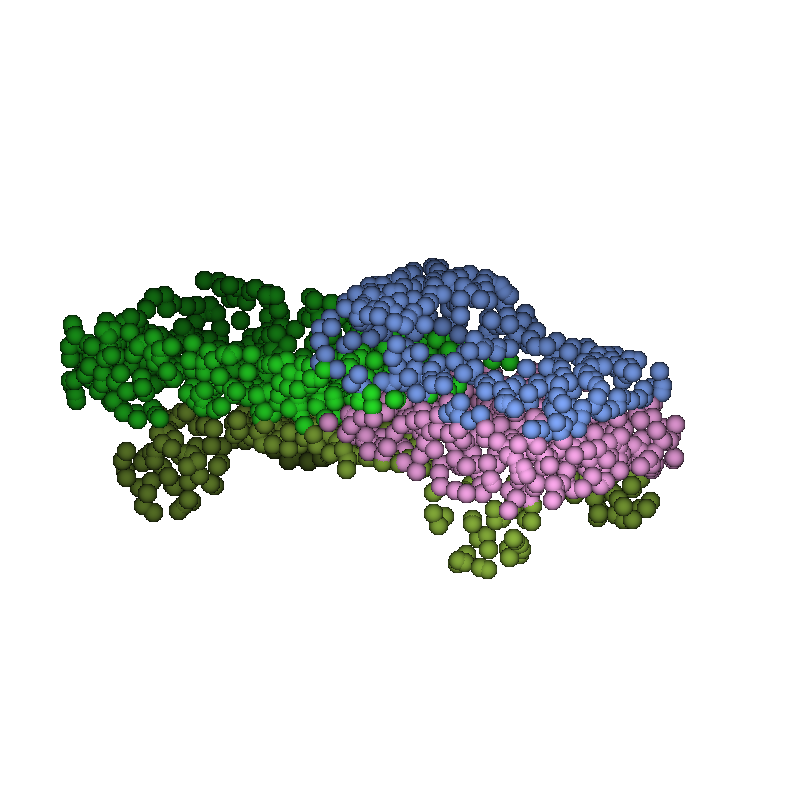}}
	{}%
\\
\vspace{6pt}
	\jsubfig{\includegraphics[height=3.49cm]{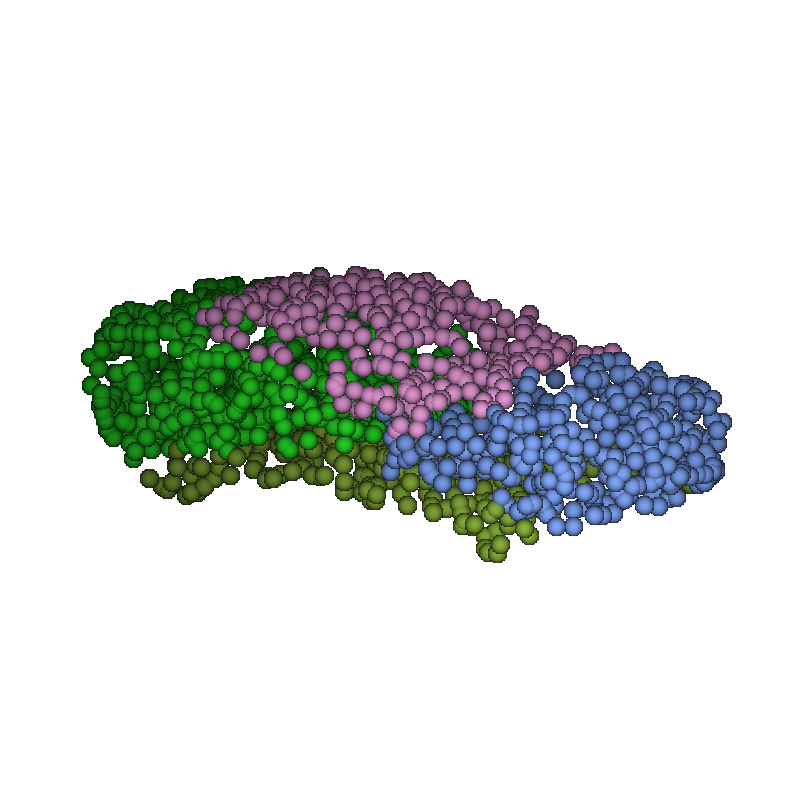}}
	{}%
	\hfill%
	\jsubfig{\includegraphics[height=3.49cm]{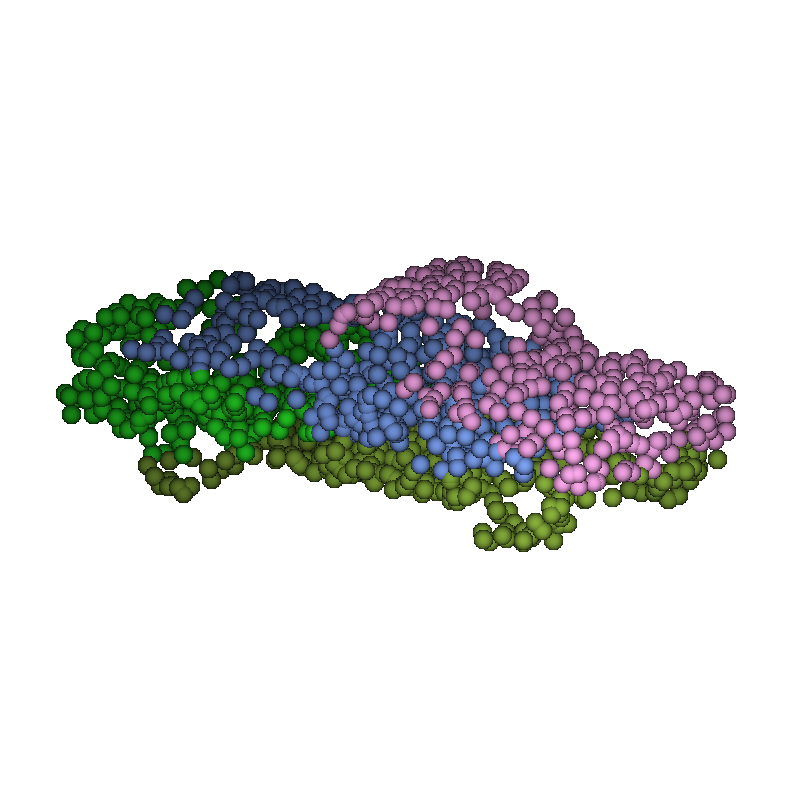}}
	{}%
	\hfill%
	\jsubfig{\includegraphics[height=3.49cm]{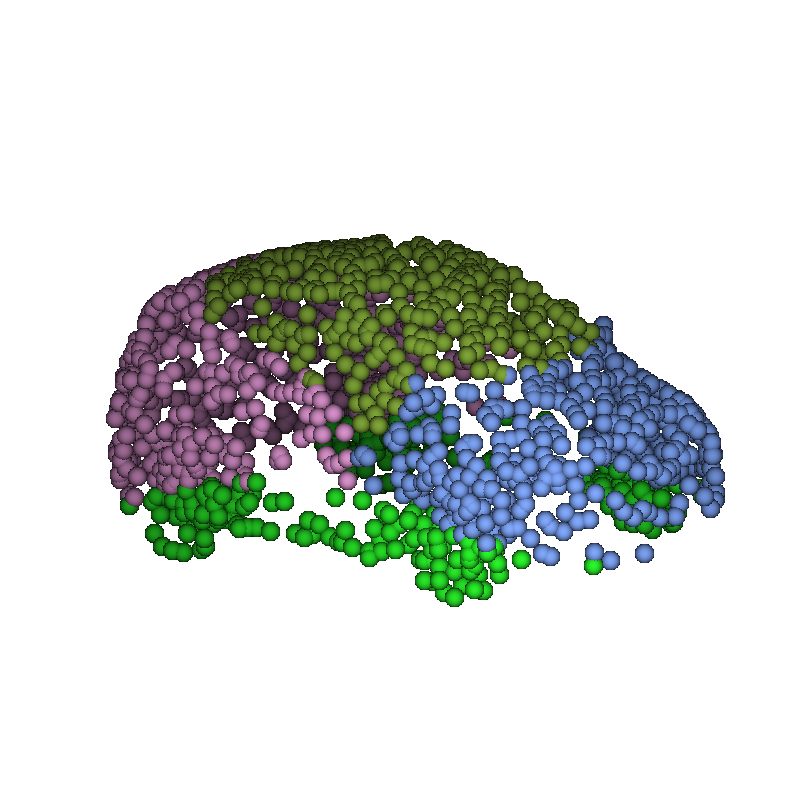}}
	{}%
	\hfill%
	\jsubfig{\includegraphics[height=3.49cm]{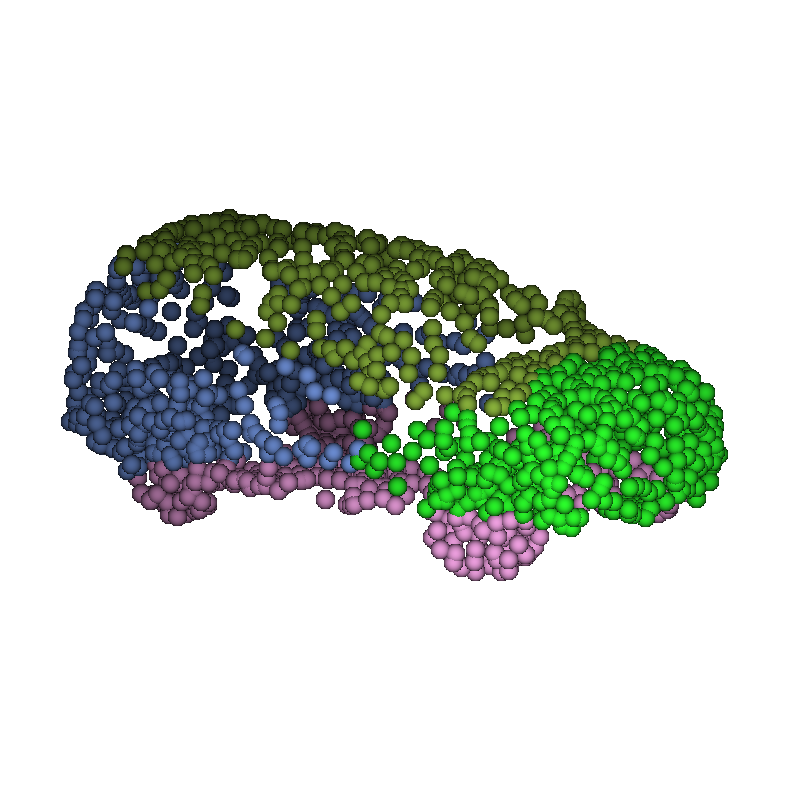}}
	{}%
\\
\vspace{6pt}
	\jsubfig{\includegraphics[height=3.49cm]{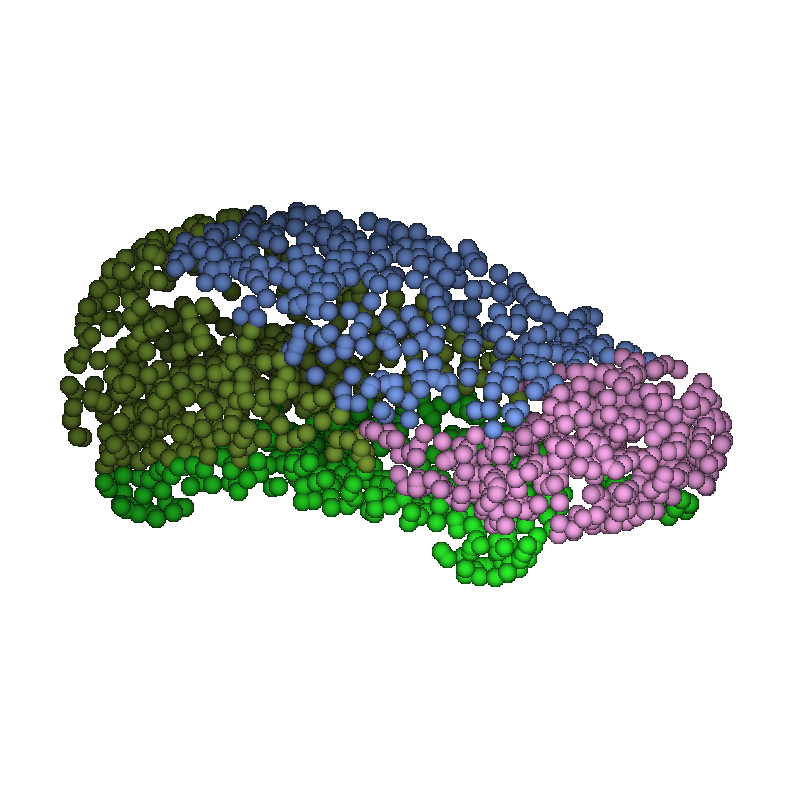}}
	{}%
	\hfill%
	\jsubfig{\includegraphics[height=3.49cm]{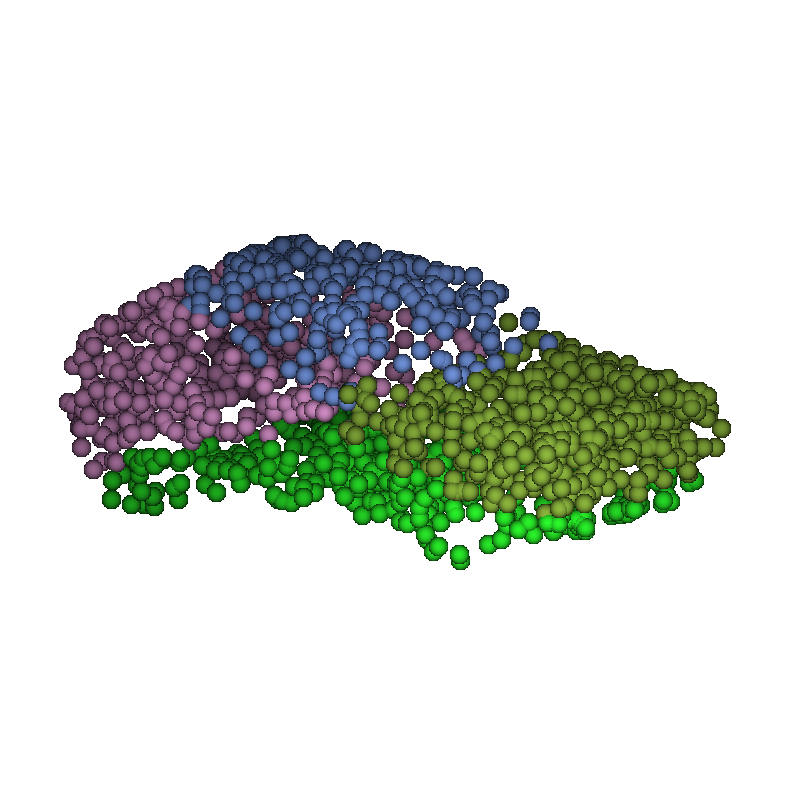}}
	{}%
	\hfill%
	\jsubfig{\includegraphics[height=3.49cm]{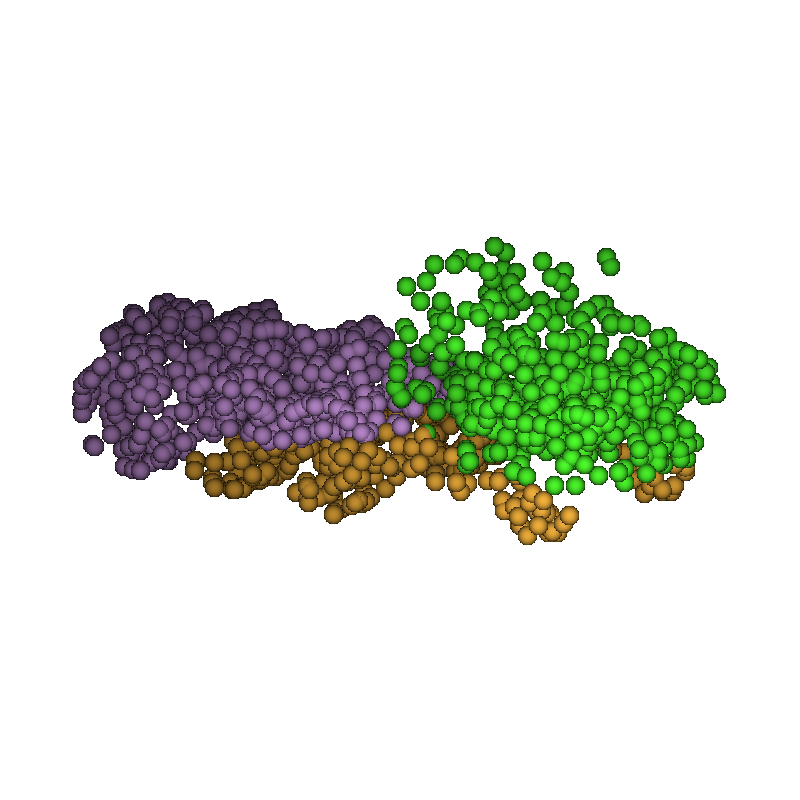}}
	{}%
	\hfill%
	\jsubfig{\includegraphics[height=3.49cm]{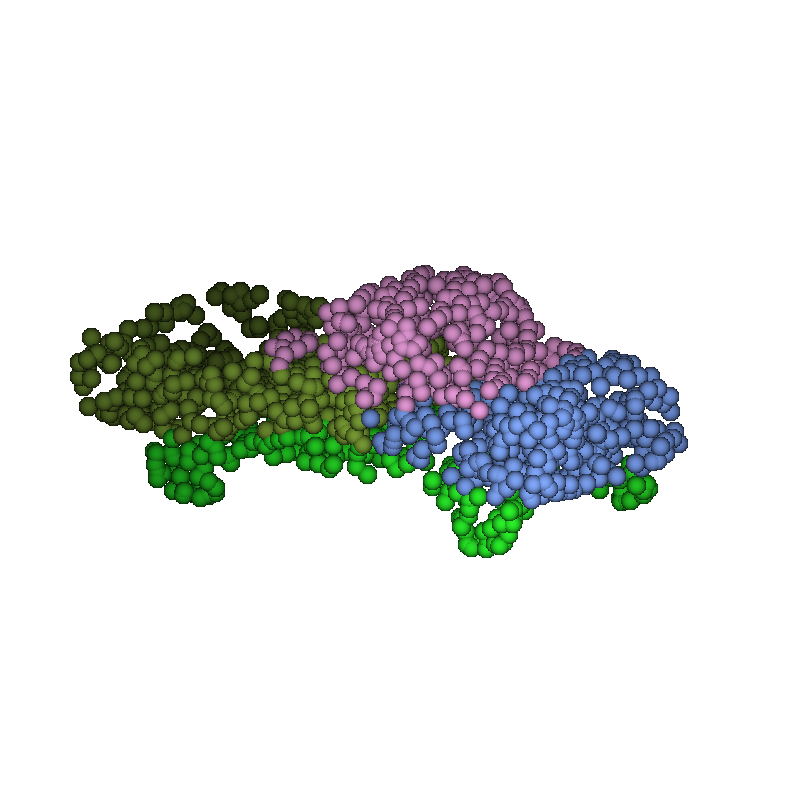}}
	{}%
\\
\vspace{6pt}
	\jsubfig{\includegraphics[height=3.49cm]{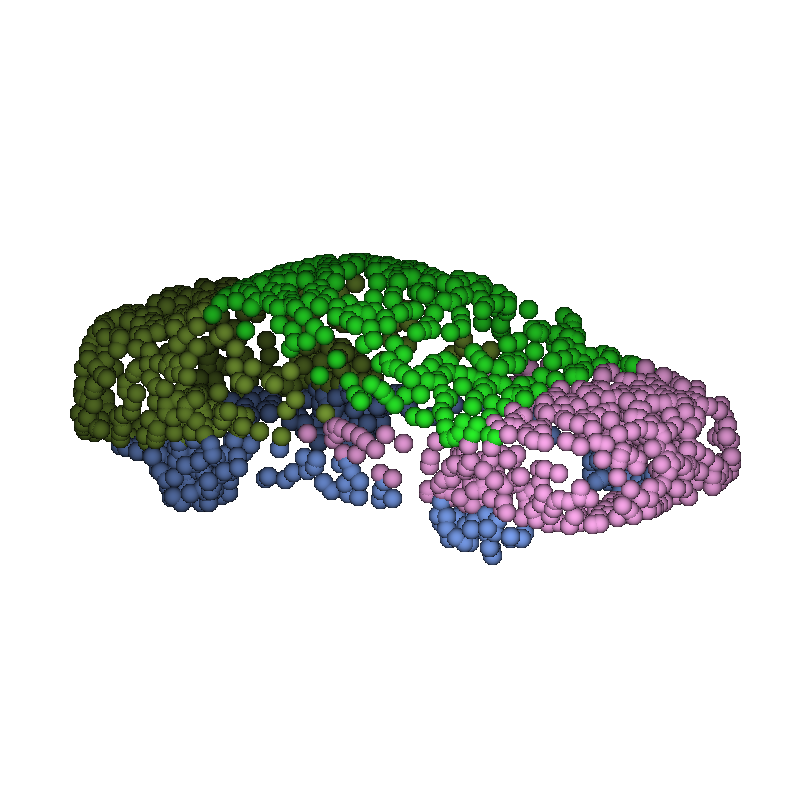}}
	{}%
	\hfill%
	\jsubfig{\includegraphics[height=3.49cm]{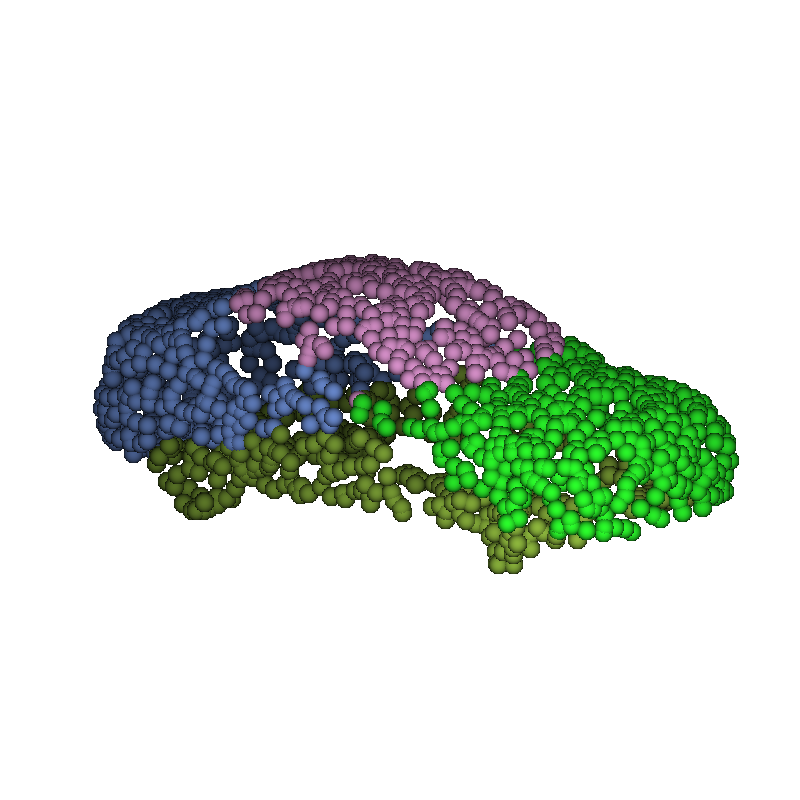}}
	{}%
	\hfill%
	\jsubfig{\includegraphics[height=3.49cm]{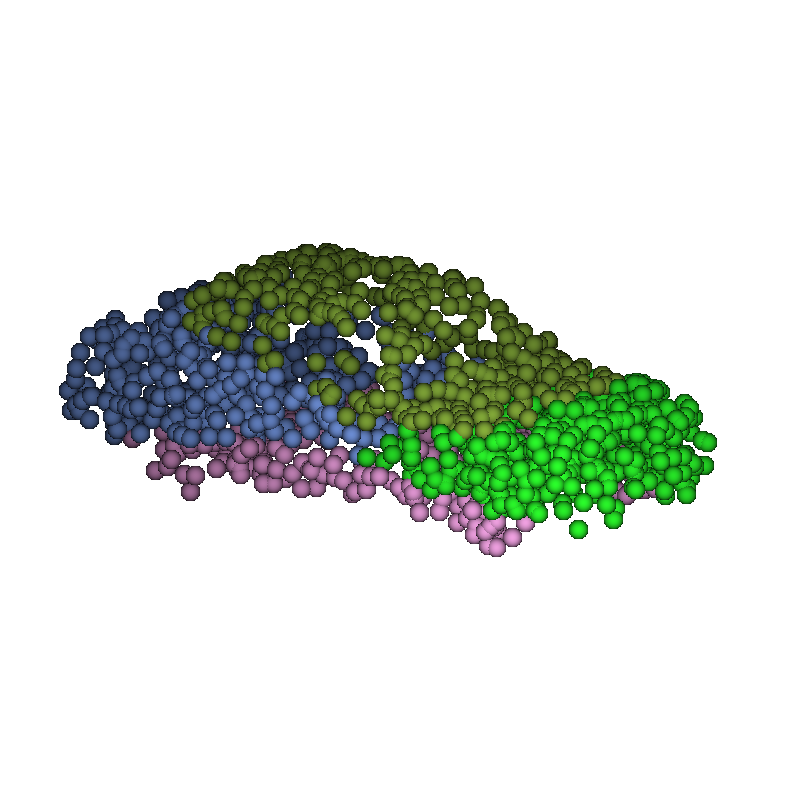}}
	{}%
	\hfill%
	\jsubfig{\includegraphics[height=3.49cm]{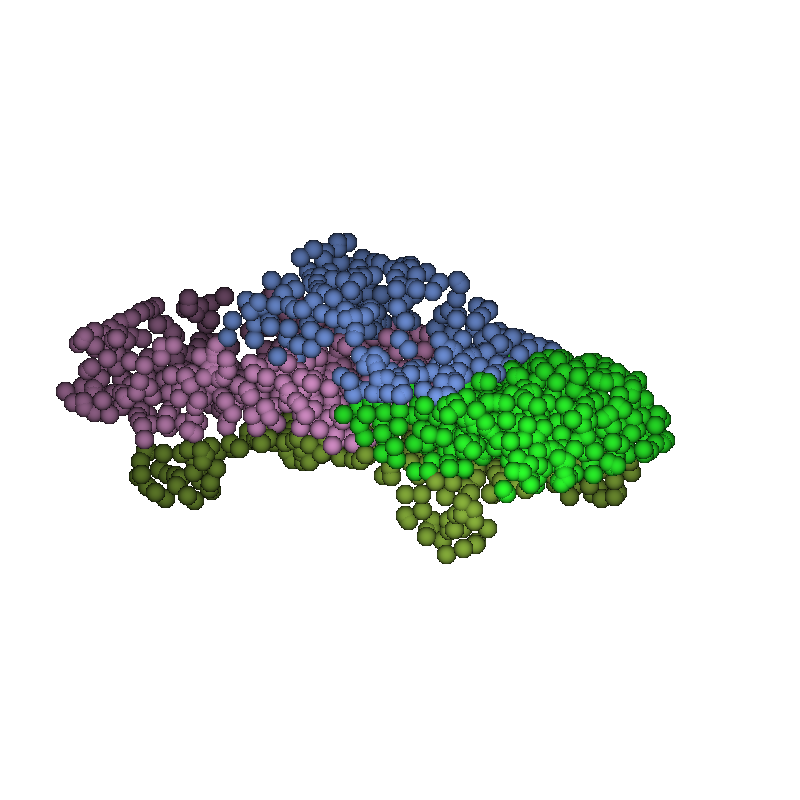}}
	{}%
\\
\vspace{6pt}
	\jsubfig{\includegraphics[height=3.49cm]{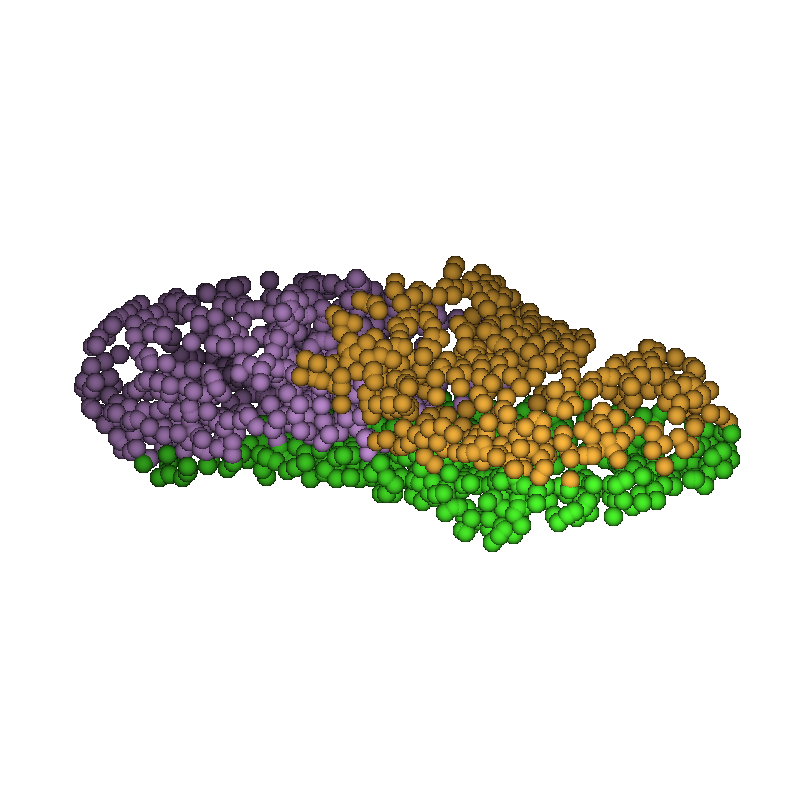}}
	{}%
	\hfill%
	\jsubfig{\includegraphics[height=3.49cm]{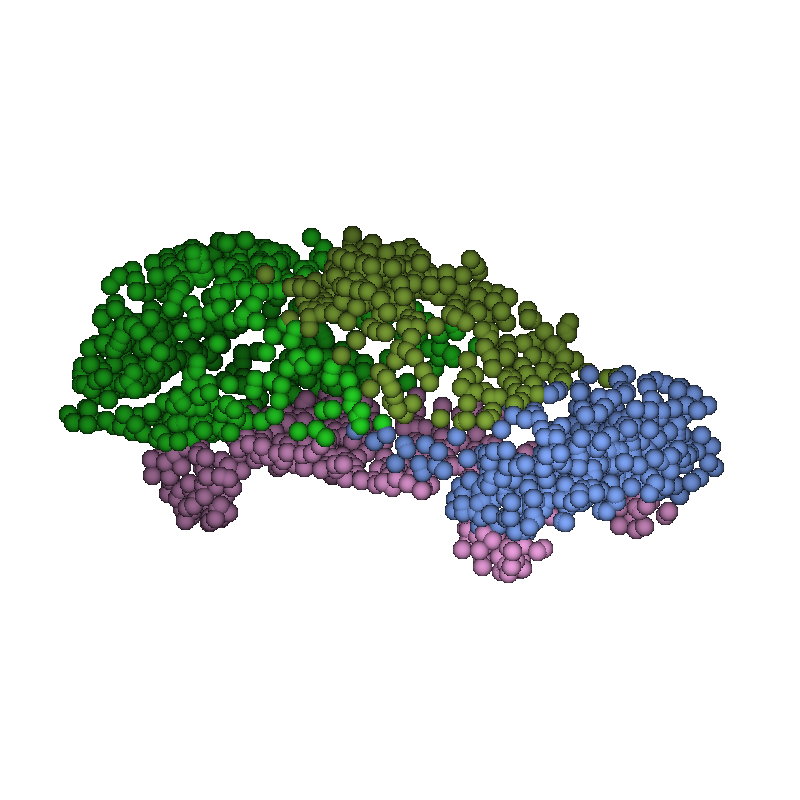}}
	{}%
	\hfill%
	\jsubfig{\includegraphics[height=3.49cm]{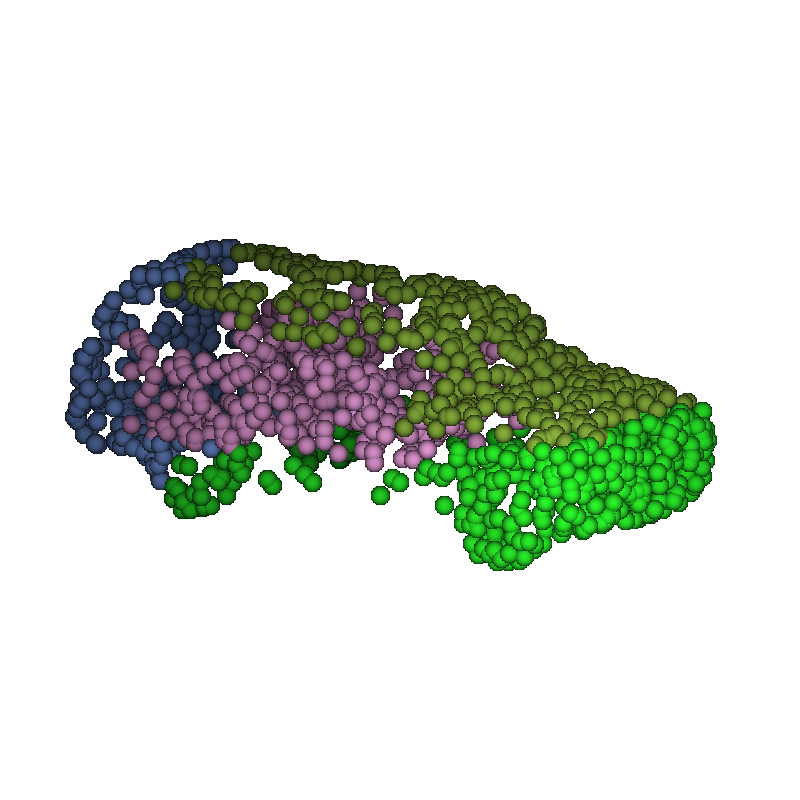}}
	{}%
	\hfill%
	\jsubfig{\includegraphics[height=3.49cm]{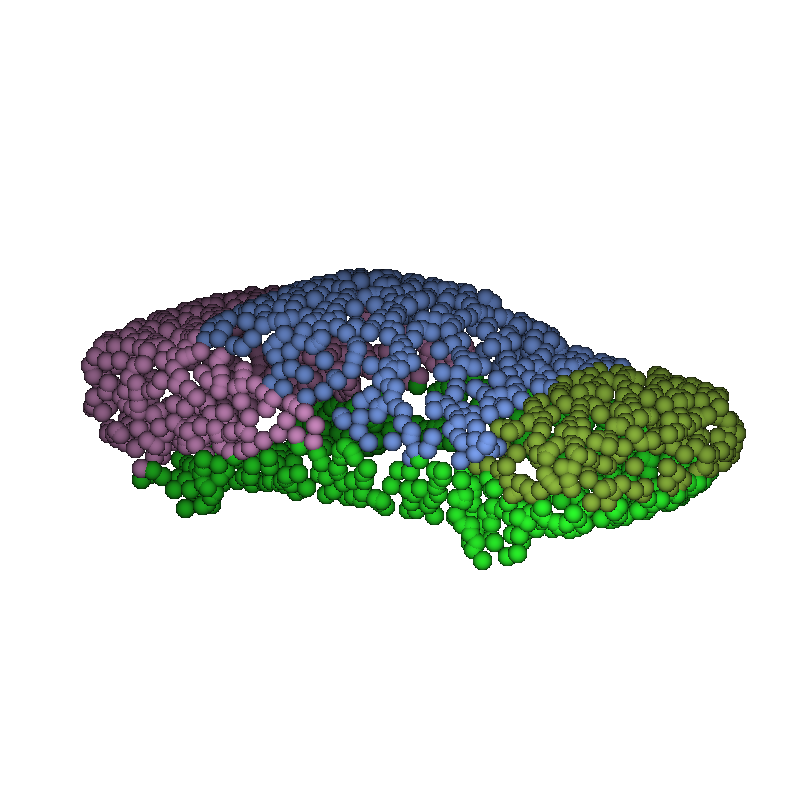}}
	{}%
	
\end{center}
\caption{Unsupervised segmentation examples on cars taken from the training data.}

\label{fig:seg_train_cars}
\end{figure*}


\begin{figure*}
\begin{center}
	\jsubfig{\includegraphics[height=3.49cm]{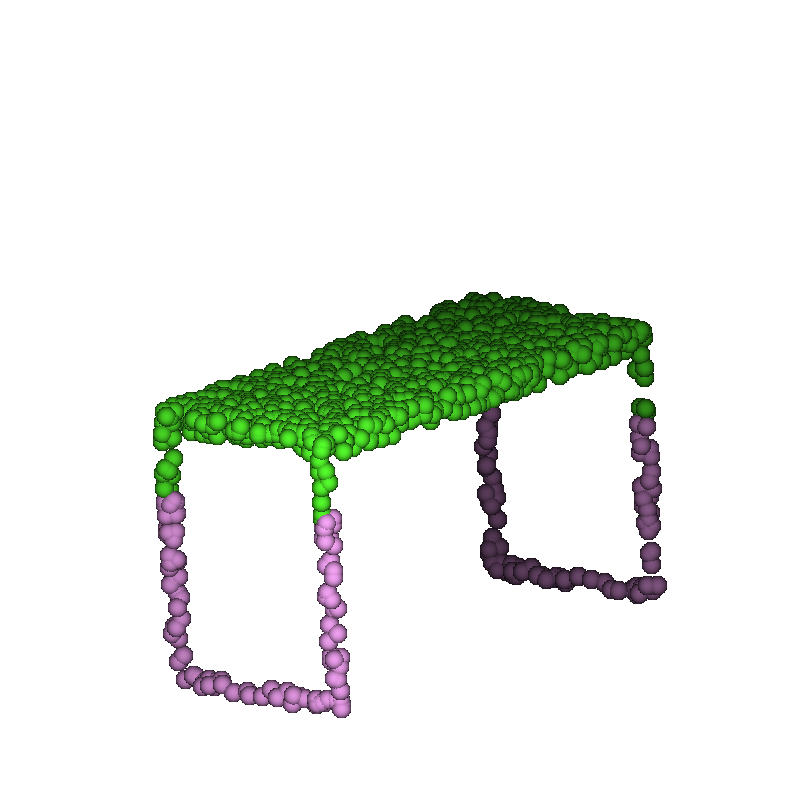}}
	{}%
	\hfill%
	\jsubfig{\includegraphics[height=3.49cm]{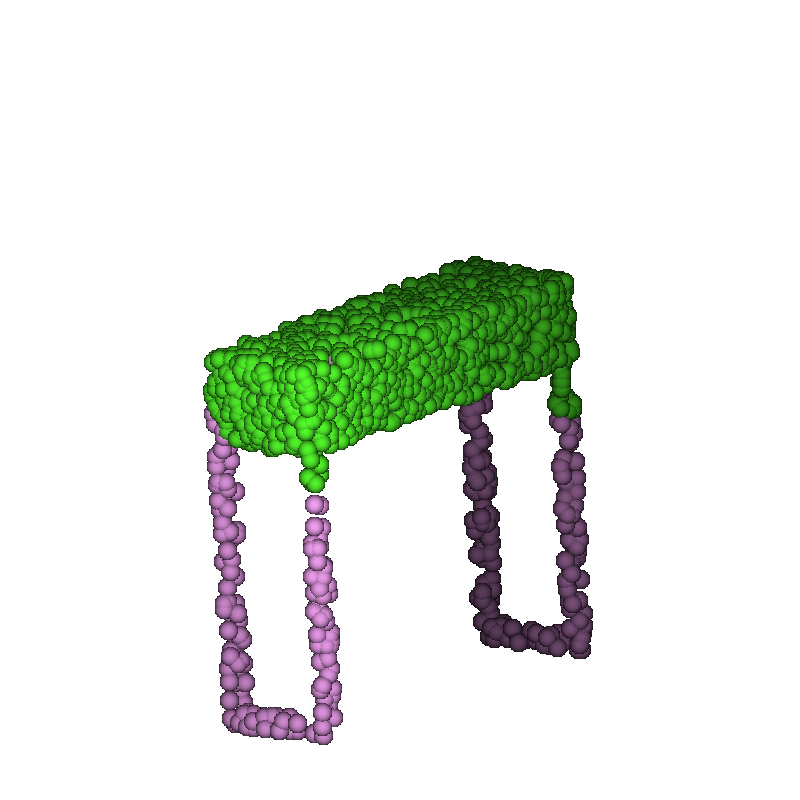}}
	{}%
	\hfill%
	\jsubfig{\includegraphics[height=3.49cm]{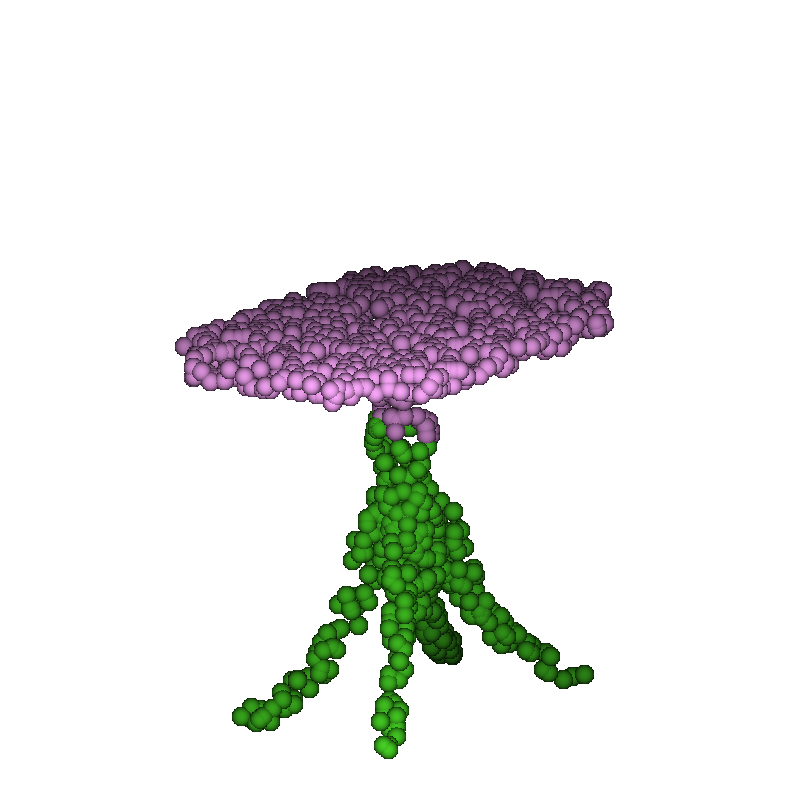}}
	{}%
	\hfill%
	\jsubfig{\includegraphics[height=3.49cm]{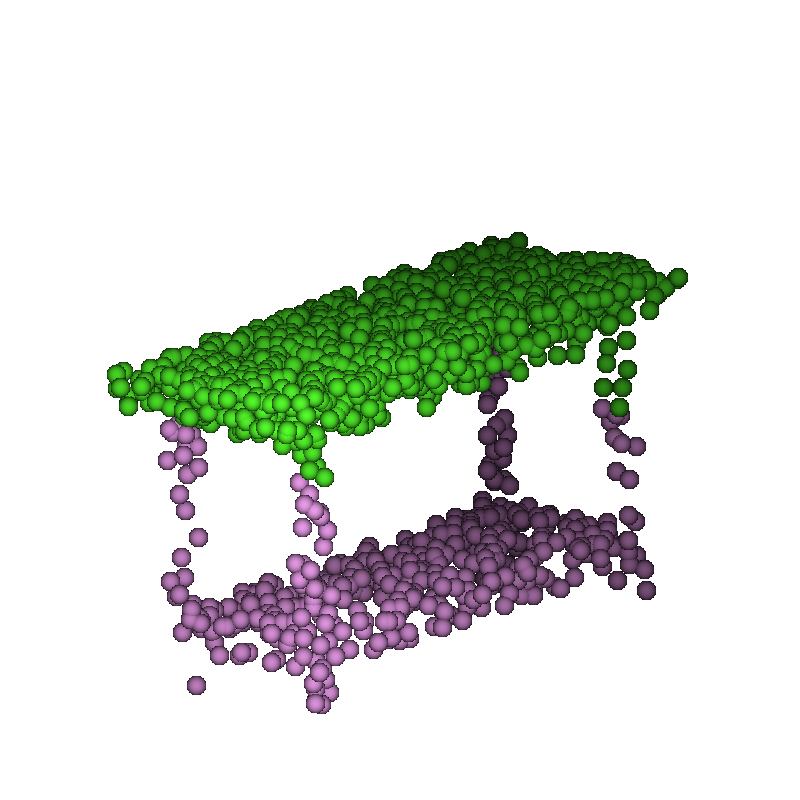}}
	{}%
\\
\vspace{6pt}
	\jsubfig{\includegraphics[height=3.49cm]{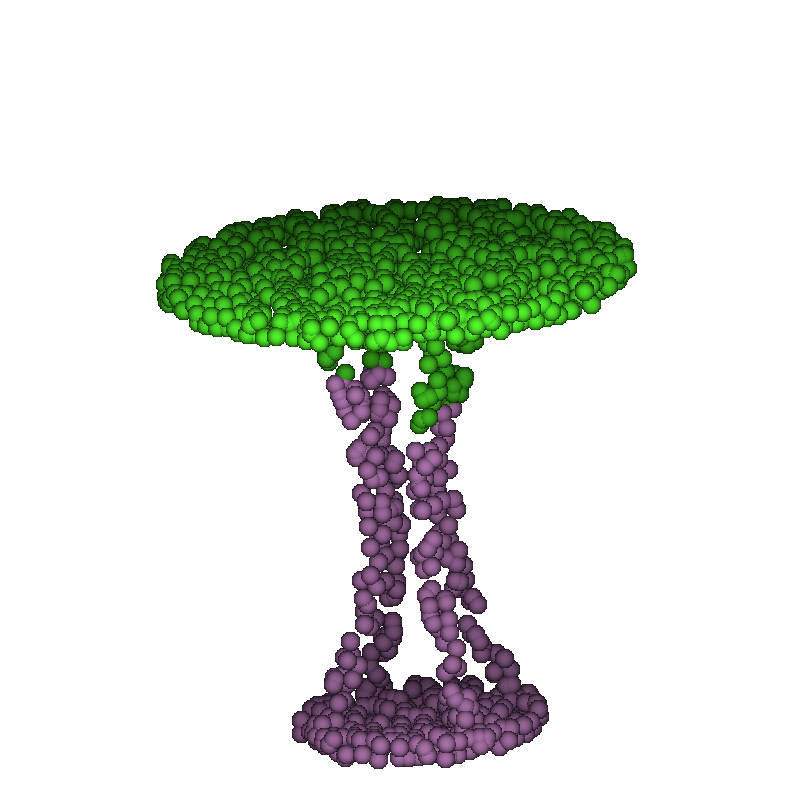}}
	{}%
	\hfill%
	\jsubfig{\includegraphics[height=3.49cm]{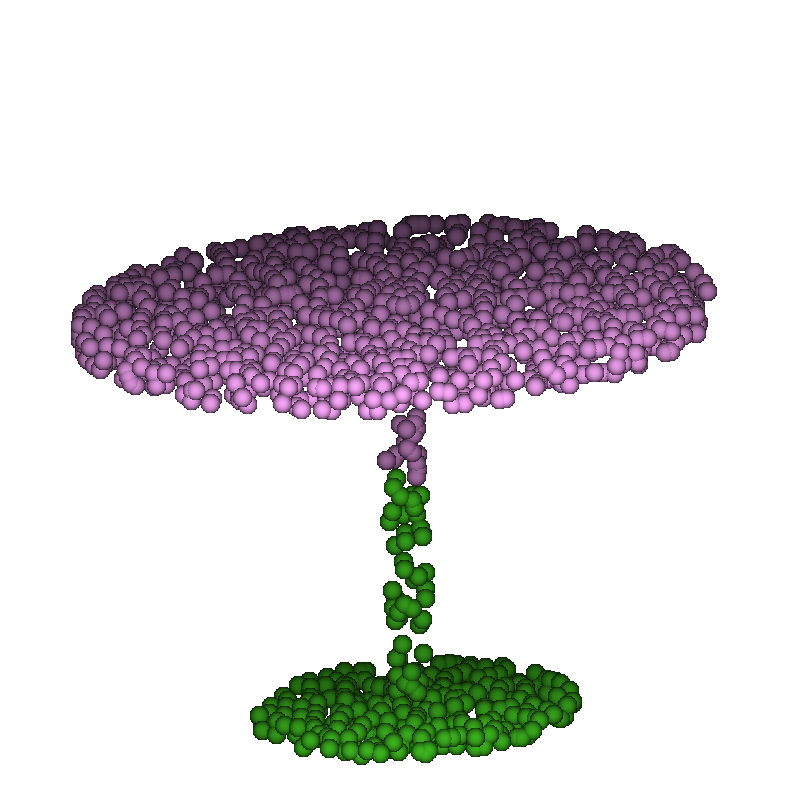}}
	{}%
	\hfill%
	\jsubfig{\includegraphics[height=3.49cm]{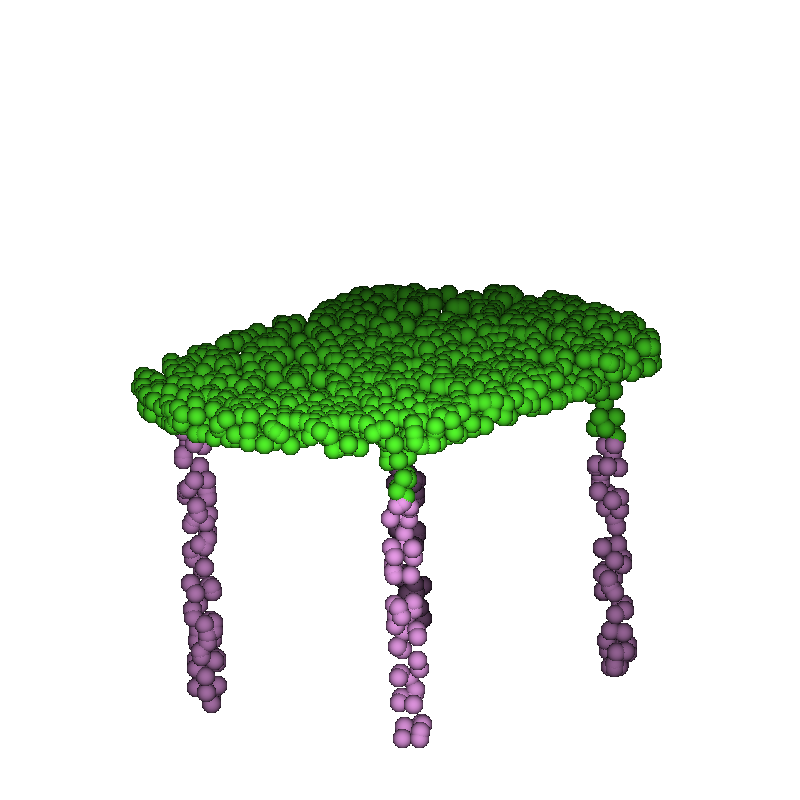}}
	{}%
	\hfill%
	\jsubfig{\includegraphics[height=3.49cm]{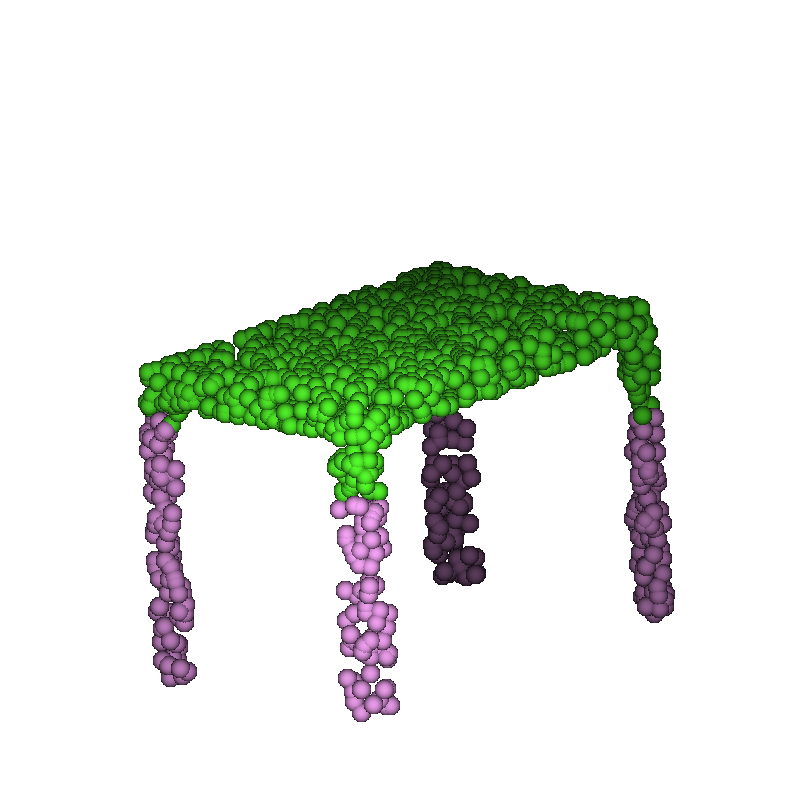}}
	{}%
\\
\vspace{6pt}
	\jsubfig{\includegraphics[height=3.49cm]{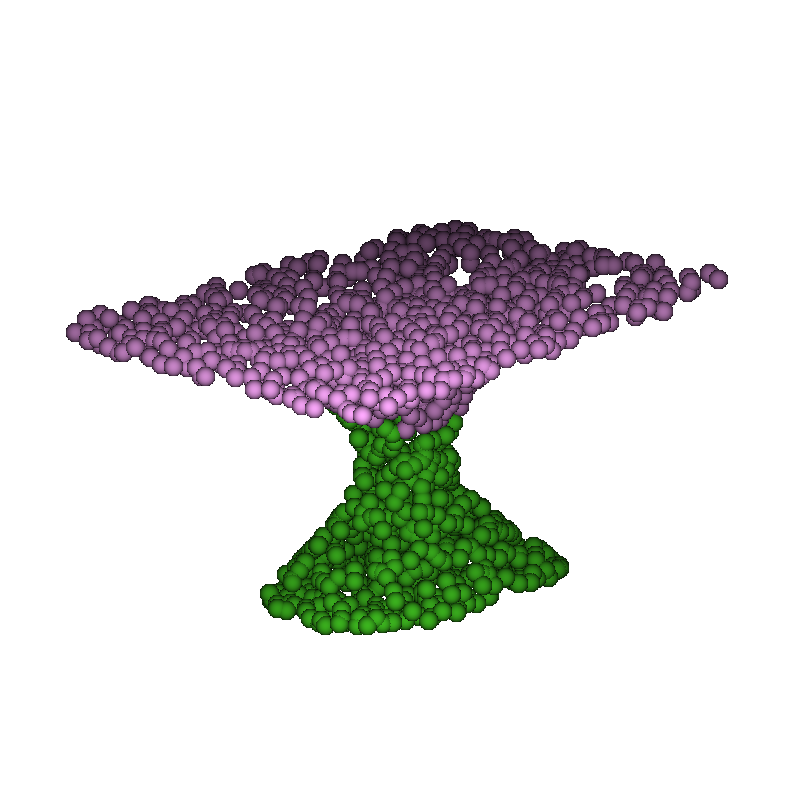}}
	{}%
	\hfill%
	\jsubfig{\includegraphics[height=3.49cm]{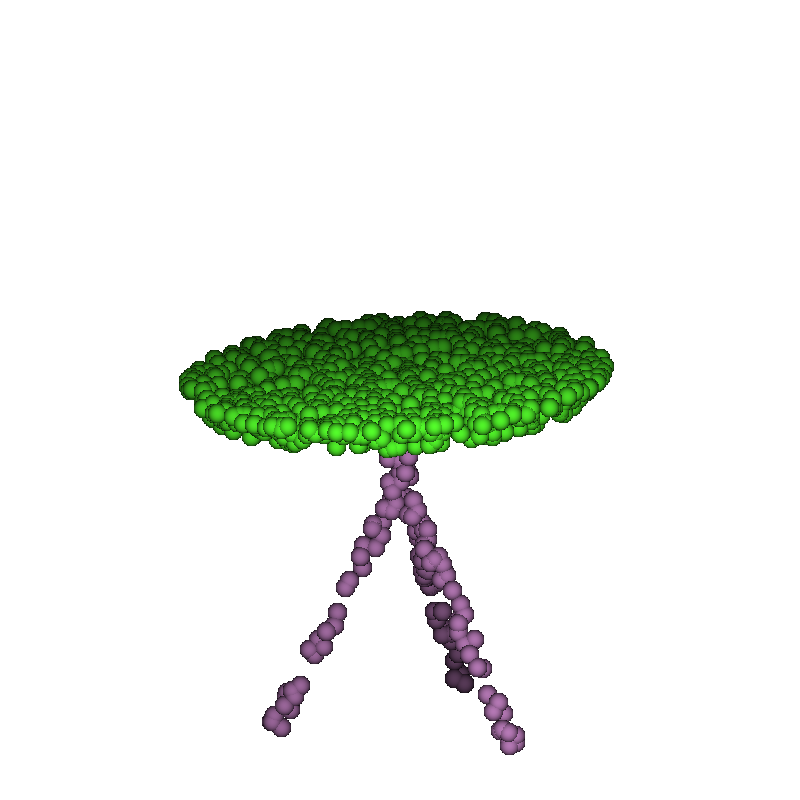}}
	{}%
	\hfill%
	\jsubfig{\includegraphics[height=3.49cm]{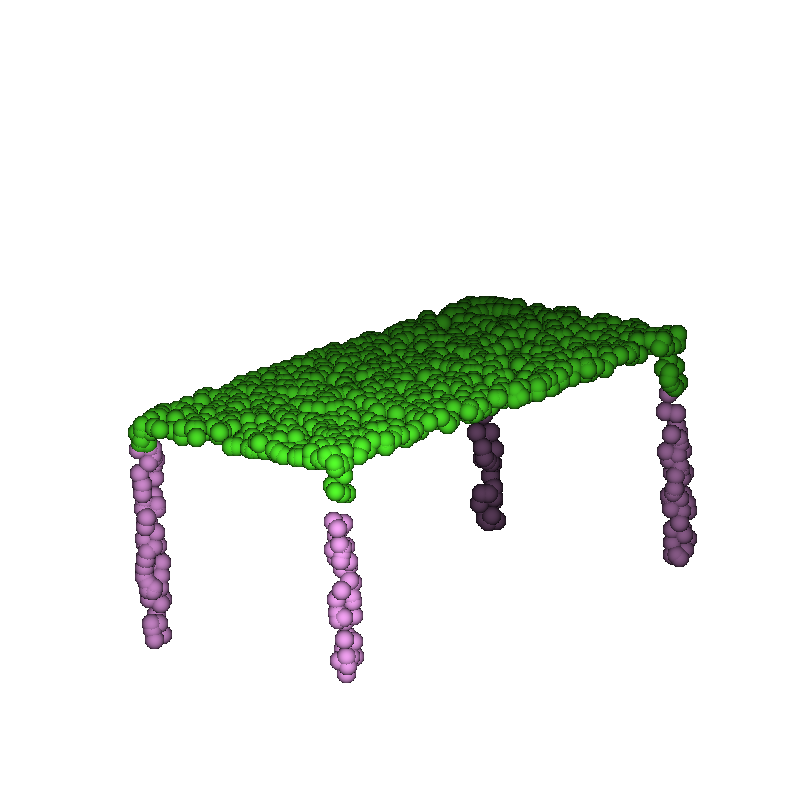}}
	{}%
	\hfill%
	\jsubfig{\includegraphics[height=3.49cm]{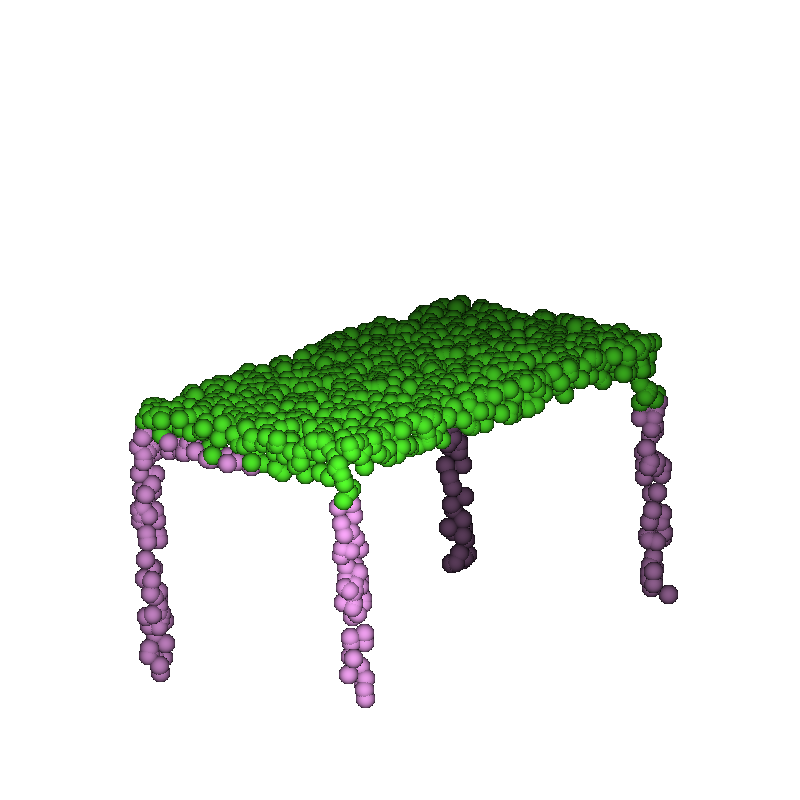}}
	{}%
\\
\vspace{6pt}
	\jsubfig{\includegraphics[height=3.49cm]{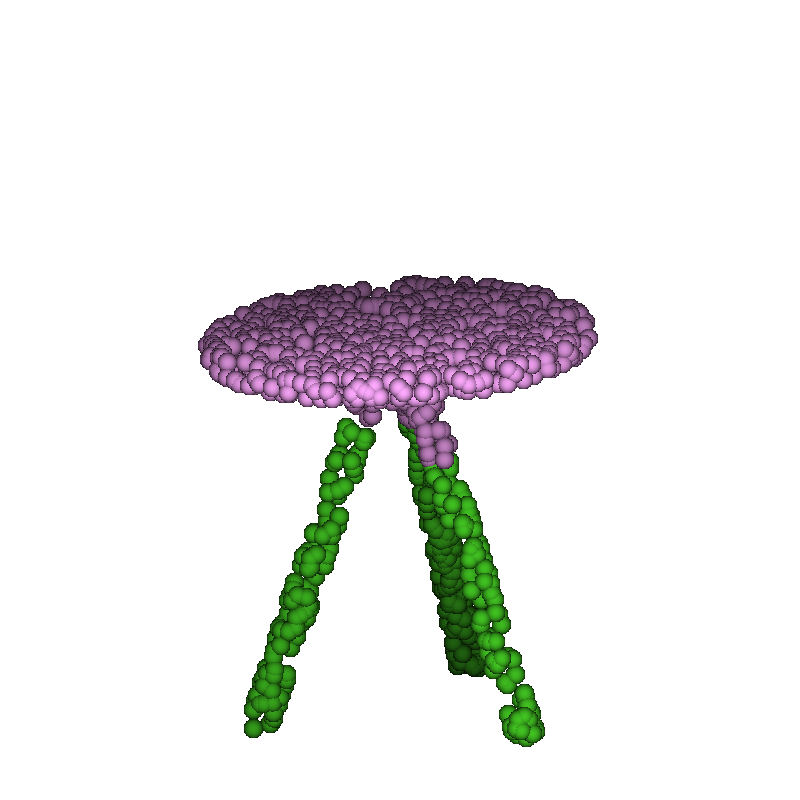}}
	{}%
	\hfill%
	\jsubfig{\includegraphics[height=3.49cm]{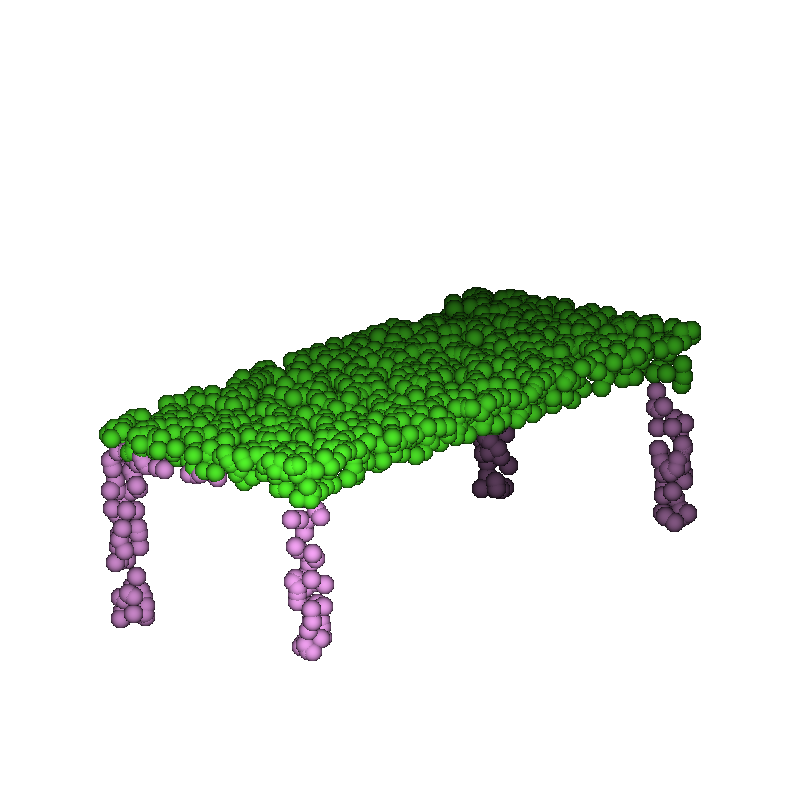}}
	{}%
	\hfill%
	\jsubfig{\includegraphics[height=3.49cm]{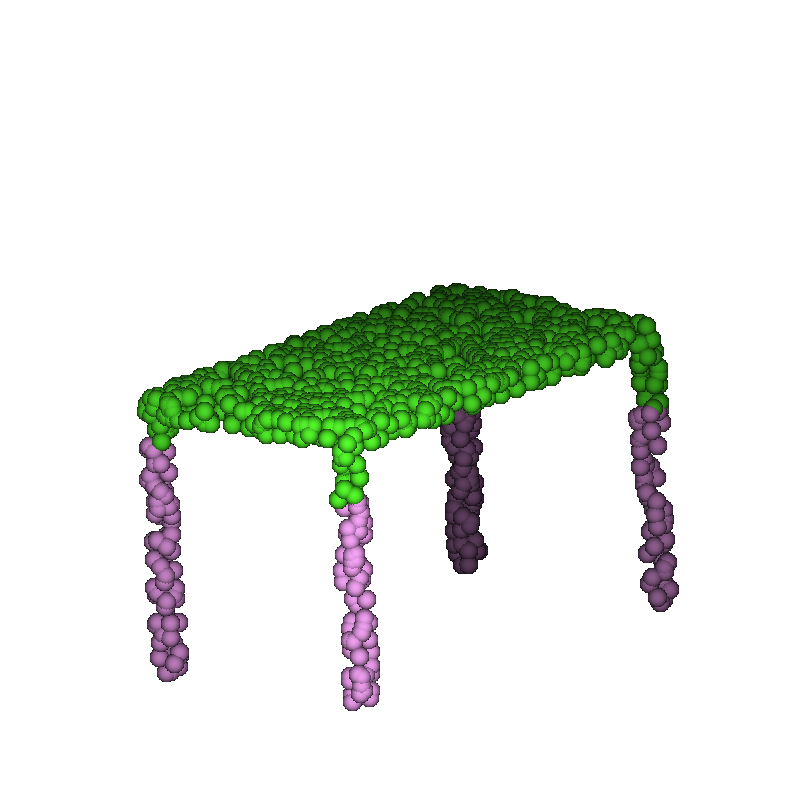}}
	{}%
	\hfill%
	\jsubfig{\includegraphics[height=3.49cm]{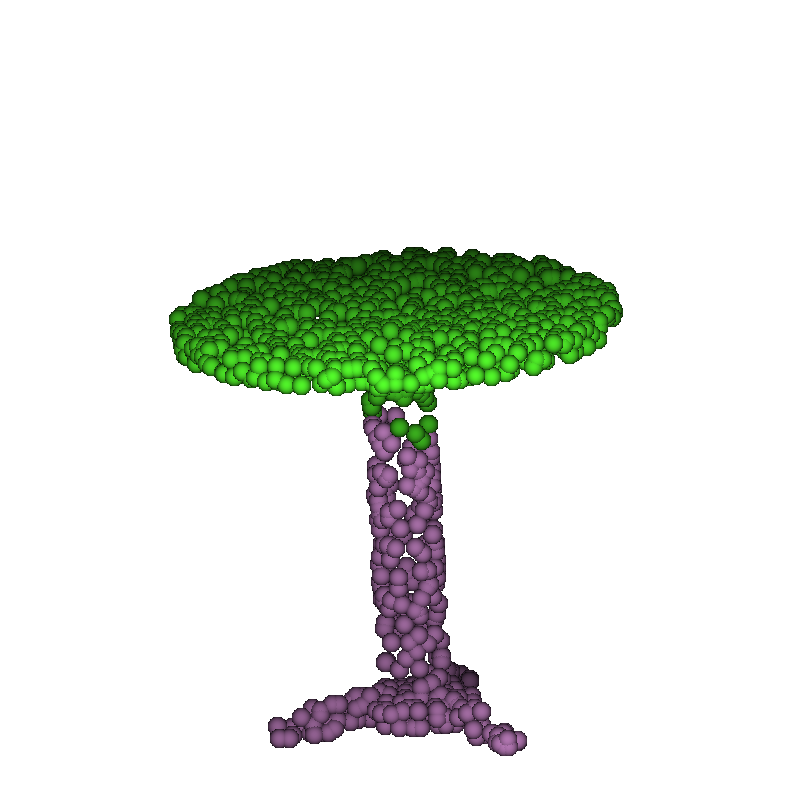}}
	{}%

\end{center}
\caption{Unsupervised segmentation examples on tables taken from the training data.}

\label{fig:seg_train_tables}
\end{figure*}

\end{document}